\documentclass[11pt,letterpaper]{article}
\usepackage[margin=2cm,headheight=14pt,headsep=10pt,footskip=18pt]{geometry}
\usepackage[T1]{fontenc}
\usepackage[sc]{mathpazo}
\usepackage{amsmath,amssymb,bm}
\usepackage{graphicx,booktabs,tabularx,array,longtable}
\usepackage[numbers,sort&compress]{natbib}
\usepackage{enumitem,titlesec,microtype,xcolor}
\usepackage[font=small,labelfont=bf,skip=5pt]{caption}
\usepackage{float,placeins,fancyhdr,needspace}
\usepackage{tocloft}
\usepackage{pdfpages}
\usepackage{tikz}
\usetikzlibrary{arrows.meta,positioning,calc}
\usepackage[hidelinks,pdfpagelabels=true]{hyperref}
\usepackage{bookmark}
\titleformat{\part}[display]{\Large\bfseries\color{navy}}{\partname \thepart}{4pt}{}
\renewcommand{\thepart}{\arabic{part}}
\titlespacing*{\part}{0pt}{0pt}{12pt}
\titlespacing*{\section}{0pt}{1.5ex}{0.8ex}
\titlespacing*{\subsection}{0pt}{1ex}{0.5ex}
\titlespacing*{\subsubsection}{0pt}{0.8ex}{0.4ex}
\titleformat{\section}{\large\bfseries}{\thesection.}{0.5em}{}
\titleformat{\subsection}{\normalsize\bfseries}{\thesubsection}{0.5em}{}
\titleformat{\subsubsection}{\normalsize\bfseries}{\thesubsubsection}{0.5em}{}
\setlist{leftmargin=1.4em,itemsep=2pt,topsep=3pt}
\definecolor{navy}{RGB}{27,56,79}
\definecolor{pale}{RGB}{240,245,248}
\newcommand{\vct}[1]{\bm{#1}}
\newcommand{\mat}[1]{\mathbf{#1}}
\newcommand{\dd}{\,\mathrm d}

\newcolumntype{Y}{>{\raggedright\arraybackslash}X}
\newcolumntype{P}[1]{>{\raggedright\arraybackslash}p{#1}}
\hypersetup{pdftitle={},pdfauthor={},pdfsubject={},pdfkeywords={},pdfcreator={},pdfproducer={},bookmarksnumbered=true,bookmarksopen=true,bookmarksopenlevel=1}
\pdftrailerid{}
\renewcommand{\partname}{Chapter}
\RequirePackage{etoolbox}
\numberwithin{figure}{section}
\numberwithin{table}{section}
\numberwithin{equation}{section}
\definecolor{coverivory}{RGB}{247,245,236}
\definecolor{covergold}{RGB}{207,175,112}
\definecolor{coverblue}{RGB}{119,179,202}
\definecolor{covermuted}{RGB}{191,208,216}

\definecolor{headingink}{RGB}{42,69,86}
\titleformat{\section}[hang]
  {\normalfont\fontsize{20}{24}\selectfont\bfseries\color{navy}\raggedright
   \hyphenpenalty=10000\exhyphenpenalty=10000}
  {\thesection.}{0.65em}{}
  [\vspace{5pt}{\color{navy!35}\titlerule[0.6pt]}]
\titlespacing*{\section}{0pt}{24pt plus 3pt minus 2pt}{12pt}

\titleformat{\subsection}[hang]
  {\normalfont\fontsize{14}{17}\selectfont\bfseries\color{navy}\raggedright
   \hyphenpenalty=10000\exhyphenpenalty=10000}
  {\thesubsection}{0.65em}{}
\titlespacing*{\subsection}{0pt}{16pt plus 2pt minus 1pt}{7pt}

\titleformat{\subsubsection}[hang]
  {\normalfont\fontsize{12}{15}\selectfont\bfseries\color{headingink}\raggedright
   \hyphenpenalty=10000\exhyphenpenalty=10000}
  {\thesubsubsection}{0.6em}{}
\titlespacing*{\subsubsection}{0pt}{11pt plus 2pt minus 1pt}{4pt}

\titleformat{\paragraph}[hang]
  {\normalfont\normalsize\bfseries\color{headingink}\raggedright}
  {\theparagraph}{0.6em}{}
\titlespacing*{\paragraph}{0pt}{10pt plus 1pt}{3pt}

\AtBeginEnvironment{thebibliography}{\setlength{\bibsep}{0pt}}
\apptocmd{\appendix}{\setlength{\defaultaddspace}{3pt}}{}{}

\renewcommand{\sectionmark}[1]{}
\fancypagestyle{plain}{%
  \fancyhf{}\fancyfoot[C]{\small\thepage}%
}

\newcommand{\bsgoverviewcontact}{}
\fancypagestyle{bsgoverviewcontacts}{%
  \fancyfoot[L]{\small\bsgoverviewcontact}%
  \fancyfoot[C]{}%
  \fancyfoot[R]{\small\thepage}%
}

\titleformat{\part}[display]
  {\normalfont\bfseries\color{navy}}
  {\fontsize{13}{17}\selectfont\partname\ \thepart}
  {16pt}{\fontsize{32}{38}\selectfont}
\titlespacing*{\part}{0pt}{0pt}{25pt}

\newenvironment{bsgpartpage}[1]{%
  \def\bsgpartmark{#1}%
  \clearpage\thispagestyle{plain}%
  \vspace*{0.85in}%
  {\small\textsc{Bodyseismography}\par}%
  \vspace{0.7in}%
}{%
  \par\vfill
  \noindent\textcolor{covergold}{\rule{0.9in}{1.2pt}}\par
  \vspace{0.55in}\markboth{\bsgpartmark}{}\clearpage
}

\providecommand{\T}{\mathsf{T}}
\DeclareMathOperator{\Null}{Null}
\DeclareMathOperator{\Row}{Row}
\DeclareMathOperator{\Col}{Col}

\DeclareMathOperator{\tr}{tr}

\begin{document}
\hypersetup{pageanchor=false}
\begin{titlepage}
\thispagestyle{empty}
\thispdfpagelabel{Cover}
\begin{tikzpicture}[remember picture,overlay]
\fill[navy] (current page.south west) rectangle (current page.north east);
\begin{scope}[shift={(current page.south west)},x=1in,y=1in]
  \draw[covergold,line width=1.1pt] (0.72,10.14) -- (1.28,10.14);
  \node[anchor=west,text=coverivory,font=\fontsize{10}{13}\selectfont] at (1.42,10.14)
    {\textsc{Cardiovascular mechanics and sensing}};
  \node[anchor=north west,text=coverivory,font=\fontsize{41}{49}\selectfont] at (0.69,9.20)
    {Bodyseismography (BSG)};
  \node[anchor=north west,text=coverivory,font=\fontsize{17.5}{23}\selectfont] at (0.72,8.43)
    {Theoretical Foundation and Numerical Modeling};
  \draw[covergold,line width=0.65pt] (0.75,7.84) -- (7.76,7.84);
  \node[anchor=north west,text=covermuted,font=\fontsize{12}{18}\selectfont,
        text width=6.6in,align=left] at (0.72,7.48)
    {Unraveling the Cardiovascular Origins of Triaxial Body Vibrations\\Through Theoretical Analysis and Numerical Validation};

  \node[anchor=north west,text=coverivory,font=\fontsize{14}{19}\selectfont,
        text width=7.01in,align=left] at (0.75,6.50)
    {Yuanyuan Zhang\textsuperscript{1,*}, Yuyan Wu\textsuperscript{2,*}, Sahas Biddappa\textsuperscript{1}, Yida Zhang\textsuperscript{1},\\
     Jiayu Chen\textsuperscript{1}, Zixuan Zeng\textsuperscript{1}, Yixuan Li\textsuperscript{1}, Jiahui Li\textsuperscript{3}, Fei Dou\textsuperscript{3},\\
     Zhenlin An\textsuperscript{3}, Younghoon Kwon\textsuperscript{4}, Hae Young Noh\textsuperscript{2}, WenZhan Song\textsuperscript{1}};
  \node[anchor=north west,text=covermuted,font=\fontsize{9.5}{12.5}\selectfont,
        text width=7.01in,align=left] at (0.75,5.48)
    {\textsuperscript{1}School of Electrical and Computer Engineering, University of Georgia\\
     \textsuperscript{2}Department of Civil and Environmental Engineering, Stanford University\\
     \textsuperscript{3}School of Computing, University of Georgia\\
     \textsuperscript{4}Department of Biochemistry, University of Washington};

  \foreach \j in {0,...,12}{
    \draw[coverblue,opacity=0.70,line width=0.45pt]
      plot[domain=0:1,samples=100,variable=\t]
      ({0.73+7.04*\t},{2.10+0.053*\j+0.80*sin(245*\t-30+1.8*\j)*exp(-6*(\t-0.51)^2)});
    \draw[coverivory,opacity=0.38,line width=0.40pt]
      plot[domain=0:1,samples=100,variable=\t]
      ({0.73+7.04*\t},{2.32+0.050*\j+0.86*sin(260*\t+25+1.7*\j)*exp(-6*(\t-0.48)^2)});
    \draw[covergold,opacity=0.75,line width=0.42pt]
      plot[domain=0:1,samples=100,variable=\t]
      ({0.73+7.04*\t},{2.52+0.048*\j+0.78*sin(255*\t+80+1.7*\j)*exp(-6*(\t-0.48)^2)});
  }

  \node[anchor=west,text=covermuted,font=\fontsize{10.5}{14}\selectfont] at (0.75,1.65)
    {\textsc{Signals\quad /\quad Physics\quad /\quad Computational models}};
  \draw[coverivory,opacity=0.3,line width=0.4pt] (0.75,1.12) -- (7.76,1.12);
  \node[anchor=west,text=coverivory,font=\fontsize{10}{13}\selectfont] at (0.75,0.79)
    {\textsc{Technical white paper}};
  \node[anchor=east,text=covermuted,font=\fontsize{10}{13}\selectfont] at (7.76,0.79)
    {September 2026};
\end{scope}
\end{tikzpicture}
\null
\end{titlepage}

\pagenumbering{roman}
\hypersetup{pageanchor=true}
\markboth{Contents}{}
\pdfbookmark[0]{Contents}{book.contents}
\begingroup
\setlength{\parskip}{0pt}
\small\linespread{1.0}\selectfont
\setlength{\cftbeforepartskip}{12pt}
\setlength{\cftbeforesecskip}{4pt}
\setlength{\cftbeforesubsecskip}{1pt}
\setlength{\cftaftertoctitleskip}{16pt}
\tableofcontents
\endgroup
\clearpage

\markboth{Overview}{}
\section*{Overview}
\phantomsection
\addcontentsline{toc}{section}{Overview}
\renewcommand{\bsgoverviewcontact}{\textbf{Contact:} WenZhan Song (\href{mailto:wsong@uga.edu}{wsong@uga.edu}) and Yuanyuan Zhang (\href{mailto:yz52357@uga.edu}{yz52357@uga.edu}).\\
\textsuperscript{*}Yuanyuan Zhang and Yuyan Wu contributed equally and share first authorship.}
\thispagestyle{bsgoverviewcontacts}
Continuous cardiovascular monitoring can extend physiological assessment beyond isolated measurements by following changes in cardiac function and circulation during daily life. Rest and sleep offer sustained opportunities for such observation, motivating sensing systems that require little user interaction. Bodyseismography (BSG) supports this goal by recording body-generated bed vibrations without attaching sensors to the skin \citep{song2024engagement}. These vibrations carry the mechanical signatures of cardiac contraction and blood movement. Understanding those signatures creates a path from detecting heartbeats to studying stroke volume (SV), cardiac output (CO), and blood pressure (BP) through unobtrusive monitoring.

\begin{figure}[H]
\centering
\includegraphics[width=\linewidth]{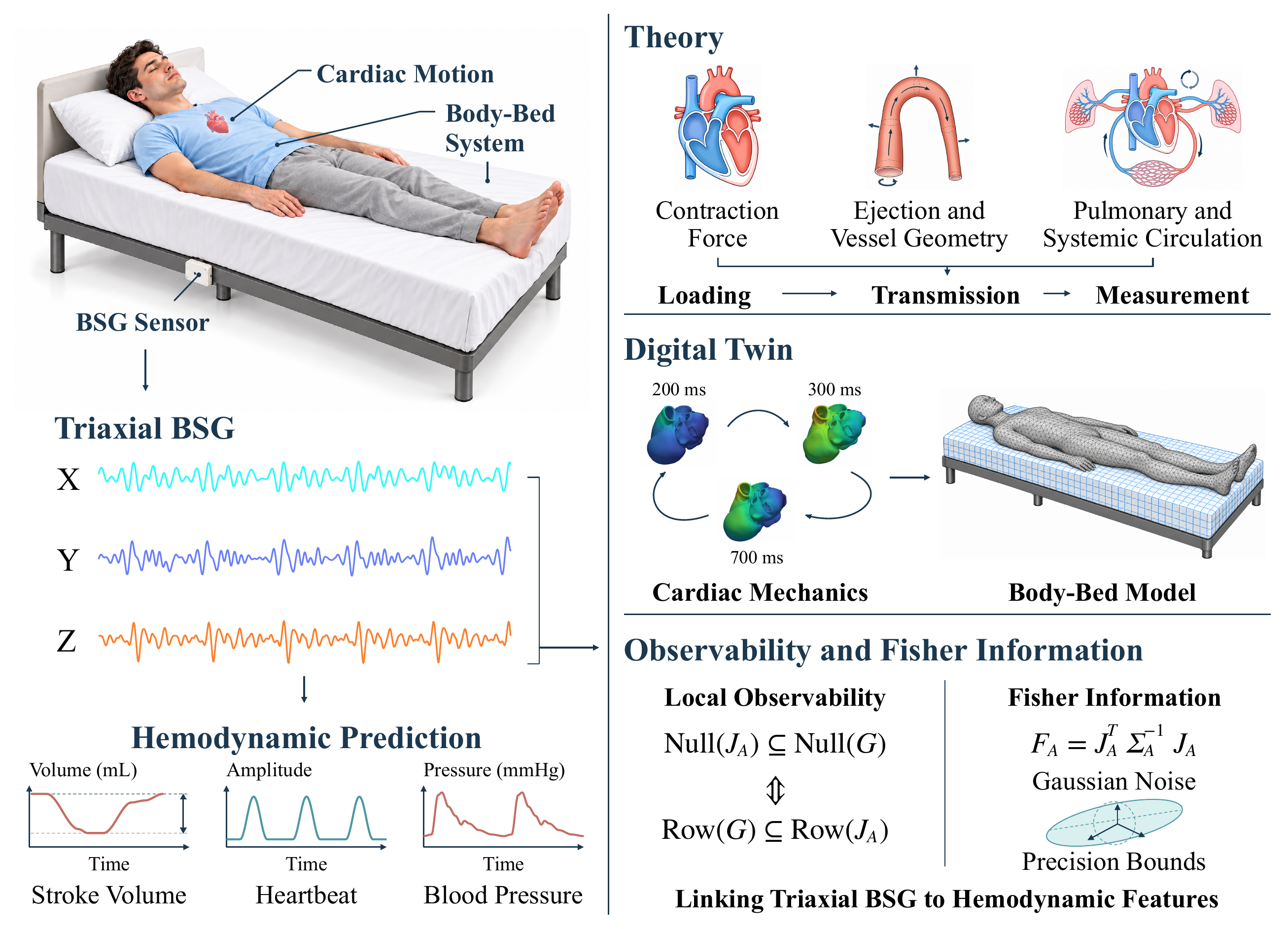}
\caption{Overview of This White Paper.}
\label{fig:chain}
\end{figure}

Realizing this potential requires a physical explanation of how cardiovascular activity becomes a measured BSG waveform. Cardiac deformation and vascular recoil contribute through different directions and transmission paths, while body and bed mechanics reshape their combined response \citep{zhang2026phybp}. A change in the recorded waveform can therefore reflect both physiology and its mechanical transmission. This white paper develops that explanation through theoretical analysis and numerical simulation, connecting BSG source generation, combined cardiac contributions, joint triaxial information for hemodynamic monitoring, and body and bed transfer functions. Figure~\ref{fig:chain} organizes this framework through measurement, theory, a digital twin, and observability analysis.

Figure~\ref{fig:chain} connects BSG measurement, cardiovascular mechanisms, numerical modeling, and target observability. At left, cardiovascular motion propagates through the body and mattress to a bed-mounted sensor, which records signed vibrations along $X$, $Y$, and $Z$. These channels preserve directional differences for studying ventricular volume change, heartbeat timing, and arterial pressure, with SV and heart rate together determining CO. At upper right, cardiac deformation, blood ejection, and circulation form \emph{one coupled cardiovascular state} with \emph{two mechanical loading pathways}: cardiac contraction and vascular recoil. These pathways connect local chest motion measured by seismocardiography (SCG) and whole-body recoil measured by ballistocardiography (BCG) to the combined BSG response. At middle right, the digital twin couples cardiac deformation to chamber pressure, volume, and flow, then predicts how the resulting excitation travels through the body and bed to the sensor. This framework allows source dynamics and transmission properties to be examined individually and together. At lower right, the observation Jacobian $J_A$ and target Jacobian $G$ describe how BSG channels and physiological targets respond to model-parameter changes. Local target observability requires every parameter change invisible to BSG to leave the target unchanged to first order. Adding sensing axes can remove such invisible directions and strengthen target constraints. Fisher information weights observation sensitivities by noise covariance $\Sigma_A$ and supports local precision bounds under Gaussian noise with fixed positive-definite covariance. The ellipsoid illustrates uncertainty across parameter directions. The prediction plots, body--bed mesh, and uncertainty ellipsoid are schematic, and the representative BSG traces include schematic extensions. The cardiac renderings are adapted from our Heart in the Loop model \citep{zhang2027heartintheloop}.

Our seismic-sensing research established this measurement approach through sleep monitoring \citep{li2018nonintrusive,clemente2019contactless}, the Helena system \citep{clemente2020helena,valero2021health}, and passive estimation of heart and respiratory rates \citep{li2020smart,song2024engagement}. Bed occupancy detection \citep{song2024self}, in-bed posture classification \citep{song2025multi}, and detection of sleep apnea, snoring, and breath-holding \citep{song2026bodyseismography,song2026attention} extend its use across sleep and respiratory monitoring.

Cardiovascular studies address BP estimation from bed vibrations \citep{song2024real,zhang2025blood}, physics-constrained triaxial estimation \citep{zhang2026phybp,zhang2026triaxial}, continuous arterial BP waveform reconstruction \citep{wu2026contact}, electrocardiogram monitoring \citep{zhang2026bsg2ecg,zhang2024radarODE,zhang2024radarODE-MTL,zhang2025high}, and tracking of SV changes \citep{zhang2027vibrosv}. Self-supervised denoising and explainable peak detection support cardiac feature extraction \citep{chen2025selfdenoiser,li2026peak,zhang2023overview,zhang2025horcrux,zhang2025robust}. Related platforms extend vibration sensing to human and animal health monitoring \citep{song2024poster}, CageDot activity and vital-sign monitoring \citep{pitafi2024cagedot,pitafi2025contactless}, and controlled validation with a programmable cardiorespiratory testbed \citep{pitafi2026testbed}.

These applications motivate a physical account of how cardiovascular changes become measurable BSG features. The analysis develops the following connected findings and their implications for BSG monitoring.

\textbf{A Physical Foundation for BSG.} Theoretical analysis and numerical simulation connect cardiovascular dynamics, mechanical excitation, body--bed propagation, and sensor observation. This framework explains how BSG is generated and provides a mechanical basis for reconstructing its waveform and interpreting its physiological content.

\textbf{Combined SCG and BCG Contributions to BSG.} Local cardiac deformation and whole-body recoil arise from a shared cardiovascular state and reach the bed through distinct mechanical paths. Their directions and timing combine in BSG, linking local cardiac motion and blood-momentum exchange within the recorded signal.

\textbf{Joint Triaxial Information for Hemodynamic Monitoring.} Sensing directions carry complementary physiological information, and combining them strengthens constraints on target changes under the modeled uncertainty. The framework investigates when joint triaxial information is necessary for SV, CO, and BP monitoring, retaining SCG alone and single-axis or triaxial BCG alone as comparisons. This connects the value of directional information to the target and measurement conditions.

\textbf{Body and Bed Transfer Functions Shape BSG.} Transmission changes waveform amplitude, timing, and spectral content, as well as the relationship between BSG features and hemodynamics. Accounting for body and bed transfer functions is therefore central to interpreting physiological changes from BSG.

\Needspace{4\baselineskip}
The three chapters develop this argument from measurement to mechanism and numerical evidence:
\begin{description}[leftmargin=0pt,labelsep=0pt,style=nextline]
\item[{Chapter 1. Preliminary}]
This chapter establishes what BSG measures and why the three sensing directions matter. It introduces the bed-mounted sensing architecture, sensor response, and measured waveforms, then relates SV, CO, and BP to cardiac timing, ejection, and vascular loading. The SCG and BCG discussions explain the complementary physiological mechanisms contributing to BSG. Measured BSG waveforms show cardiac features shifting between directions and mattress-dependent changes in amplitude and timing. These observations motivate treating cardiovascular source generation and body and bed transmission together when interpreting BSG.
\item[{Chapter 2. Theoretical Foundation of BSG}]
This chapter develops the physical and mathematical basis of BSG through physiological parameters, hemodynamic targets, vascular recoil, cardiac-contraction loading, body--bed mechanics, signal observation, and target observability. The common cardiovascular state generates both loading pathways and the physiological targets, so their relationships follow from shared dynamics. Numerical comparisons of single-axis, dual-axis, and triaxial observations show that the benefit of combining directions depends on the physiological target and the modeled uncertainty. Joint axes strengthen separation from mechanical variation, while concurrent physiological changes can imitate target-related responses. This analysis identifies how directional information, calibration, and uncertainty determine which changes in SV, CO, and BP can be distinguished from BSG.
\item[{Chapter 3. Numerical Modeling and Validation of BSG}]
This chapter turns the framework into computational models of cardiac source generation, mechanical excitation, and body--bed transmission. Four-chamber mechanics couples wall deformation to circulation and reproduces reference chamber pressure--volume behavior and local wall-motion timing. The excitation model maps cardiac motion and vascular reactions to mechanical inputs. Analytical, finite-element, and finite impulse response models then examine how those inputs become recorded BSG. Calibrated mechanical models reconstruct measured BSG morphology, connecting the physical description of source generation and transmission to the observed waveforms.

Controlled simulations show that horizontal and vertical features retain complementary information about SV after transmission. In a representative bed configuration, combining these features tracks the direction of simulated SV changes more consistently than either feature alone. Transmission models fitted to measured recordings also show that body and bed mechanics can alter, and sometimes reverse, a feature's association with simulated SV. Together, these findings explain how joint directional information and body and bed transfer functions support the physiological interpretation of BSG.
\end{description}

\clearpage
\pagenumbering{arabic}
\setcounter{part}{0}
\begin{bsgpartpage}{Preliminary}
\part{Preliminary}
\label{part:preliminary}
The measured signal is the starting point. This chapter defines triaxial BSG, introduces its sensing architecture, and relates its SCG and BCG components to cardiovascular mechanics and hemodynamics.

The examples identify directional morphology and physiological feature relationships that support target-specific estimation. They provide the context for the analytical derivations in Sections~\ref{sec:framework} and~\ref{sec:observability} and the computational representations in Chapter~\ref{part:dt}.
\end{bsgpartpage}

\section{Triaxial BSG and Its Hemodynamic Information}
\label{sec:definitions}
\subsection{Definition, Sensor Characteristics, and Measured Waveforms}
\label{sec:prelim_sensor}
\textbf{Bodyseismography (BSG), as used in this white paper, is the simultaneous recording of body-generated mechanical vibrations along three orthogonal axes through a supporting bed.} Its cardiac component contains the transmitted contributions associated with both traditional seismocardiography (SCG) and ballistocardiography (BCG). SCG describes local chest-wall motion, conventionally acceleration, associated with cardiac mechanical activity. BCG describes cardiovascular-induced whole-body recoil, measured as displacement, velocity, acceleration, or support force according to the apparatus. A bed-mounted sensor records the combined mechanical response after transmission through the body, contact interfaces, mattress, and bed \citep{zhang2026phybp}.

The SCG and BCG contributions share a cardiovascular source and reach the bed through different transmission operators. Their overlapping mechanical responses form the recorded BSG, together with respiratory motion and other disturbances. Separating the two contributions requires information about their source and transmission, rather than an arithmetic addition of independently measured chest and whole-body traces.

Use bed coordinates $X$ across the bed, $Y$ along the head--foot direction, and $Z$ normal to the mattress. The measured signal is the signed vector $\vct y_{\mathrm{BSG}}(t)=[y_X(t),y_Y(t),y_Z(t)]^{\mathsf T}$. Anatomical, bed, and hardware axes are related by explicit rotations. Historical longitudinal BCG is associated with $Y$. Vertical-only BCG here denotes the bed-normal $Z$ measurement, which differs from gravity-vertical on an inclined bed.

\begin{figure}[!htbp]
\centering
\includegraphics[width=0.85\linewidth]{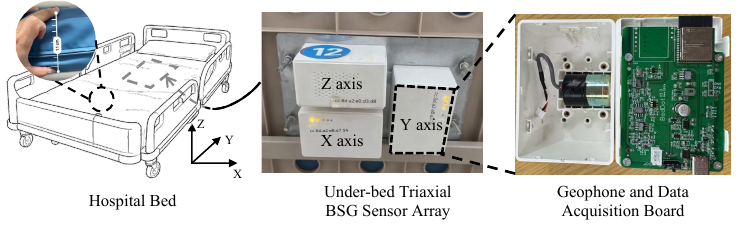}
\caption{Triaxial BSG Acquisition. Left: bed and sensing axes. Center: sensor array. Right: sensing element and acquisition board \citep{zhang2026phybp}.}
\label{fig:device}
\end{figure}

\textbf{Our BSG Device.} Our BSG device uses three RacoTech RGI-4.5Hz Low Frequency Geophone Elements \citep{rgi45hz}, mounted beneath the bed frame at different orientations to record three motion components. The sensing elements and data acquisition board form a sensing system with a lower cutoff frequency of approximately 0.7\,Hz \citep{zhang2026phybp}. Our previous contactless-monitoring work describes the sensing, digitization, and onboard-computing architecture \citep{song2024engagement}. Figure~\ref{fig:device} (right) illustrates the sensing element and acquisition board of our BSG device. Relative motion between a coil and a magnetic core generates a voltage. The acquisition board conditions and digitizes this small signal. Within the calibrated velocity-response band, the output follows bed velocity. This makes our BSG device sensitive to the weak cardiac vibrations that remain after propagation through the body and mattress. The under-bed placement requires no sensor attachment to the skin, while the three signed channels retain directional morphology that a single projection can suppress.

\subsubsection{Operating Principle and Advantages for BCG Sensing}
\label{sec:geophone_principle}
\begin{figure}[!htbp]
\centering
\includegraphics[width=0.90\linewidth]{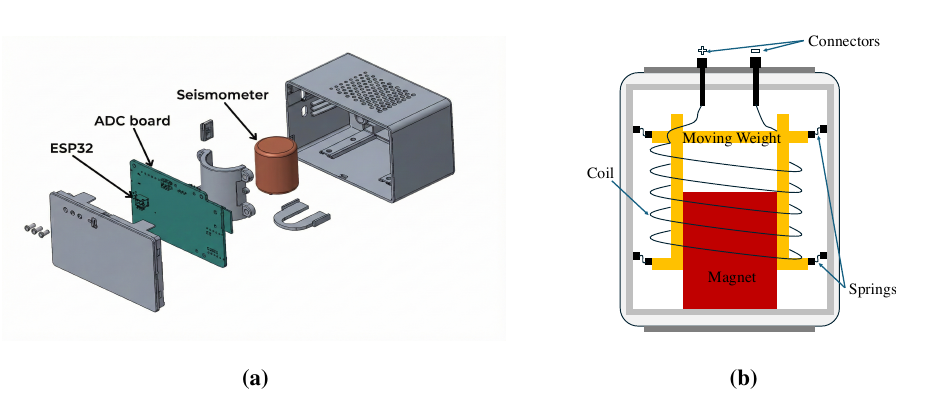}
\caption{BSG Device and Sensing Principle. (a) Device components. (b) Moving-coil conversion of bed vibration to voltage \citep{zhang2026bsg2ecg}.}
\label{fig:geophone_principle}
\end{figure}

\textbf{From Bed Motion to Voltage.} Figure~\ref{fig:geophone_principle}(b) shows the inertial mechanism of our BSG device. The housing and magnet move with the bed frame, while the coil and its supporting mass are suspended by springs. Inertia causes the suspended assembly to lag behind the housing. Relative motion of the coil through the magnetic field induces a voltage, whose magnitude is proportional to their relative velocity. The coil connections preserve the sign of that motion \citep{desy_geophones}. For a linear single-axis approximation,
\[
 m\ddot q+c\dot q+kq=-m\ddot x_b,
 \qquad e=-G\dot q,
\]
where $x_b$ is housing displacement along the sensing axis, $q$ is suspended-mass displacement relative to the housing, $m$ is its effective mass, $k$ is spring stiffness, $c$ includes mechanical and electrical damping, and $G$ is the electromagnetic conversion constant. The voltage polarity is chosen so that $e$ approaches $G\dot x_b$ in the approximately flat velocity-response region. The acquisition board amplifies, filters, and digitizes this voltage. The springs and damping shape the response near mechanical resonance, so the calibrated response of the complete device determines which bed-motion components are retained \citep{desy_geophones,song2024engagement}.

\begin{figure}[!htbp]
\centering
\includegraphics[width=\linewidth]{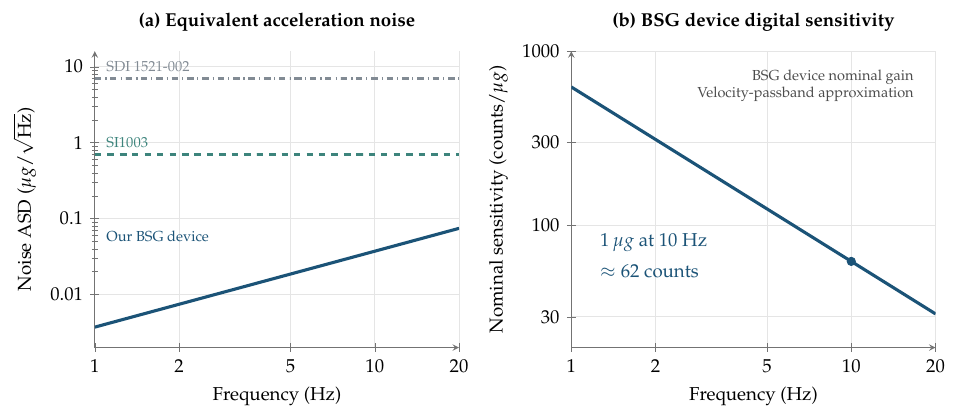}
\caption{Sensor Noise and Sensitivity. Over 1--20\,Hz, our BSG device shows (a) lower equivalent acceleration noise than the two MEMS accelerometers and (b) increasing digital sensitivity at lower frequencies \citep{safran_si1003,sdi1521}.}
\label{fig:sensor_characteristics}
\end{figure}

\textbf{Why This Architecture Suits Weak BCG Motion.} After cardiac recoil passes through the body and mattress, the bed-frame vibration can be extremely small. Within the calibrated velocity-response band, sinusoidal motion satisfies $|v|=|a|/(2\pi f)$. Our BSG device therefore emphasizes slow dynamic motion relative to an acceleration-proportional readout for a fixed acceleration amplitude. Its practical advantage over the two microelectromechanical systems (MEMS) accelerometers considered here is the lower equivalent input-noise estimate in the cardiac band, quantified in Figure~\ref{fig:sensor_characteristics} \citep{song2024engagement,safran_si1003,sdi1521}. The analytical noise comparison uses the spectral models specified below.

\textbf{Installation and Directional Information.} Pressure-sensitive and piezoelectric BCG films sense local loading or deformation at the mattress interface. For example, a published piezoelectric-film system places the sensing strip beneath the chest and adjusts its position to each participant's height \citep{li2024sleepbcg}. Load-cell systems instead measure forces transmitted through a supporting structure \citep{albukhari2019bed}. Our BSG device attaches to the bed frame and senses its motion without requiring a film beneath a particular body location or a sensor in the bed's load-bearing path. Three oriented channels also preserve signed transverse, longitudinal, and normal components, rather than reducing the observation to one support-force or pressure trace \citep{zhang2026phybp,zhang2026bsg2ecg}. These features favor unobtrusive monitoring and recovery of directional cardiac morphology. Mechanical coupling, environmental vibration, and calibrated frequency response determine how this directional information appears in the measured signal.

\textbf{Digital Sensitivity of Our BSG Device.} Our BSG device produces a larger output for a lower-frequency acceleration of the same magnitude, within its velocity passband. For sinusoidal motion, $|v|=|a|/(2\pi f)$. An end-to-end velocity sensitivity $K_v$ therefore gives an acceleration-equivalent digital sensitivity $K_v g_0/(2\pi f)$ in counts/$g$, where $g_0=9.80665\,\mathrm{m/s^2}$. The nominal digital sensitivity of our BSG device is approximately $4.00\times10^8$ counts/(m/s), with a 24-bit converter, 100 samples/s, and estimated $-3$\,dB limits of 0.7--44\,Hz. Under the nominal passband approximation, a $1\,\mu g$ component at 10\,Hz produces about 62 counts. Figure~\ref{fig:sensor_characteristics}(b) shows this frequency-dependent gain for our BSG device. End-to-end calibration relates digital counts to bed motion, with analog noise and filtering setting the effective resolution.

\textbf{Instrument Noise Comparison.} Low instrument noise is valuable for BSG because body--bed transmission can strongly attenuate cardiac motion. Figure~\ref{fig:sensor_characteristics}(a) compares analytical noise models for our BSG device and two MEMS accelerometers over a common 1--20\,Hz band, using the BSG specifications and published MEMS specifications \citep{safran_si1003,sdi1521}. The BSG model assumes white velocity noise inferred from a root-mean-square (RMS) detection threshold of $0.08\,\mu\mathrm{m/s}$, specified as 10\,dB above the noise. The MEMS models use typical white-noise densities of 0.7 and $7\,\mu g/\sqrt{\mathrm{Hz}}$ for SI1003 and SDI 1521-002, respectively. The model of our BSG device gives approximately $0.19\,\mu g$ RMS equivalent acceleration noise. The two MEMS white-noise models give approximately $3.1$ and $30.5\,\mu g$ RMS. Under these spectral assumptions, the modeled equivalent acceleration noise is lower for our BSG device than for either MEMS readout. This supports its use for small dynamic motion in the cardiac band, while SI1003 and SDI 1521-002 provide direct, DC-capable acceleration measurement. Device-level noise characterization includes the acquisition board and calibration. Recorded signal-to-noise ratio (SNR) additionally reflects environmental vibration and body--bed transmission.

\textbf{Frequency Response and Three-Axis Information.} The nominal 4.5\,Hz mechanical resonance of our BSG device \citep{rgi45hz} and the conditioned system passband describe different stages of the instrument. Electronic conditioning can extend the low-frequency dynamic response, with calibrated gain and phase defining the usable respiratory and cardiac bands. The $1/f$ sensitivity relation applies within this dynamic passband and excludes zero frequency, so removal of a high-pass stage alone does not provide DC acceleration measurement. Three-axis recording adds geometrical information by preserving transverse, longitudinal, and normal motion, including cardiac features that weaken on another axis. \textbf{Calibrated dynamic response and directional coverage provide the sensing basis for our BSG measurement.} The analytical target-recovery criteria in Section~\ref{sec:observability} incorporate calibrated noise and directional observation.

\begin{figure}[!htbp]
\centering
\includegraphics[width=\linewidth]{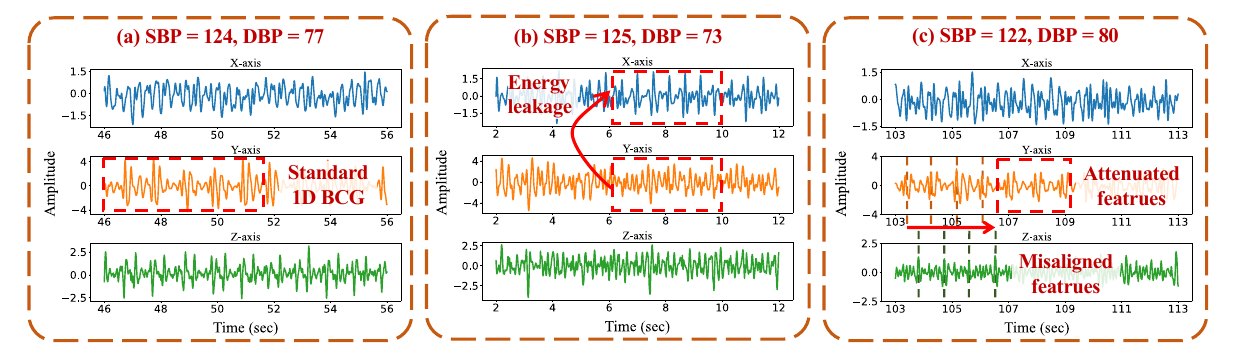}
\caption{Directional BSG Features. (a) Longitudinal morphology. (b) Transverse features. (c) Mattress-dependent attenuation and timing shifts \citep{zhang2026phybp}. Blood pressure is in mmHg.}
\label{fig:prelim_bsg}
\end{figure}

\textbf{Triaxial Recording Retains Directional Cardiac Features.} Our BSG device records signed $X$, $Y$, and $Z$ motion to preserve cardiac morphology as its dominant direction changes. Under similar blood-pressure conditions, Figure~\ref{fig:prelim_bsg}(a) shows conventional BCG features concentrated in the longitudinal $Y$ channel, while Figure~\ref{fig:prelim_bsg}(b) shows features distributed toward the transverse $X$ channel and less distinct in $Y$. Posture can redistribute cardiac vibration among all three directions. Figure~\ref{fig:prelim_bsg}(c) shows attenuation and timing shifts with a soft/thick mattress. \textbf{Directional recording captures features that a longitudinal projection can suppress, while the transmission model accounts for changes imposed by the supporting bed.} Together, these observations motivate modeling cardiac excitation and body--bed propagation within the same measurement framework \citep{zhang2026phybp}.

\subsection{Hemodynamic Quantities Represented by the Mechanical Signal}
Hemodynamics concerns blood pressure (BP), flow, and their interaction with cardiac and vascular mechanics. Stroke volume (SV) is the volume ejected per beat. Cardiac output (CO) is the volume delivered per unit time. With consistent time units and cycle-averaged quantities,
\begin{equation}
\mathrm{SV}=\int_{\mathrm{ejection}} Q_{\mathrm{AV}}(t)\dd t,
\qquad \mathrm{CO}=\mathrm{HR}\times\mathrm{SV},
\qquad \mathrm{MAP}-\mathrm{CVP}\simeq\mathrm{CO}\times\mathrm{SVR}.
\label{eq:prelim_hemodynamics}
\end{equation}
Here $Q_{\mathrm{AV}}$ is forward aortic flow, HR is heart rate, MAP is mean arterial pressure, CVP is central venous pressure, and SVR is systemic vascular resistance (also termed total peripheral resistance, TPR). The first expression assumes negligible aortic regurgitation. The final relation is a mean-flow approximation. Systolic and diastolic pressures additionally depend on pulsatile flow, compliance, and wave reflection. Recoil reflects momentum exchange, CO measures volume flow, and BP depends on vascular loading as well as flow. A physiological model links these distinct quantities when interpreting BSG as an indirect measurement of hemodynamics \citep{zhang2026phybp}.

\subsection{SCG: Cardiac Timing, Stroke Volume, and Diastolic Function}
Myocardial contraction, chamber deformation, valve transitions, and blood acceleration generate vibrations transmitted to the chest. SCG fiducials can mark the timing of mechanical events, but their correspondence depends on location, axis, and waveform morphology \citep{sorensen2018,taebi2017}. Aortic-opening and aortic-closure landmarks provide an estimate of ventricular ejection time (VET). An electrocardiogram (ECG) reference additionally permits estimation of the pre-ejection period (PEP), from ventricular electrical activation to aortic opening. These intervals locate pressure development and ejection in time. Estimating ejected volume additionally requires information about the flow magnitude and waveform.

\begin{figure}[!htbp]
\centering
\input{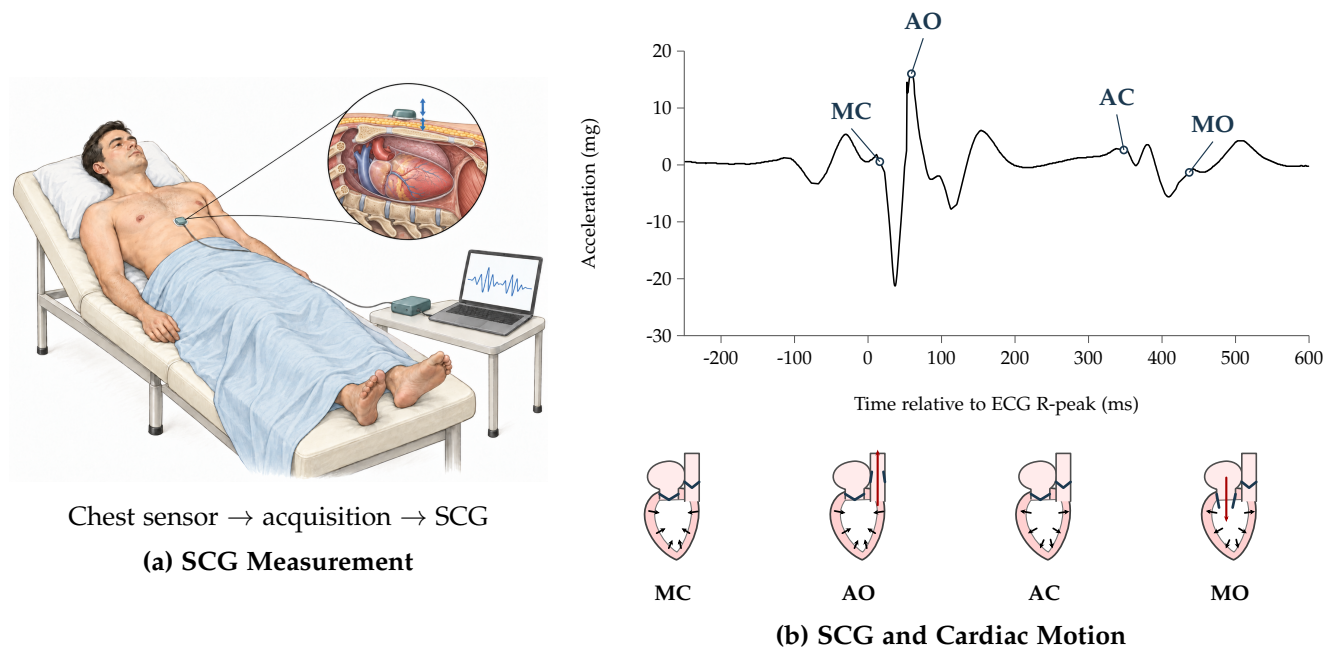}
\caption{SCG and Cardiac Events. (a) Chest measurement. (b) Mean SCG with valve-event landmarks and cardiac motion \citep{sorensen2018}.}
\label{fig:prelim_scg}
\end{figure}

Figure~\ref{fig:prelim_scg} links the chest measurement to four valve events: mitral closure (MC), aortic opening (AO), aortic closure (AC), and mitral opening (MO). In the idealized left-heart cycle, MC--AO is isovolumetric contraction, AO--AC is ejection, and AC--MO is isovolumetric relaxation. Filling follows MO. The black arrows summarize the radial component of ventricular wall motion during contraction and during relaxation and filling. The isovolumetric phases involve changes in shape and wall deformation at approximately constant cavity volume. Red arrows show blood flow through the open valve. SCG measures the resulting transmitted chest acceleration.

The displayed curve is a published mean SCG obtained by time-aligning and resampling recordings from 34 subjects. The four labels identify SCG landmarks associated with the valve events. Earlier features are associated with atrial contraction and peak atrial inflow, while additional features during ejection and filling are associated with peak systolic outflow and early ventricular filling \citep{sorensen2018}.

Combined electrical and mechanical timing features support baseline SV estimation in congenital heart disease. In 45 patients, chest SCG and ECG were recorded around routine cardiovascular magnetic resonance imaging. A ridge-regression model using predominantly systolic timing features achieved a held-out SV root-mean-square error (RMSE) of 11.48\,mL and $R^2=0.76$, with a reported percentage error of 28\%. Combining electrical and mechanical features outperformed either modality alone. The evaluation concerns baseline SV across patients, establishing a population-specific multifeature relationship \citep{ganti2022sv}.

SCG also carries information about relaxation and filling. In a study recruiting 45 healthy subjects, diastolic SCG peak-to-peak amplitude correlated with early diastolic mitral annular velocity $e'$ measured by tissue Doppler echocardiography ($r=0.713$, $p<0.001$). Usable SCG data were available from 38 subjects supine and 39 during $30^\circ$ head-up tilt. Tilting reduced reference SV and altered the $E_d$ amplitude, while other diastolic amplitudes did not change significantly. These findings identify relaxation and loading as physiological contributors to diastolic SCG morphology \citep{agam2022diastolic}.

The distinction follows from the ejection integral in Eq.~\eqref{eq:prelim_hemodynamics}: timing constrains the interval of integration, while flow magnitude and shape determine its area. Preload, contractility, afterload, and heart rate can change these quantities together. Chest geometry and tissue transmission further affect the measured amplitude. Quantitative SV or BP estimation therefore combines timing and amplitude features with the physiological and transmission variables that determine them.

\subsection{BCG: Blood Momentum, Ejection, and Arterial Loading}
A major source of confusion in interpreting BSG is the use of BCG as a generic label for cardiac mechanical signals recorded through a bed. Recent studies apply this label to recordings used for heart and respiration monitoring \citep{carlson2021bed,li2024sleepbcg}, sleep staging \citep{perezavila2024sleep}, sleep apnea screening \citep{zhang2025apnea}, atrial fibrillation detection \citep{sandelin2025af}, and BP estimation \citep{xing2024bedbp}. Many such systems directly sense cardiac-synchronous changes in normal contact force or mattress deformation, which correspond to vertical loading on a horizontal bed. Li et al.\ describe chest and abdominal vibrations compressing an under-mattress piezoelectric film, while Zhang et al.\ explicitly describe the measured quantity as vertical force generated by heartbeat and breathing \citep{li2024sleepbcg,zhang2025apnea}. Equating these contact-force recordings with whole-body recoil conflates the measured quantity with its physiological mechanism.

The ballistic root of \emph{ballistocardiography} points to ejection and recoil. In the classical framework established by Starr et al.\ in 1939, BCG records the body's mechanical reaction to blood ejection and subsequent changes in blood momentum \citep{starr1939ballistocardiogram}. Their suspended-table apparatus recorded longitudinal motion relative to a stationary frame, relating that motion to ejection and the redirection of blood in the great vessels \citep{starr1940standards}. This physical definition concerns recoil of the body, observed as displacement, velocity, acceleration, or the corresponding support reaction force according to the apparatus \citep{albukhari2019bed}. Starr's later essays developed this Newtonian interpretation of cardiac performance, relating the recorded recoil to the dynamics of blood ejection \citep{starr1954essay,starr1965progress,starr1967newtonian}.

\begin{figure}[!htbp]
\centering
\includegraphics[width=\linewidth]{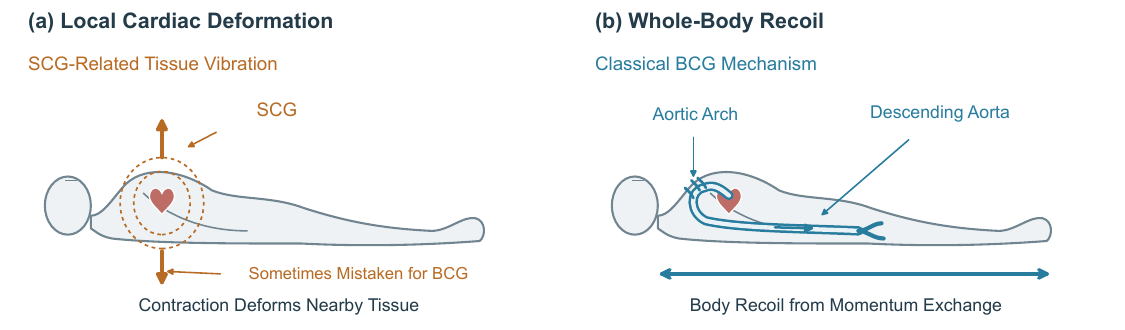}
\caption{Local Vibration and Body Recoil. (a) Contraction deforms nearby tissue. (b) Aortic blood momentum produces whole-body recoil \citep{sorensen2018,starr1940standards,kim2016mechanism}.}
\label{fig:bcg_definition}
\end{figure}

\textbf{Whole-body recoil and local cardiac vibration share a cardiac origin but differ in their force-generation and transmission mechanisms.} Myocardial contraction deforms the heart and surrounding tissues, producing local chest-wall motion associated with SCG \citep{sorensen2018,taebi2017}. The BCG recoil pathway instead reflects momentum exchange as blood accelerates, decelerates, and changes direction through the cardiovascular system, including the ascending aorta, aortic arch, and descending aorta \citep{kim2016mechanism}. Figure~\ref{fig:bcg_definition} contrasts these two mechanisms.

The systolic I--J--K morphology of classical BCG is associated with ejection-driven momentum exchange, with later components reflecting subsequent deceleration and filling \citep{starr1940standards,starr1954genesis,rabineau2021closedloop}. Circulation models formalize these links while accounting for vascular geometry and boundary conditions \citep{guidoboni2019cardiovascular,rabineau2021closedloop}.

\begin{figure}[!htbp]
\centering
\begin{minipage}[b]{0.54\linewidth}
\centering\includegraphics[width=\linewidth]{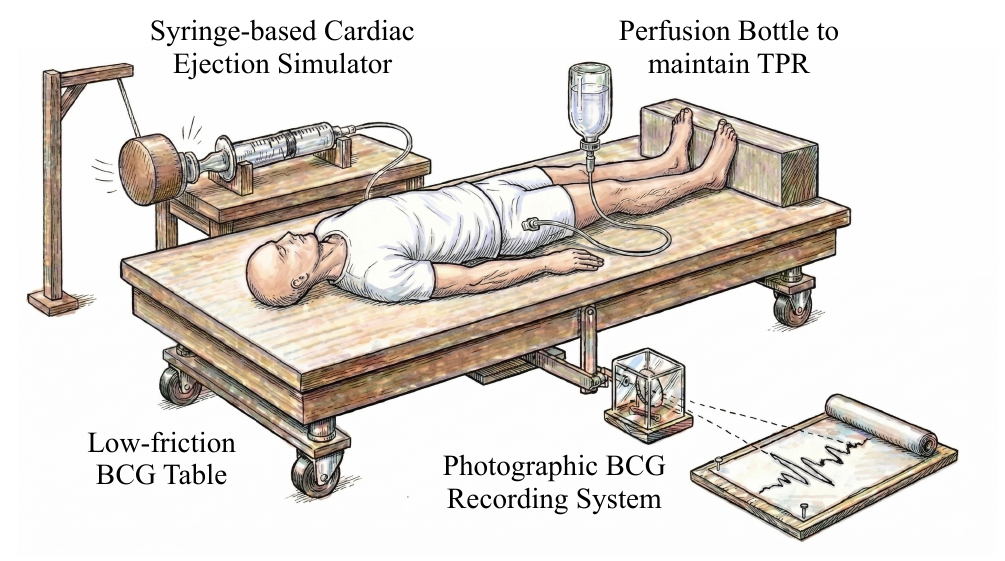}\par
\small\textbf{(a) Controlled Ejection and Body Recoil}
\end{minipage}\hfill
\begin{minipage}[b]{0.43\linewidth}
\centering\includegraphics[width=\linewidth]{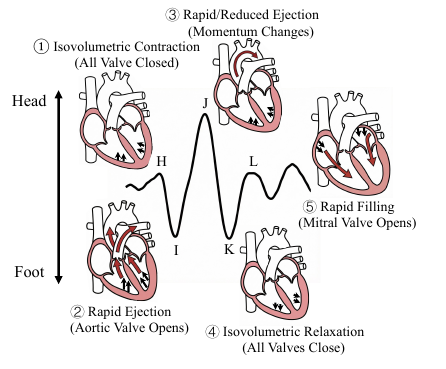}\par
\small\textbf{(b) Conventional BCG Morphology}
\end{minipage}
\caption{BCG Reflects Cardiovascular Momentum Exchange. (a) Simulated ejection produces body recoil. (b) Conventional BCG morphology across the cardiac cycle \citep{zhang2026phybp,starr1959studies}.}
\label{fig:prelim_bcg}
\end{figure}

\begin{figure}[!htbp]
\centering
\includegraphics[width=\linewidth]{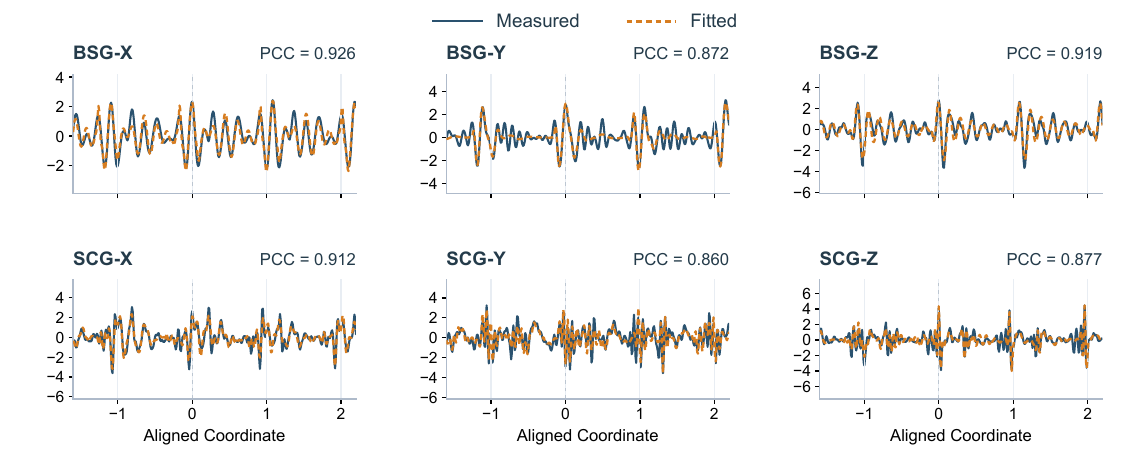}
\caption{BSG and SCG. Measured (solid blue) and fitted (dashed orange) waveforms along $X$, $Y$, and $Z$. BSG is shown above and SCG below. PCC denotes the Pearson correlation coefficient between measured and fitted waveforms over the displayed interval.}
\label{fig:signal_comparison}
\end{figure}

Starr's necropsy experiments reproduced ejection mechanically while recording body motion, linking BCG morphology to the strength and time course of ejection under controlled cardiac and peripheral conditions \citep{starr1950standardization,starr1953peripheral,starr1954genesis,starr1959studies}. The same experimental program tested BCG-based SV estimates against injected volumes and examined the relationship of pulse pressure to SV and CO \citep{starr1955strokevolume,starr1954pulsepressure}. Its analysis of higher cardiac dynamics distinguished ejected volume from the velocity and acceleration of ejection \citep{starr1958dynamics}.

Respiratory experiments connected cyclic BCG amplitude changes to altered cardiac filling \citep{starr1946respiratory}. Longitudinal studies related BCG changes to aging and disease development, and examined associations of baseline amplitude or waveform abnormalities with subsequent morbidity and mortality \citep{starr1952aging,starr1961twentyyear,starr1964prognostic}. A discussion of smoking-induced ECG and BCG changes further considered the response of these signals to cardiovascular stress \citep{starr1960smoking}. Together, these studies connect recoil dynamics to physiological loading and clinical course, providing historical context for interpreting the BCG contribution to BSG.

The pressure-gradient model of Kim et al.\ further predicts a J-wave amplitude approximately proportional to aortic pulse pressure multiplied by the descending aortic cross-sectional area. The model used pressure recordings from 21 cardiac surgical patients without concurrent BCG measurements. This supplies a forward mechanical explanation for pressure-related BCG changes whose magnitude also depends on vascular geometry and pressure-wave timing \citep{kim2016mechanism}.

BCG amplitude tracks exercise-associated changes in CO in human recordings. Using a fiber-optic micro-vibration sensing system (FO-MVSS), Zhan et al.\ compared resting and postexercise measurements in 11 participants with normal cardiac function. Increases in I--J amplitude were positively associated with increases in echocardiographic CO (Spearman $r_s=0.923$, $p<0.001$). The target here is CO, which combines heart rate and SV through $\mathrm{CO}=\mathrm{HR}\times\mathrm{SV}$ and varies with the accompanying exercise response \citep{zhan2024cardiac}.

Simultaneous SCG and BCG measurements reveal feature-specific sensitivity to exercise loading. In 15 patients with heart failure and reduced ejection fraction, submaximal exercise increased reference SV and CO and several BCG energy/power measures. The exercise-to-rest CO ratio correlated with the BCG maximum-power ratio ($r=0.65$, $p=0.008$) and the SCG maximum-power ratio ($r=0.54$, $p=0.04$), while the corresponding SV-ratio relationships were nonsignificant. These results distinguish mechanical features associated with changing CO from those needed to estimate its separate HR and SV components \citep{herkert2021kinocardiograph}.

The force history depends on the rate of momentum change, whereas SV depends on the time integral of flow. Two ejection waveforms can have the same integral and different acceleration profiles. Arterial geometry, compliance, peripheral loading, and the support system further alter the observed peaks. These dependencies make ejection shape, vascular loading, and support mechanics the variables that connect BCG morphology to SV and pressure.

Figure~\ref{fig:signal_comparison} presents time-aligned BSG and SCG waveforms along the $X$, $Y$, and $Z$ directions, highlighting recurring cardiac vibrations and the distinct morphology observed at the bed and chest. The overlaid fits illustrate how models based on the source and transmission framework in Chapter~\ref{part:theory}, together with the numerical representations in Chapter~\ref{part:dt}, can reproduce these measured waveforms. BSG is fitted with the body--bed FEM, while SCG is fitted with the analytical model that includes local thorax motion. Each complete cardiac cycle is fitted independently before cropping to the displayed interval, with amplitudes retained in the source records' display units.

\FloatBarrier
\subsection{From Hemodynamic Sensitivity to Inverse Estimation}
\label{sec:prelim_limits}
The ejection experiments, physiological models, and human recordings establish complementary links between cardiovascular state and mechanical features. \textbf{Quantitative inverse estimation uses these links to distinguish the target physiological change from other source and transmission effects.} CO, pulse pressure, diastolic function, and SV are distinct targets with different feature dependencies. Table~\ref{tab:prelim_evidence} assigns each body of evidence to the corresponding inference task.

\begin{table}[!htbp]
\centering\small
\caption{Hemodynamic Information in SCG/BCG Features and the Scope of Inverse Estimation. Mechanistic predictions, measured associations, and indirect comparisons support different levels of inference.}
\label{tab:prelim_evidence}
\begin{tabularx}{\linewidth}{P{0.19\linewidth}P{0.36\linewidth}Y}
\toprule
Signal and condition & Reported observation & Scope of inference\\
\midrule
Modeled BCG from aortic pressure, 21 cardiac surgical patients \citep{kim2016mechanism} &
Predicted J amplitude is approximately aortic pulse pressure times descending aortic area. Simultaneous BCG was unavailable. &
Forward prediction of pressure-driven recoil under specified geometry and pressure-wave timing.\\
\addlinespace
BCG before and after exercise, 11 participants with normal cardiac function \citep{zhan2024cardiac} &
Changes in I--J amplitude correlated with changes in echocardiographic CO (Spearman $r_s=0.923$, $p<0.001$). &
CO-related mechanical sensitivity reflects the combined contributions of HR and SV.\\
\addlinespace
Diastolic SCG and echocardiography, supine and head-up tilt \citep{agam2022diastolic} &
Peak-to-peak amplitude correlated with $e'$ ($r=0.713$, $p<0.001$). Tilt reduced SV, while only the $E_d$ amplitude changed significantly. &
Sensitivity to relaxation and loading, expressed through $e'$ and the response to tilt.\\
\addlinespace
SCG after 10 days of horizontal bed rest, 10 healthy men \citep{solbiati2021bedrest} &
AO--IVC amplitude rose from a median of 0.02 to 0.03\,$g$ ($p<0.05$), contrasting with SV reductions in prior bed-rest literature. &
An indirect comparison against the expected amplitude--SV trend. SV was not measured concurrently.\\
\addlinespace
SCG-to-finger-PPG transit time during lower-body negative pressure \citep{sorensen2017challenges} &
In 7 selected participants of 18 enrolled, SBP fell from $125\pm15$ to $112\pm17$\,mmHg and transit time from $237\pm18$ to $232\pm20$\,ms. Both changes were nonsignificant. &
The mean trend opposes the assumed inverse BP--transit-time relation. Records were selected for falling SBP and usable signals, defining a BP--timing comparison.\\
\addlinespace
SCG and BCG during exercise in heart failure \citep{herkert2021kinocardiograph} &
SV, CO, BCG energy, and BCG maximum power increased significantly. Corresponding whole-cycle SCG increases were nonsignificant. &
Feature-specific exercise responses support separate evaluation of whole-cycle SCG and BCG power measures.\\
\addlinespace
BCG, ECG, and M-mode ultrasound in arrhythmia \citep{zhan2025dataset} &
An illustrated premature ventricular contraction had an ECG event without a detected BCG response and with little mitral-valve motion. &
Dissociation between electrical activation and detectable mechanical response informs mechanical beat detection.\\
\bottomrule
\end{tabularx}
\par\smallskip
\raggedright
AO--IVC denotes the SCG amplitude difference between the aortic-opening (AO) and isovolumetric-contraction (IVC) landmarks \citep{solbiati2021bedrest}. PPG denotes photoplethysmogram, and SBP denotes systolic blood pressure. The SCG-to-finger-PPG transit time runs from SCG AO to the finger PPG onset, identified by tangent intersection \citep{sorensen2017challenges}.
\end{table}

The observation model explains the ambiguity. Let $\vct h$ denote hemodynamics and $\vct\eta$ the remaining source and measurement parameters. For a feature $f$,
\begin{equation}
 f=F(\vct h,\vct\eta),\qquad
 \Delta f\simeq
 \frac{\partial F}{\partial\vct h}\Delta\vct h+
 \frac{\partial F}{\partial\vct\eta}\Delta\vct\eta.
\label{eq:prelim_confounds}
\end{equation}
The two terms in Eq.~\eqref{eq:prelim_confounds} separate hemodynamic sensitivity from the effects of other source and measurement variables. Physiological and transmission changes can reinforce, cancel, or reverse a feature trend. Even with fixed transmission, multiple hemodynamic variables can affect one feature. For example, an amplitude $f=g\,s$ can arise from different combinations of source magnitude $s$ and mechanical gain $g$. SV recovery therefore requires observations or constraints that distinguish plausible SV values across these combinations.

Absolute BP requires additional physiological information: the same CO can correspond to different mean pressures when SVR/TPR differs, as in Eq.~\eqref{eq:prelim_hemodynamics}, and pulsatile pressure also depends on arterial compliance and wave reflection. Body--bed constraints characterize mechanical transmission, while the cardiovascular model supplies the physiological relations needed to estimate SV, absolute BP, and TPR.

\textbf{SCG and BCG provide complementary hemodynamic information that a calibrated source--transmission model can organize for target-specific estimation.} Triaxial BSG retains directional observations from both pathways, while physiological and mechanical constraints distinguish the target from other waveform influences. The analytical observability analysis in Section~\ref{sec:observability} derives these conditions. Chapter~\ref{part:dt} implements their source and transfer structure in a coupled digital-twin framework.

\FloatBarrier
\clearpage
\begin{bsgpartpage}{Theoretical Foundation of BSG}
\part[Theoretical Foundation of BSG]{Theoretical Foundation\\of BSG}
\label{part:theory}
\textbf{From Cardiovascular Physiology to Tri-Axial Bed Vibration}\par
\emph{A General Theory of Target Observability, Mechanical Nuisance, and Conditional Three-Axis Necessity}

This chapter develops the BSG theoretical framework in seven connected sections: physiological parameters, hemodynamic targets, vascular recoil, cardiac-contraction loading, body--bed mechanics, signal observation, and target observability. The argument begins with a shared physiological source and follows its transformation into bed vibration before asking which hemodynamic targets remain distinguishable.

Bodyseismography links cardiovascular dynamics to vibrations transmitted through the body and its supporting bed. Two mechanical pathways arise from a common physiological state: vascular recoil and local myocardial transmission. Vascular recoil is associated with the whole-body reaction mechanism emphasized in ballistocardiography (BCG), while myocardial transmission contributes to the local chest-wall vibration measured by seismocardiography (SCG). These associations do not imply that either measured signal isolates a single physiological source.

The framework couples pressure, volume, and thorax motion through a reciprocal myocardial interface and reciprocal body--bed contact. Vascular recoil is evaluated from the same circulation solution under an explicit one-way loading approximation. The numerical analysis compares target recovery across seven axis combinations and examines how mechanical compensation and source uncertainty affect scalar targets and waveform observations. These calculations connect the physiological and transmission mechanisms to the conditions under which three-axis BSG adds target information.

\end{bsgpartpage}

\section{Framework and Physiological Parameters}
\label{sec:framework}
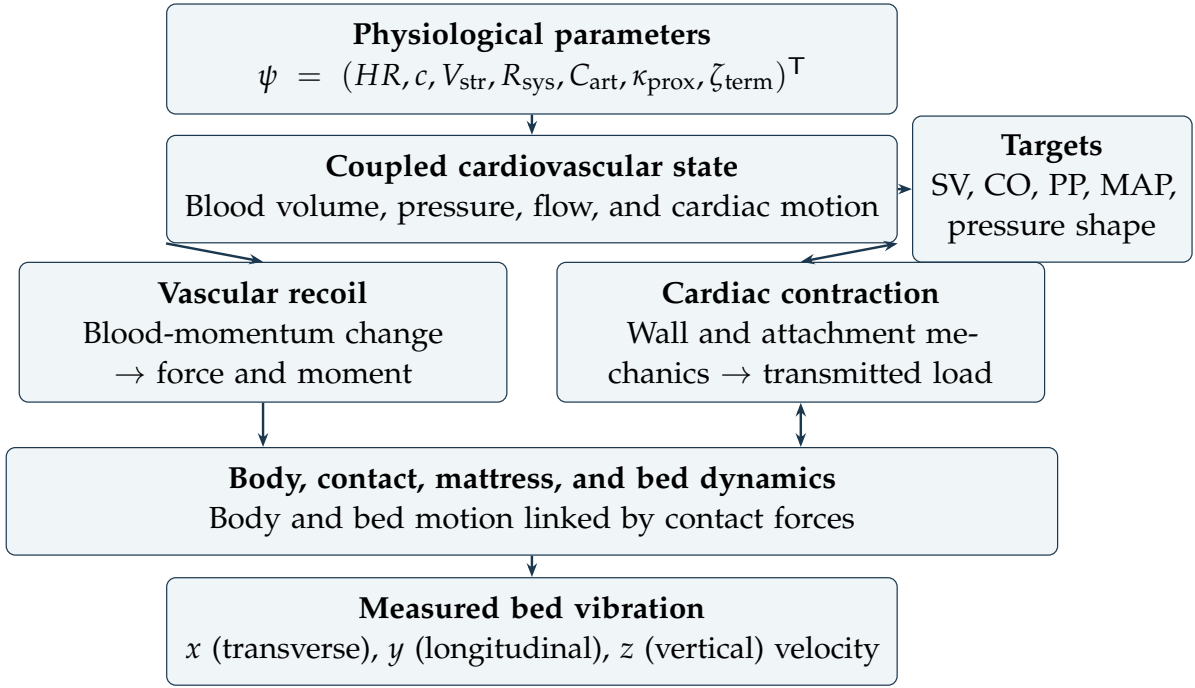
\begin{figure}[!htbp]
\centering\resizebox{0.90\linewidth}{!}{\begin{tikzpicture}[font=\small,>={Stealth[length=1.6mm,width=1.2mm]},
 box/.style={draw=navy,rounded corners=3pt,fill=pale,align=center,inner sep=6pt},
 arr/.style={->,thick,draw=navy}]
\node[box,text width=8.0cm] (psi) at (0,0)
 {\textbf{Physiological parameters}\\
 $\psi=(HR,c,V_{\rm str},R_{\rm sys},C_{\rm art},\kappa_{\rm prox},\zeta_{\rm term})^\T$};
\node[box,text width=8.0cm] (state) at (0,-1.5)
 {\textbf{Coupled cardiovascular state}\\
 Blood volume, pressure, flow, and cardiac motion};
\node[box,text width=2.8cm] (target) at (6.0,-1.5)
 {\textbf{Targets}\\SV, CO, PP, MAP, pressure shape};
\node[box,text width=5.2cm] (vasc) at (-3.1,-3.15)
 {\textbf{Vascular recoil}\\Blood-momentum change $\rightarrow$ force and moment};
\node[box,text width=5.2cm] (card) at (3.1,-3.15)
 {\textbf{Cardiac contraction}\\Wall and attachment mechanics $\rightarrow$ transmitted load};
\node[box,text width=11.7cm] (body) at (0,-5.10)
 {\textbf{Body, contact, mattress, and bed dynamics}\\
 Body and bed motion linked by contact forces};
\node[box,text width=8.0cm] (obs) at (0,-6.60)
 {\textbf{Measured bed vibration}\\$x$ (transverse), $y$ (longitudinal), $z$ (vertical) velocity};
\draw[arr] (psi)--(state);
\draw[arr] (state)--(target);
\draw[arr] (state.south west)--(vasc.north);
\draw[<->,thick,draw=navy] (state.south east)--(card.north);
\draw[arr] (vasc.south)--(body.north -| vasc.south);
\draw[<->,thick,draw=navy] (card.south)--(body.north -| card.south);
\draw[arr] (body)--(obs);
\end{tikzpicture}}
\caption{Shared physiology and two mechanical loading pathways. Two-way arrows denote cardiac--circulatory and cardiac--thorax coupling. Vascular recoil is applied as a one-way mechanical input in the selected reduction. Mechanical and residual-contraction parameters are included in the same coupled solution. Targets are evaluated from the state; the recorded vibration is its downstream mechanical observation.}
\label{fig:framework}
\end{figure}

Bed-recorded bodyseismography (BSG) measures cardiovascular vibration after transmission through the body, its contact with the mattress, and the bed structure. The framework retains two loading pathways: \emph{vascular recoil}, associated with blood-momentum changes, and \emph{cardiac contraction}, associated with cardiac deformation and attachment reactions. Both arise from one coupled physiological state (Figure~\ref{fig:framework}). The principal targets are stroke volume (SV), cardiac output (CO), and pulse pressure (PP); mean arterial pressure (MAP) and pressure-waveform shape can also be evaluated. These quantities are computed from the physiological state rather than prescribed as force amplitudes.

\textbf{Reading the notation.} Time $t$ is measured in seconds. A dot denotes a time derivative, so $\dot V$ is the rate of volume change and $\ddot{\mathbf q}$ is mechanical acceleration. A superscript $\T$ transposes a row into a column or interchanges a matrix's rows and columns. Bold symbols denote physical vectors; named state/parameter arrays and matrices are identified explicitly below. Subscript $0$ denotes a stated reference value unless a different meaning is given locally. In differential expressions, $\dd$ denotes an infinitesimal increment; $\partial$ denotes a partial derivative with the other arguments fixed.

The core physiological parameter vector is explicitly
\begin{equation}
\boxed{\psi=\big(HR,\ c,\ V_{\rm str},\ R_{\rm sys},\ C_{\rm art},\
\kappa_{\rm prox},\ \zeta_{\rm term}\big)^\T.}
\label{eq:psi}
\end{equation}
Table~\ref{tab:psi} defines all seven coordinates. Pressure, flow, ventricular volume, and active tension are evolving states or constitutive outputs, rather than additional entries in this vector.

\begin{table}[!htbp]
\centering\small
\caption{Core physiological coordinates and their roles in the forward model.}
\label{tab:psi}
\begin{tabularx}{\linewidth}{@{}P{0.14\linewidth}P{0.28\linewidth}Y@{}}
\toprule Coordinate & Definition & Entry into the model\\\midrule
$HR$ & Heart rate, beats/min & Sets the cycle duration $T_b=60/HR$ in seconds and activation recurrence; changes filling time, ejection timing, and CO.\\[3pt]
$c$ & Dimensionless contractility factor & Scales active reference tension (stress), $T_{\rm ref}=cT_{{\rm ref},0}$, in Pa, where $T_{{\rm ref},0}$ is its baseline value, in the wall-mechanics realization; influences pressure generation, ejection, and attachment loading.\\[3pt]
$V_{\rm str}$ & Stressed blood volume, mL & Blood volume above the specified unstressed compartment inventory. Controls filling/preload through the closed circulation; it is not SV or end-diastolic volume.\\[3pt]
$R_{\rm sys}$ & Effective systemic resistance, Pa\,s/m$^3$ & Controls resistive pressure losses in the declared systemic network and thereby afterload and mean pressure.\\[3pt]
$C_{\rm art}$ & Effective arterial compliance, m$^3$/Pa & Controls arterial storage, $\dd V_{\rm art}=C_{\rm art}\dd p_{\rm art}$ locally, where $V_{\rm art}$ and $p_{\rm art}$ are arterial blood volume and pressure; influences pulse amplitude and diastolic decay.\\[3pt]
$\kappa_{\rm prox}$ & Dimensionless proximal stiffness factor & Scales the specified proximal arterial stiffness law relative to its baseline. Changes the local pressure--volume response and pulse transmission.\\[3pt]
$\zeta_{\rm term}$ & Dimensionless terminal impedance factor & Scales the specified terminal vascular load relative to its baseline. Changes the pressure--flow boundary relation and downstream loading.\\
\bottomrule
\end{tabularx}
\end{table}

For the last two coordinates, the scale convention is $K_{\rm prox}=\kappa_{\rm prox}K_{{\rm prox},0}$ and $Z_{\rm term}(\omega)=\zeta_{\rm term}Z_{{\rm term},0}(\omega)$, with the baseline laws held fixed. Here $K_{\rm prox}$ is the dimensional stiffness coefficient in the specified proximal arterial law; its units and those of $K_{{\rm prox},0}$ depend on that law (Pa/m$^3$ for a pressure--volume slope). $Z_{\rm term}$ is terminal pressure-to-flow impedance (Pa\,s/m$^3$), and $\omega$ is angular frequency (rad/s). The factors $\kappa_{\rm prox}$ and $\zeta_{\rm term}$ are dimensionless. The vascular network must specify which elements are controlled by systemic resistance, arterial compliance, proximal stiffness, and terminal loading. The same constitutive coefficient must not be independently assigned to two coordinates. These are effective model parameters, not automatically identifiable anatomical measurements.

Other uncertain quantities are kept separate:
\begin{equation}
\vartheta=(\psi,\eta,\nu_{\rm cc},c_{\rm ref}).
\label{eq:theta}
\end{equation}
The vector $\vartheta$ collects all free parameters: the seven physiological coordinates $\psi$, the mechanical/calibration coordinates $\eta$, residual contraction coordinates $\nu_{\rm cc}$, and the pressure-reference offset $c_{\rm ref}$. These coordinates retain their physical units until explicitly scaled in Section~\ref{sec:observability}. Here $\eta$ includes cardiac geometry/orientation and attachment coefficients, body mass distribution and deformation, contact geometry/stiffness/damping, bed dynamics, and any uncalibrated sensor gain or timing. The residual cardiac-contraction coordinates $\nu_{\rm cc}$ allow calcium rise time, calcium decay time, activation span (durations in s), and cross-bridge kinetics (rate coefficients in s$^{-1}$, or declared dimensionless multipliers) to vary beyond the single factor $c$. The selected kinetic parameterization must state which of these are free. An optional discrepancy family must be declared separately. The pressure-reference offset $c_{\rm ref}$ has pressure units (Pa, or mmHg for reporting) and is retained when absolute pressure is a target. Quantities calibrated and fixed for an experiment are not free coordinates.

The complete state vector $X(t)$ contains the time-varying circulation pressures, flows and volumes, cardiac coordinates, body/bed coordinates, and required velocities. In contrast, $\vartheta$ specifies the parameters of one calculation. The coupled equations and their noise-free outputs are written
\begin{equation}
\mathcal R(\dot X,X,t;\vartheta)=0,\qquad
 y_A^{(0)}=f_A(\vartheta),\qquad g=\mathfrak g(\vartheta),
\label{eq:state}
\end{equation}
Here $\mathcal R$ is the vector of governing-equation residuals; $\mathcal R=0$ means that all balances and constraints are satisfied. The map $f_A$ returns the predicted vibration $y_A^{(0)}$, where superscript $(0)$ means noise-free. The map $\mathfrak g$ returns the selected scalar or vector target $g$. The subscript $A$ specifies the recorded axes: $x$ (transverse), $y$ (longitudinal), and $z$ (vertical); $xy$, $xz$, $yz$, and $xyz$ denote stacked combinations. The right-handed model frame has $+x$ toward the subject's left, $+y$ toward the feet, and $+z$ anterior, upward for a supine subject. Reproduced numerical panels label the same axes X, Y, and Z. Native sensor polarity requires calibration; it is not established by axis names alone. The system also obeys blood-volume, periodic, and phase constraints. Writing $g=\mathfrak g(\psi)$ is shorthand only when all other parameters are fixed or their feedback is neglected explicitly.

Throughout this formulation, \emph{cardiac contraction} names the transmitted cardiac pathway across the full cycle, including relaxation and passive wall/attachment effects. Its load is distinct from cellular active tension. Section~\ref{sec:signals} defines how bodyseismography (BSG), ballistocardiography (BCG), and seismocardiography (SCG) observe the same coupled motion.

\section{How Physiology Determines the Target Information}
\label{sec:targets}
The circulation couples filling, contraction, ejection, arterial storage, and venous return. Its basic volume and flow relations are
\begin{align}
\dot V&=Q_{\rm MV}-Q_{\rm Ao}, & V+\sum_iV_i&=V_{\rm tot},\label{eq:volume}\\
C_i\dot p_i&=Q_{{\rm in},i}-Q_{{\rm out},i}, &
L_e\dot Q_e+R_eQ_e&=p_{{\rm up},e}-p_{{\rm down},e}.
\label{eq:circulation}
\end{align}
In these equations, $V$ is ventricular cavity volume, $Q_{\rm MV}$ is mitral-valve inflow, and $Q_{\rm Ao}$ is aortic-valve outflow. Index $i$ labels the other blood compartments: $V_i$ is their blood volume and $V_{\rm tot}$ is the conserved total blood volume for a fixed parameter set. The symbol $\sum_i$ adds over those compartments. For compartment $i$, $p_i$ is internal pressure, $C_i$ local compliance, and $Q_{{\rm in},i}$ and $Q_{{\rm out},i}$ total inflow and outflow. Index $e$ labels a vascular segment: $Q_e$ is segment flow, $L_e$ hydraulic inertance, $R_e$ hydraulic resistance, and $p_{{\rm up},e}$ and $p_{{\rm down},e}$ its upstream and downstream pressures.

\textbf{Units.} In mechanical calculations, volumes are in m$^3$, flow rates in m$^3$/s, pressures in Pa, compliances in m$^3$/Pa, resistances in Pa\,s/m$^3$, and inertances in Pa\,s$^2$/m$^3$. Clinical volume and pressure reporting can use mL and mmHg, with $1\,\mathrm{mL}=10^{-6}\,\mathrm{m^3}$ and $1\,\mathrm{mmHg}\approx133.322\,\mathrm{Pa}$.

The storage expression assumes fixed external pressure and a locally linear compliance; nonlinear pressure--volume laws can replace it. Valve flow depends on the instantaneous pressure difference. Ventricular pressure is supplied by the cardiac mechanics in Section~\ref{sec:cardiac} and is solved jointly with these equations.

At a specified vascular measurement site $s$, let $p_s(t)$ be the pressure waveform at that site. The principal targets are
\begin{align}
SV&=\int_{t_{\rm AO}}^{t_{\rm AC}}Q_{\rm Ao}(t)\dd t,
&CO&=\frac{HR\,SV}{1000},\label{eq:svco}\\
PP&=\max_t p_s(t)-\min_t p_s(t),
&MAP&=\frac{1}{T_b}\int_0^{T_b}p_s(t)\dd t.
\label{eq:pressuretargets}
\end{align}
Here $t_{\rm AO}$ and $t_{\rm AC}$ are the times of aortic-valve opening and closure within a beat. The integral adds the ejected flow over that interval. Expressing SV in mL and HR in beats/min gives CO in L/min through the factor $1000$. If the flow integral is evaluated in SI units, its result is first converted from m$^3$ to mL. Both $\max_t$ and $\min_t$ are taken over one complete beat of duration $T_b$. Systolic blood pressure is $SBP=\max_t p_s(t)$, diastolic blood pressure is $DBP=\min_t p_s(t)$, PP is their difference, and MAP is the time-averaged pressure; all have pressure units. The ejection integral assumes the stated valve-event convention and negligible regurgitation; a signed-flow or forward-only convention must otherwise be specified. Central aortic pressure and peripheral arterial blood pressure (ABP) are different targets and require an explicit transfer relationship when compared.

The seven coordinates act through this shared solution. Heart rate changes the period and time available for filling. Contractility changes active tension and hence pressure, ejected volume, and cardiac loading. Stressed volume changes filling conditions. Systemic resistance changes pressure losses and afterload; compliance changes arterial storage. Proximal stiffness and terminal impedance alter the pulse's transmission and load. Their effects interact, so a change in one coordinate cannot generally be assigned a universal monotone change in SV or PP.

For a small parameter perturbation, $\delta$ denotes the resulting first-order change in a quantity. For example, $CO=HR\,SV/1000$ gives
\begin{equation}
\delta CO=\frac{SV}{1000}\delta HR+\frac{HR}{1000}\delta SV.
\label{eq:cochange}
\end{equation}
Thus the explicit heart-rate effect and the induced change in filling/ejection both matter. Relations such as $PP\approx SV/C_{\rm art}$ require additional assumptions about ejection and arterial runoff and are not the target definition used here.

Waveform targets preserve information that scalar summaries discard. Define pulsatile pressure $p_c(t)=p_s(t)-MAP$ and, for $PP>0$, range-normalized shape $u(t)=[p_s(t)-DBP]/PP$. Here $p_c$ has pressure units and $u$ is dimensionless, with a range from zero to one. The first removes the mean; the second also removes amplitude. Neither retains the absolute-pressure offset. These target definitions, the pressure site, and the time/phase alignment must remain fixed when comparing sensing axes.

\begin{figure}[!htbp]
\centering\includegraphics[width=\linewidth,height=0.79\textheight,keepaspectratio]{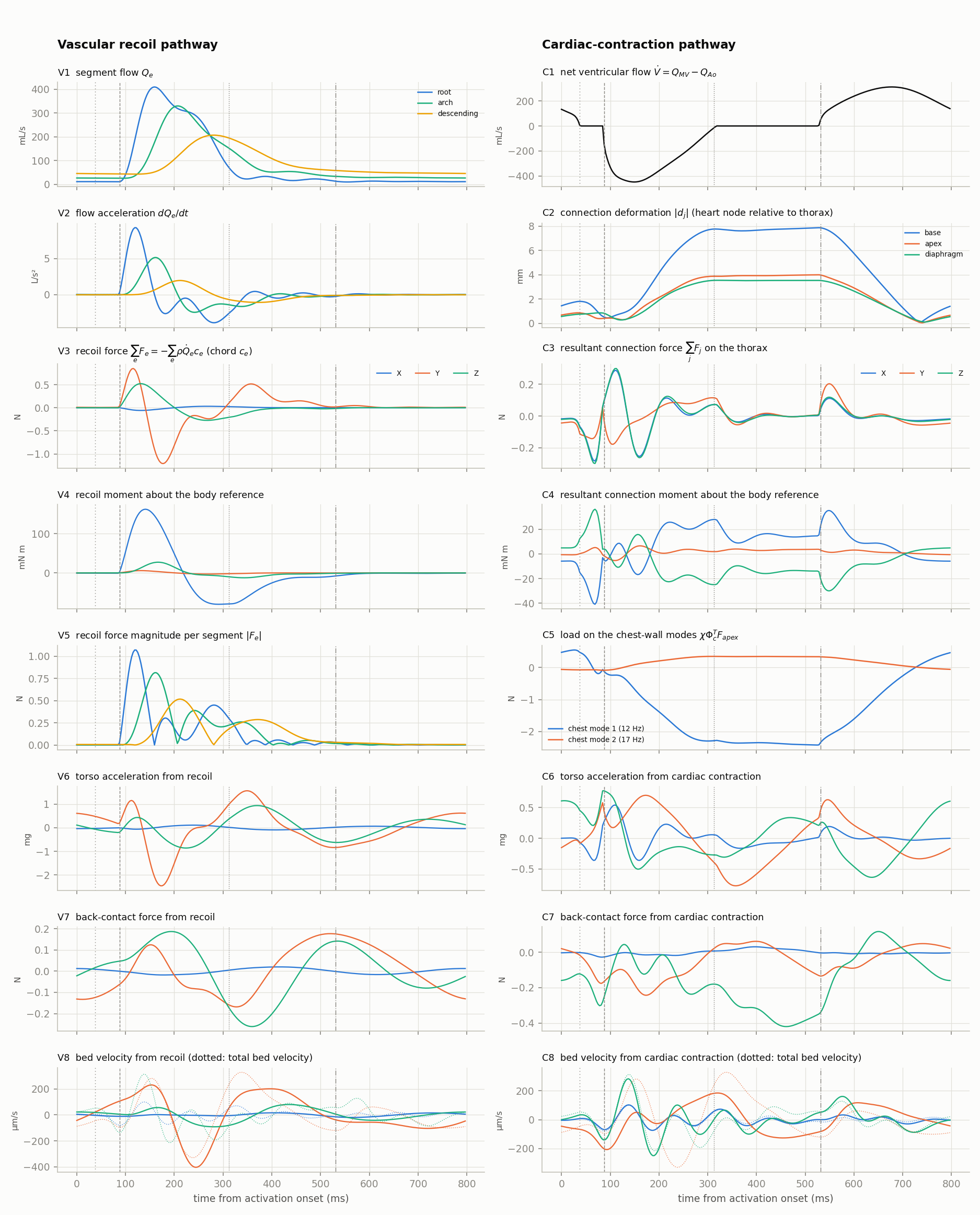}
\caption{BSG Generation by Two Cardiovascular Pathways. Simulated vascular recoil (left) and cardiac contraction (right) at 75 beats/min. Dotted curves show total bed velocity, and vertical lines mark valve events.}
\label{fig:time-domain-chain}
\end{figure}

\paragraph{From Blood Flow to the Measured Time History}
Figure~\ref{fig:time-domain-chain} follows both loading pathways through one simulated beat. The vascular column begins with $Q_1(t),Q_2(t),Q_3(t)$ and their time derivatives. Arterial storage and flow dynamics delay and reshape the pulse along the three consecutive arterial segments. For segment $e$,
\begin{equation}
\dot Q_e(t)=\frac{\mathrm dQ_e(t)}{\mathrm dt}
=\frac{\Delta p_e(t)-R_eQ_e(t)}{L_e}.
\label{eq:flow-time-derivative}
\end{equation}
Here $\Delta p_e=p_{{\rm up},e}-p_{{\rm down},e}$, with $R_e,L_e$ from Eq.~\eqref{eq:circulation}. The derivative $\dot Q_e$, in m$^3$/s$^2$, is positive during increasing flow and negative during decreasing flow; it is not blood-particle acceleration. It vanishes at a smooth interior flow maximum, so a flow peak and a recoil-force peak need not coincide.

The segment flow derivatives generate recoil forces and moments (Section~\ref{sec:recoil}); cardiac wall motion generates attachment loads (Section~\ref{sec:cardiac}). Both act on the body, whose motion relative to the bed generates contact forces and bed velocity (Section~\ref{sec:mechanics}). The aligned traces expose changes in timing, polarity, and shape. They are a simulated forward calculation, not a measured separation of the two sources.

\FloatBarrier
\section{Vascular Recoil Force and Moment}
\label{sec:recoil}
\subsection{Segment Selection and Three-Dimensional Geometry}
The numerical illustration divides one anatomical aortic centreline into three consecutive arc-length cells. Segment 1 contains the ascending aorta and proximal arch; segment 2 contains the distal arch and proximal descending thoracic aorta; segment 3 contains the distal descending thoracic and abdominal aorta. Their flows are $Q_1,Q_2,Q_3$, and their hydraulic lengths are $\ell_e=(0.12,0.18,0.25)$ m. The corresponding areas are $A_e=(5.0,4.5,4.0)\times10^{-4}$ m$^2$. A terminal resistance--compliance--resistance (RCR) load closes this reduced network. Three cells are a modeling choice, not a demonstrated minimum or three independent vessels.

Coordinates follow $x$ (transverse), $y$ (longitudinal), and $z$ (vertical), with positive directions leftward, footward, and anterior. The origin $O$ is the body reference point. The aortic valve is placed at the heart-base connection, $(x,y,z)=(0.8,-11.8,2.6)$ cm. The ascending aorta runs headward for 5 cm; a semicircular arch of approximately 5.3 cm radius turns toward the left and posteriorly; the descending aorta then runs footward at approximately $x=2$ cm and $z=-8$ cm. The total centreline length is 0.55 m. These positions are declared anatomical assumptions, not subject imaging.

\begin{figure}[!htbp]
\centering\includegraphics[width=\linewidth]{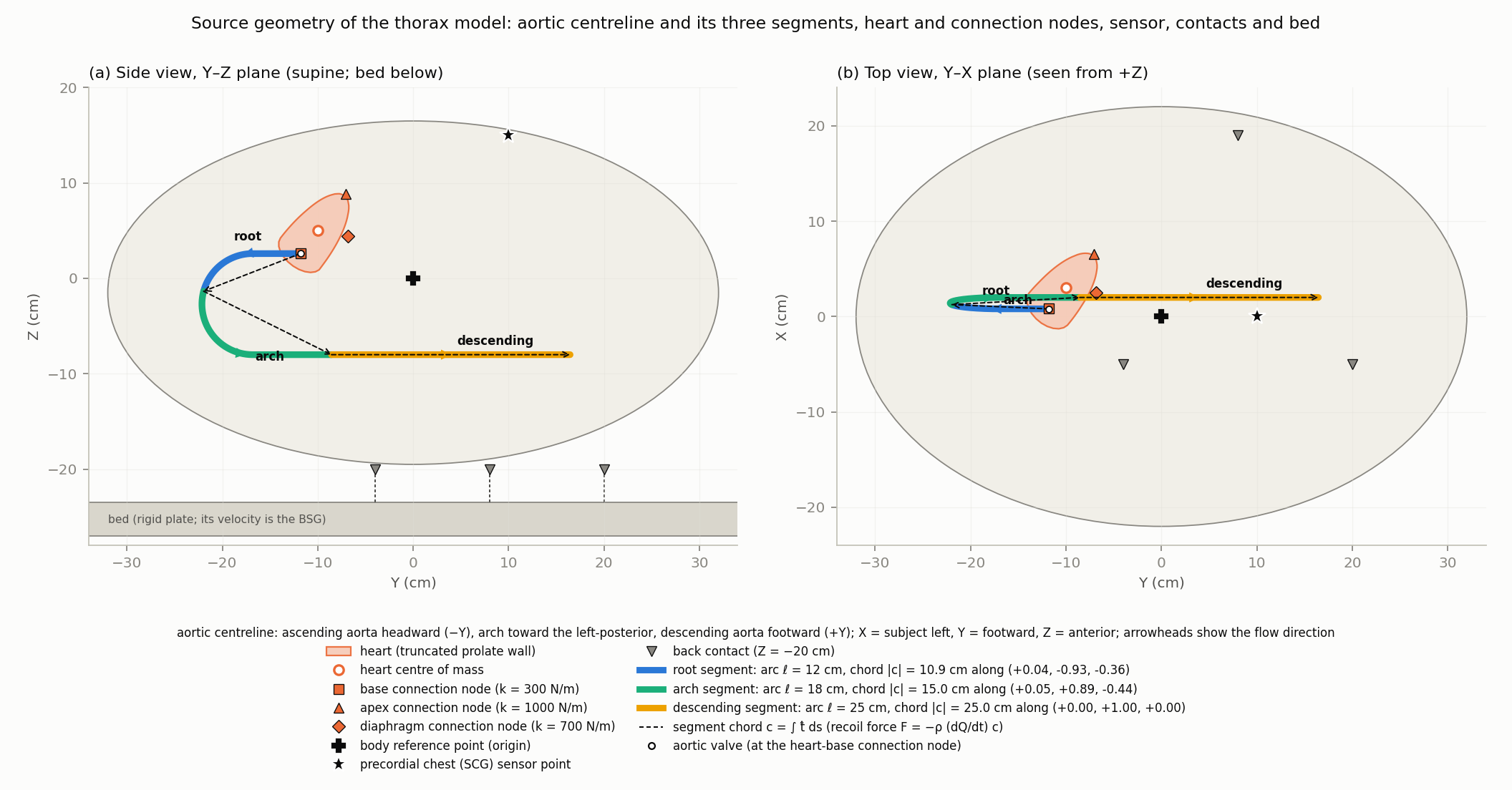}
\caption{Anatomically arranged source geometry in the body frame. Left: longitudinal--vertical ($y$--$z$) side view of the supine body. Right: longitudinal--transverse ($y$--$x$) view from above. Coloured curves show the three consecutive aortic cells; arrows give positive flow, and dashed chords set each cell's resultant recoil direction. The heart, base/apex/diaphragm attachments, precordial SCG point, back contacts, and bed share the same frame. $+y$ points footward, so the ascending aorta runs toward $-y$ and the descending aorta toward $+y$. The centreline and heart placement are model assumptions. X/Y/Z in the panels denote $x/y/z$; all positions are in cm.}
\label{fig:vascular-geometry}
\end{figure}

For segment $e$, let $s$ be arc length (m), $\mathbf r_e(s)$ its position relative to $O$ (m), and $\hat{\mathbf t}_e(s)=\mathrm d\mathbf r_e/\mathrm ds$ its dimensionless unit tangent in the positive-flow direction. The chord is the endpoint displacement,
\begin{equation}
\mathbf c_e=\int_0^{\ell_e}\hat{\mathbf t}_e(s)\dd s
=\mathbf r_e(\ell_e)-\mathbf r_e(0).
\label{eq:vascular-tangents}
\end{equation}
It has units m and need not have magnitude $\ell_e$. The baseline chords, rounded to the displayed precision, are
\begin{center}\small
\begin{tabular}{c c c c}
\toprule Segment & Arc length (cm) & Chord magnitude (cm) & Chord direction $(x,y,z)$\\\midrule
1 & 12 & 10.9 & $(0.04,-0.93,-0.36)$\\
2 & 18 & 15.0 & $(0.05,0.89,-0.44)$\\
3 & 25 & 25.0 & $(0,1,0)$\\\bottomrule
\end{tabular}
\end{center}
The root and descending chords therefore point predominantly along opposite longitudinal directions. The arch supplies a vertical component; the transverse chord components are at most approximately 5\%. Source-orientation perturbations rotate the entire centreline about the valve, preserving its connection to the heart. Figure~\ref{fig:vascular-geometry} shows two orthogonal projections of this three-dimensional layout.

\subsection{Signed Force and Moment from the Same Centreline}
Assume fixed geometry and one spatially uniform flow history $Q_e(t)$ per cell. With blood density $\rho_b$ (kg/m$^3$), its retained linear momentum is $\mathbf P_e=\rho_bQ_e\mathbf c_e$ (kg\,m/s). The reaction on the surrounding mechanics and the hydraulic flow equation are
\begin{equation}
\mathbf F_e=-\rho_b\dot Q_e\mathbf c_e,\qquad
L_e=\frac{\rho_b\ell_e}{A_e},\qquad
L_e\dot Q_e=\Delta p_e-R_eQ_e.
\label{eq:recoil}
\end{equation}
The force $\mathbf F_e$ has units N. Hydraulic inertance $L_e$ depends on arc length, whereas resultant force depends on the chord. For a straight segment, $\mathbf c_e=\ell_e\hat{\mathbf t}_e$, recovering the straight-segment expression. For a curved segment, replacing the chord by the arc length times an arbitrary tangent generally changes the force. For a nonzero chord,
\begin{equation}
\frac{\mathbf F_e}{\|\mathbf F_e\|}=
\begin{cases}
-\mathbf c_e/\|\mathbf c_e\|,&\dot Q_e>0,\\
+\mathbf c_e/\|\mathbf c_e\|,&\dot Q_e<0,
\end{cases}
\qquad \mathbf F_e=0\quad\text{if }\dot Q_e=0.
\label{eq:recoil-direction}
\end{equation}
Here $\|\cdot\|$ is Euclidean length. Recoil opposes the change of blood momentum, rather than always opposing flow. Its component on axis $k\in\{x,y,z\}$ is $F_{e,k}=-\rho_b\dot Q_ec_{e,k}$. In particular, accelerating headward root flow pushes the body footward; later acceleration of descending flow pushes it headward.

Moment is calculated from the distributed centreline, not an independently assigned point arm:
\begin{equation}
\mathbf m_e=\int_0^{\ell_e}\mathbf r_e(s)\times\hat{\mathbf t}_e(s)\dd s,
\qquad \mathbf M_e=-\rho_b\dot Q_e\mathbf m_e,
\qquad
W_v=\begin{bmatrix}\sum_e\mathbf F_e\\\sum_e\mathbf M_e\end{bmatrix}.
\label{eq:wrench}
\end{equation}
The geometric moment vector $\mathbf m_e$ has units m$^2$ and is not a mass; $\mathbf M_e$ has units N\,m about $O$. The symbol $\times$ is the cross product in the declared right-handed frame. The six-entry wrench $W_v$ contains resultant force and moment. Its generalized load is $B_vW_v$, where $B_v$ maps six wrench entries to the retained mechanical coordinates. This source follows the same flow solution as SV and pressure, so target magnitude alone does not prescribe force. Momentum and pressure-gradient descriptions provide mechanistic precedents \citep{kim2016mechanism,rabineau2021closedloop}.

\subsection{Scope of the Reduction}
Figure~\ref{fig:time-domain-chain} shows that segment timing and opposing chord directions cause cancellation in the resultant. At the illustrated baseline, recoil-force RMS is approximately $0.02$, $0.35$, and $0.16$ N on $x,y,z$, respectively, with a predominant pitching moment about $x$. These are source loads before body--bed transmission.

The model retains unsteady momentum on a fixed, coarse centreline. Intracardiac blood momentum, supra-aortic branches, peripheral and venous momentum, boundary momentum fluxes, and vessel motion are not resolved. A spatial-convergence study is still required. A force already represented by the momentum balance must not be added again as a separate pressure load. The resultant wrench reproduces rigid-body virtual work; deformation-mode forcing also needs a declared spatial projection.

The selected vascular pathway is one-way: it omits body/vessel acceleration feeding back into the flow equations, while cardiac--thorax feedback remains coupled. For rigid vessel motion, the omitted pressure term can be screened as $p_{{\rm fb},e}=\rho_b(\mathbf c_e\cdot\mathbf a_O+\mathbf m_e\cdot\boldsymbol\alpha_O)$, where $\mathbf a_O$ and $\boldsymbol\alpha_O$ are body translational and angular accelerations. The baseline ratio of its RMS to the segment driving-pressure RMS is approximately $0.0010$, $0.0008$, and $0.0006$. This is a small-load diagnostic, not a demonstrated bound on target-recovery or axis-ranking error.
\FloatBarrier

\section{Cardiac-Contraction Loading}
\label{sec:cardiac}
Cardiac contraction generates active wall stress, changes cavity geometry and pressure, and transmits forces through surrounding tissue. Let $\mathbf r$ collect internal cardiac coordinates, $V(\mathbf r)$ be ventricular cavity volume (m$^3$), and $\mathbf q_T$ collect thorax coordinates. The subscript $T$ here means thorax. Transmural pressure $p=P_{\rm LV}-P_{\rm ext}$ is left-ventricular internal pressure $P_{\rm LV}$ minus external pressure $P_{\rm ext}$, in Pa.

\textbf{Definition of $\mathbf z_{\rm cc}$.} The vector $\mathbf z_{\rm cc}(t)$ is the generalized mechanical load transmitted by the cardiac-contraction attachment pathway \emph{to the thorax}. Subscript cc means cardiac contraction; the letter $z$ labels a load vector and does not denote the vertical axis. It has $n_T$ entries, one for each of the $n_T$ retained thorax coordinates in $\mathbf q_T$. An entry conjugate to translation is a force (N), one conjugate to rotation is a moment (N\,m), and one conjugate to a deformation amplitude is the corresponding modal load. Its defining work relation is
\begin{equation}
\delta W_T=\mathbf z_{\rm cc}^\T\delta\mathbf q_T,
\label{eq:cc-definition}
\end{equation}
where $\delta W_T$ is virtual work delivered to the thorax (J), and $\delta\mathbf q_T$ is a compatible virtual thorax displacement. Thus $\mathbf z_{\rm cc}$ is not restricted to a three-component force or to one fixed direction.

A \emph{generalized coordinate} describes a retained translation, rotation, deformation, or cavity-volume degree of freedom. Its associated generalized force is defined so that force times coordinate increment is work (J): a translation in m pairs with force in N, a rotation in rad with moment in N\,m, and a volume in m$^3$ with pressure in Pa. Thus a generalized-force vector can contain components with different units. A reduced wall-mechanics system is
\begin{align}
M_H\ddot{\mathbf r}+C_H\dot{\mathbf r}+\nabla_{\mathbf r}U_{\rm pas}
 &=p\nabla_{\mathbf r}V+\mathbf Q_{\rm act}-B_H^\T\mathbf F_\Gamma,
\label{eq:heart}\\
(\nabla_{\mathbf r}V)^\T\dot{\mathbf r}&=Q_{\rm MV}-Q_{\rm Ao},\nonumber\\
\mathbf d=B_H\mathbf r-B_T\mathbf q_T,\qquad
\mathbf F_\Gamma&=K_\Gamma^{\rm link}\mathbf d+C_\Gamma^{\rm link}\dot{\mathbf d},\label{eq:link}\\
\boxed{\mathbf z_{\rm cc}=B_T^\T\mathbf F_\Gamma.}&&\label{eq:ccforce}
\end{align}
Here $M_H$ and $C_H$ are cardiac inertia and damping matrices; subscript $H$ denotes the heart. $U_{\rm pas}$ is passive wall strain energy (J), and $\nabla_{\mathbf r}$ is the column of derivatives with respect to the cardiac coordinates. Thus $\nabla_{\mathbf r}U_{\rm pas}$ is the passive restoring load and $p\nabla_{\mathbf r}V$ is the cavity-pressure load. The vector $\mathbf Q_{\rm act}$ is the active generalized cardiac load; it is a mechanical force array, distinct from the scalar blood flow $Q$. The second equation states that geometric cavity-volume change equals net valve inflow.

The matrices $B_H,B_T$ map cardiac and thorax coordinates to a common stack of attachment-point displacements in m, measured relative to the reference configuration. With $N_\Gamma$ modeled attachments, $\mathbf F_\Gamma\in\mathbb R^{3N_\Gamma}$ stacks their three-dimensional forces, and $B_T$ has $3N_\Gamma$ rows and $n_T$ columns. Here $N_\Gamma$ is the number of attachment points. Consequently, $\mathbf d$ is the relative attachment displacement, $\dot{\mathbf d}$ its velocity, and $\mathbf F_\Gamma$ the stacked link forces (N). The label $\Gamma$ denotes the attachment interfaces; $K_\Gamma^{\rm link}$ and $C_\Gamma^{\rm link}$ are their stiffness (N/m) and damping (N\,s/m) matrices. The vector $\mathbf z_{\rm cc}$ is the resulting generalized thorax load, with one entry per thorax coordinate; subscript cc means cardiac contraction. The heart receives $-B_H^\T\mathbf F_\Gamma$ and the thorax receives $+B_T^\T\mathbf F_\Gamma$. This paired reaction allows thorax/contact motion to affect the cardiac state. Each inertia or damping matrix has one row and column per coordinate in its subsystem; multiplying it by the corresponding acceleration or velocity gives a generalized load. For purely translational coordinates, inertia, damping, and stiffness have units kg, N\,s/m, and N/m. For mixed coordinates, each matrix entry carries the units needed to preserve work and power. Some cardiac coordinates can be massless, so pressure and motion may be determined by a differential--algebraic system.

When the retained thorax coordinates include rigid translation and rotation about $O$, their entries in $\mathbf z_{\rm cc}$ are the resultant force and moment
\begin{equation}
\mathbf F_{\rm cc}=\sum_{j=1}^{N_\Gamma}\mathbf f_{\Gamma,j},\qquad
\mathbf M_{\rm cc}=\sum_{j=1}^{N_\Gamma}\mathbf a_{\Gamma,j}\times\mathbf f_{\Gamma,j}.
\label{eq:cc-resultants}
\end{equation}
Here $j$ labels an attachment, $\mathbf f_{\Gamma,j}$ is its force on the thorax (N), and $\mathbf a_{\Gamma,j}$ is the vector from $O$ to that attachment (m). Deformation coordinates receive their own projections through $B_T^\T$, rather than a copy of the resultant force. The force directions follow the attachment geometry, relative motion, and stiffness/damping law in Eq.~\eqref{eq:link}; they are not fixed by the vascular chords. In Figure~\ref{fig:time-domain-chain}, panel C3 shows the $x/y/z$ force resultant, C4 its moment, and C5 its local chest-mode loads. Together these illustrate distinct components of the generalized load.

For the dimensionless unit fibre direction $\mathbf f$, active stress has the form $\sigma_{\rm act}=T_a\mathbf f\otimes\mathbf f$. Here $T_a$ is active fibre tension expressed as a stress (Pa), $\sigma_{\rm act}$ is the active stress tensor (Pa), and $\otimes$ forms the outer product of the fibre direction with itself. In a small-strain reduced basis, its generalized force satisfies
\begin{equation}
\mathbf Q_{\rm act}^\T\delta\mathbf r
=-\int_{\Omega_H}\sigma_{\rm act}:\delta\varepsilon\dd\Omega.
\label{eq:activework}
\end{equation}
In Eq.~\eqref{eq:activework}, $\Omega_H$ is the cardiac-wall volume, $\dd\Omega$ a volume element (m$^3$), and $\varepsilon$ the dimensionless strain tensor. The variations $\delta\mathbf r$ and $\delta\varepsilon$ are compatible virtual changes in cardiac coordinates and strain. The colon denotes tensor contraction, the sum of products of corresponding tensor entries. Both sides therefore represent virtual work in J; the sign follows the stated contractile-force convention.

The wall-mechanics realization evaluates $T_a$ from calcium-dependent contraction dynamics with reference tension scaled by $c$ \citep{land2017contraction}. It retains volume, an internal shape coordinate, and heart translation, with reduced fibre geometry and passive wall mechanics \citep{arts1991relation,guccione1991passive}. Base, apex, and inferior-wall links represent transmission to the mediastinum, anterior chest, and diaphragm. Pressure follows the wall/flow balance; a separate elastance pressure generator is not imposed on top of it.

The load $\mathbf z_{\rm cc}$ includes the consequences of active contraction, passive deformation, and attachment motion. Isovolumic shape changes can alter attachment loads without directly changing cavity volume. Consequently, SV, CO, or a pressure waveform alone does not uniquely specify the transmitted cardiac-contraction force. Figure~\ref{fig:time-domain-chain} follows net ventricular flow through attachment deformation, resultant loading, and the contact pathway; Figure~\ref{fig:modalities} compares the transmitted contribution at the body, chest, and bed.

\section{From the Two Loads to Body and Bed Vibration}
\label{sec:mechanics}
Within the chosen cardiovascular reduction, the two pathways supply the forcing of the retained body--bed system:
\begin{equation}
\boxed{\mathbf Q_{\rm ext}(t)=B_vW_v(t)+S_T^\T\mathbf z_{\rm cc}(t).}
\label{eq:twoforces}
\end{equation}
Here $\mathbf Q_{\rm ext}$ is the total cardiovascular generalized load on the body--bed subsystem. The word external refers to that subsystem boundary. The coordinate vector $\mathbf q$ contains body, thorax/deformation, and bed motion; internal cardiac coordinates remain in the cardiac subsystem. The selection matrix $S_T$ gives $\mathbf q_T=S_T\mathbf q$ and has one row per thorax coordinate and one column per full mechanical coordinate. Its transpose distributes the cardiac-contraction load back to the full system while preserving work. Both pathways must use disjoint loading boundaries. In particular, root/attachment tractions already assigned to one pathway cannot be counted in the other. Equation~\eqref{eq:twoforces} closes the modeled cardiac forcing once the spatial projections are specified. It is not a claim that all physiological or environmental excitation is represented.

At contact $j$, let $B_{b,j}\mathbf q$ and $B_{p,j}\mathbf q$ be the body- and bed-side displacements. With $L_j=B_{b,j}-B_{p,j}$,
\begin{equation}
\boldsymbol\delta_j=L_j\mathbf q,\qquad
\mathbf f_j=K_{c,j}\boldsymbol\delta_j+C_{c,j}\dot{\boldsymbol\delta}_j.
\label{eq:contact}
\end{equation}
Index $j$ labels a body--bed contact. The matrices $B_{b,j}$ and $B_{p,j}$ map the full coordinate vector to the three-dimensional contact displacements in m; subscripts $b$ and $p$ mean body and bed/plate. Their difference $L_j$ maps to relative displacement $\boldsymbol\delta_j$ (m). This bold displacement is distinct from the variation symbol $\delta$. The vector $\mathbf f_j$ is contact force (N); $K_{c,j}$ and $C_{c,j}$ are contact stiffness (N/m) and damping (N\,s/m) matrices. The subscript $c$ here means contact. The bed receives $+\mathbf f_j$ and the body receives $-\mathbf f_j$. For passive symmetric contact matrices, assembly yields
\begin{align}
K_0&=K_{\rm free}+\sum_jL_j^\T K_{c,j}L_j,
& C&=C_{\rm free}+\sum_jL_j^\T C_{c,j}L_j,\nonumber\\
M\ddot{\mathbf q}+C\dot{\mathbf q}+K_0\mathbf q
&=B_vW_v+S_T^\T\mathbf z_{\rm cc}.
\label{eq:mechanics}
\end{align}
The matrix $M$ is the full mechanical inertia matrix. $K_{\rm free}$ and $C_{\rm free}$ are structural stiffness and damping before contact assembly. The sums add every contact's contribution, giving stiffness $K_0$ and damping $C$; $K_0$ excludes the cardiac attachment stiffness. All these matrices have one row and column per entry of $\mathbf q$, with the generalized-coordinate units described in Section~\ref{sec:cardiac}. This assembly includes the conjugate reaction on every body/thorax coordinate contributing to contact motion. For explicit bed coordinates $\mathbf q_p$, the bed or plate subsystem can equivalently be written
\begin{equation}
M_p\ddot{\mathbf q}_p+C_p\dot{\mathbf q}_p+K_p\mathbf q_p
=\sum_j\widetilde B_{p,j}^\T\mathbf f_j,
\label{eq:plate}
\end{equation}
where $\mathbf q_p$ contains only bed/plate coordinates. The matrices $M_p,C_p,K_p$ are its inertia, damping, and stiffness, and $\widetilde B_{p,j}$ maps these bed-only coordinates to the displacement at contact $j$. It differs from $B_{p,j}$, which acts on the full system. The bed support conditions are included in $K_p,C_p$. A modal plate, beam, or finite-element representation determines these matrices. Contact force is therefore the input to the bed; the bed's motion also changes the contact force through Eq.~\eqref{eq:contact}.

The cardiac attachment forces in Eq.~\eqref{eq:mechanics} are evaluated from the coupled wall mechanics and relative attachment motion in Section~\ref{sec:cardiac}. Their stiffness and damping contributions are included through $\mathbf z_{\rm cc}$ and must not also be added independently to $K_0$ and $C$. A fixed transfer function can describe a linear mechanical subsystem with a prescribed load history, but it does not replace the coupled physiological and mechanical solution.

The BSG instrument observes bed velocity along $x$ (transverse), $y$ (longitudinal), and $z$ (vertical):
\begin{equation}
 y_A(t)=\mathcal H_A\!\left[C_AS_p\dot{\mathbf q}(t)\right]+n_A(t).
\label{eq:measurement}
\end{equation}
Here $y_A$ and additive noise $n_A$ have units m/s after velocity calibration. The matrix $S_p$ maps full mechanical coordinates to the sensor point's translational displacement; $C_A$ selects and orients the recorded axes. The operator $\mathcal H_A$ applies calibrated acquisition gain and temporal filtering. With zero noise this gives $f_A(\vartheta)$. In the periodic model, $\mathbf v_p(\omega)=\mathbf a_p(\omega)/(i\omega)$ for $\omega\ne0$, with zero mean velocity. Here $\mathbf a_p=S_p\ddot{\mathbf q}$, $\mathbf v_p=S_p\dot{\mathbf q}$, $i^2=-1$, and $\omega$ is angular frequency. This conversion weights each harmonic by inverse frequency, so acceleration and velocity can rank axes differently. Unknown physical gain, polarity, and timing require calibration or additional uncertainty coordinates.

\begin{figure}[!htbp]
\centering\includegraphics[width=0.92\linewidth]{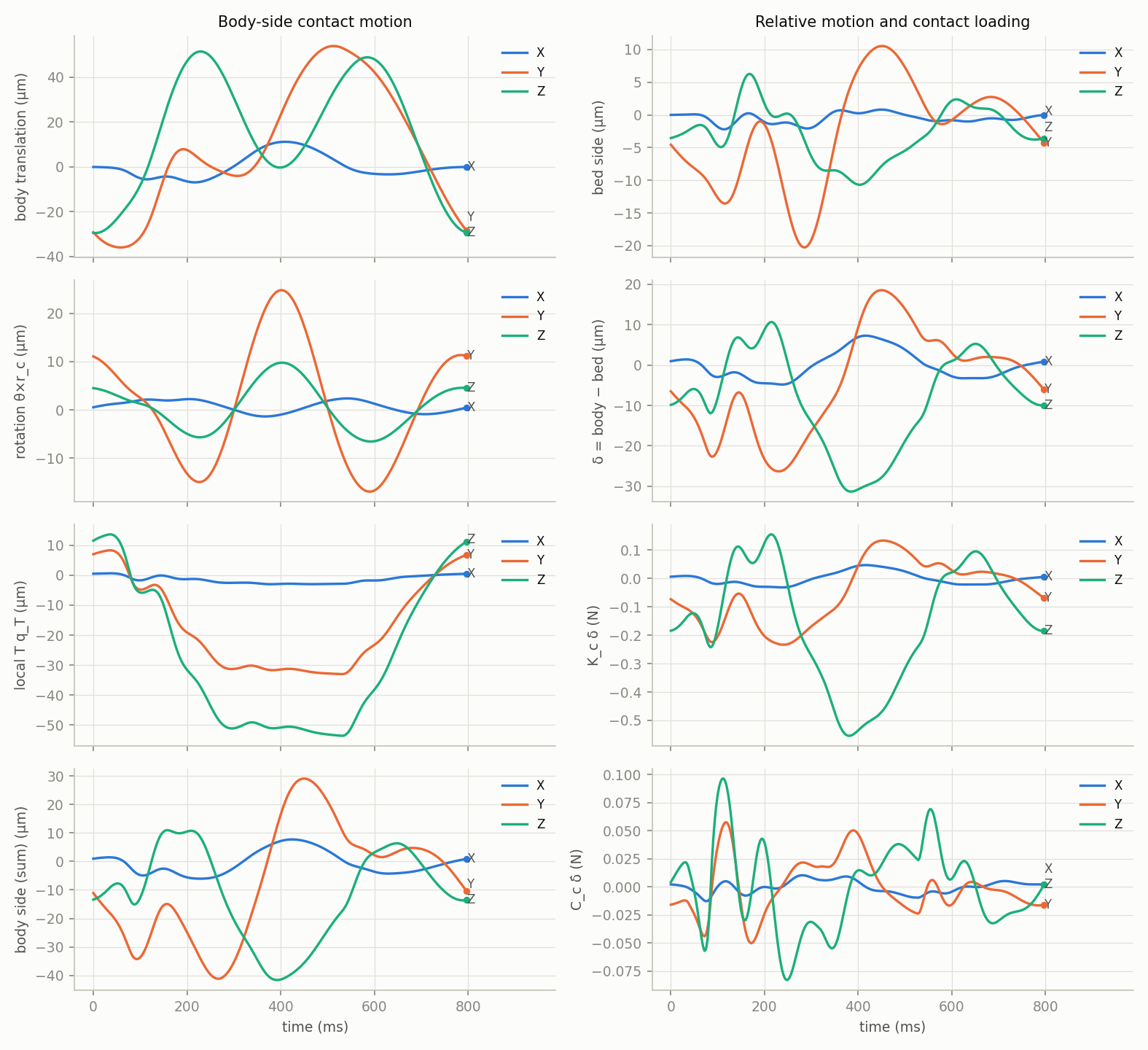}
\caption{Contact transmission in a coupled simulated beat. Left: body translation, small-rotation displacement, local thorax displacement, and their sum at a contact. Right: bed displacement, relative displacement, spring force, and damping force. In the panels, $\boldsymbol\theta$ is the small body-rotation vector (rad), $\mathbf r_c$ a contact lever arm (m), and $T\mathbf q_T$ a local thorax displacement map. The symbols are distinct from the full parameter vector $\vartheta$ and the axis labels. The body and bed displacement sums enter Eq.~\eqref{eq:contact}; displacement is in $\mu$m, force in N, and time in ms. X/Y/Z denote $x/y/z$. This is a representative contact-level illustration, not an experimental source separation.}
\label{fig:contact-transmission}
\end{figure}

\paragraph{Contact Transmission and Source Mixing}
Figure~\ref{fig:time-domain-chain} traces each source through body acceleration and contact force to bed velocity. Figure~\ref{fig:contact-transmission} resolves how body translation, rotation, and local chest deformation combine at a contact. A bed axis mixes both loading pathways; its name does not identify a source fraction.

\FloatBarrier

\section{BSG, SCG, and BCG as Observations of One Coupled Motion}
\label{sec:signals}
The three modalities differ by observation location and kinematic quantity. Here BSG is bed-sensor velocity, SCG is precordial chest-wall acceleration, and BCG is the translational acceleration of the torso rigid-body reference. They are not names for three independent forces. For the same solved mechanical state,
\begin{align}
\mathbf y_{\rm BCG}&=\ddot{\mathbf x}_b,\nonumber\\
\mathbf y_{\rm SCG}&=\ddot{\mathbf x}_b+
 \boldsymbol\alpha_b\times\mathbf r_{\rm SCG}+\Phi_c\ddot{\mathbf q}_c,\nonumber\\
\mathbf y_{\rm BSG}&=S_p\dot{\mathbf q}.
\label{eq:modalities}
\end{align}
Here $\mathbf x_b$ is body translation (m), $\boldsymbol\alpha_b$ body angular acceleration (rad/s$^2$), and $\mathbf r_{\rm SCG}$ the body-reference-to-chest-sensor vector (m). The local chest-mode coordinates $\mathbf q_c$ and the mode-shape matrix $\Phi_c$ are defined so that $\Phi_c\mathbf q_c$ is chest displacement (m). Thus BCG and SCG have units m/s$^2$, while BSG has units m/s. The SCG expression assumes small rotation and omits quadratic rotational terms. The ideal BSG expression precedes the acquisition operator in Eq.~\eqref{eq:measurement}.

BCG reflects momentum exchange among body, heart, and represented blood. In the unrestrained limit,
\begin{equation}
\ddot{\mathbf x}_{b,\rm free}
=-\frac{1}{M_b}\frac{\mathrm d}{\mathrm dt}
 \left(\sum_e\rho_bQ_e\mathbf c_e+m_H\dot{\mathbf u}_H\right).
\label{eq:freebcg}
\end{equation}
Here $M_b$ is torso mass, $m_H$ heart mass (65 and 0.5 kg in this realization), and $\mathbf u_H$ heart centre-of-mass position. The bracket is the represented blood-plus-heart momentum. Bed contact adds an external force and changes this free-body response. SCG additionally contains rotation and local chest deformation driven by the apex connection. Ventricular volume and shape affect that motion; in particular, $\ddot V=\dot Q_{\rm MV}-\dot Q_{\rm Ao}$ links valve-flow transitions to ventricular volume acceleration. Neither volume acceleration alone nor a force amplitude fully specifies SCG \citep{gurev2012mechanisms}.

\begin{figure}[!htbp]
\centering\includegraphics[width=\linewidth]{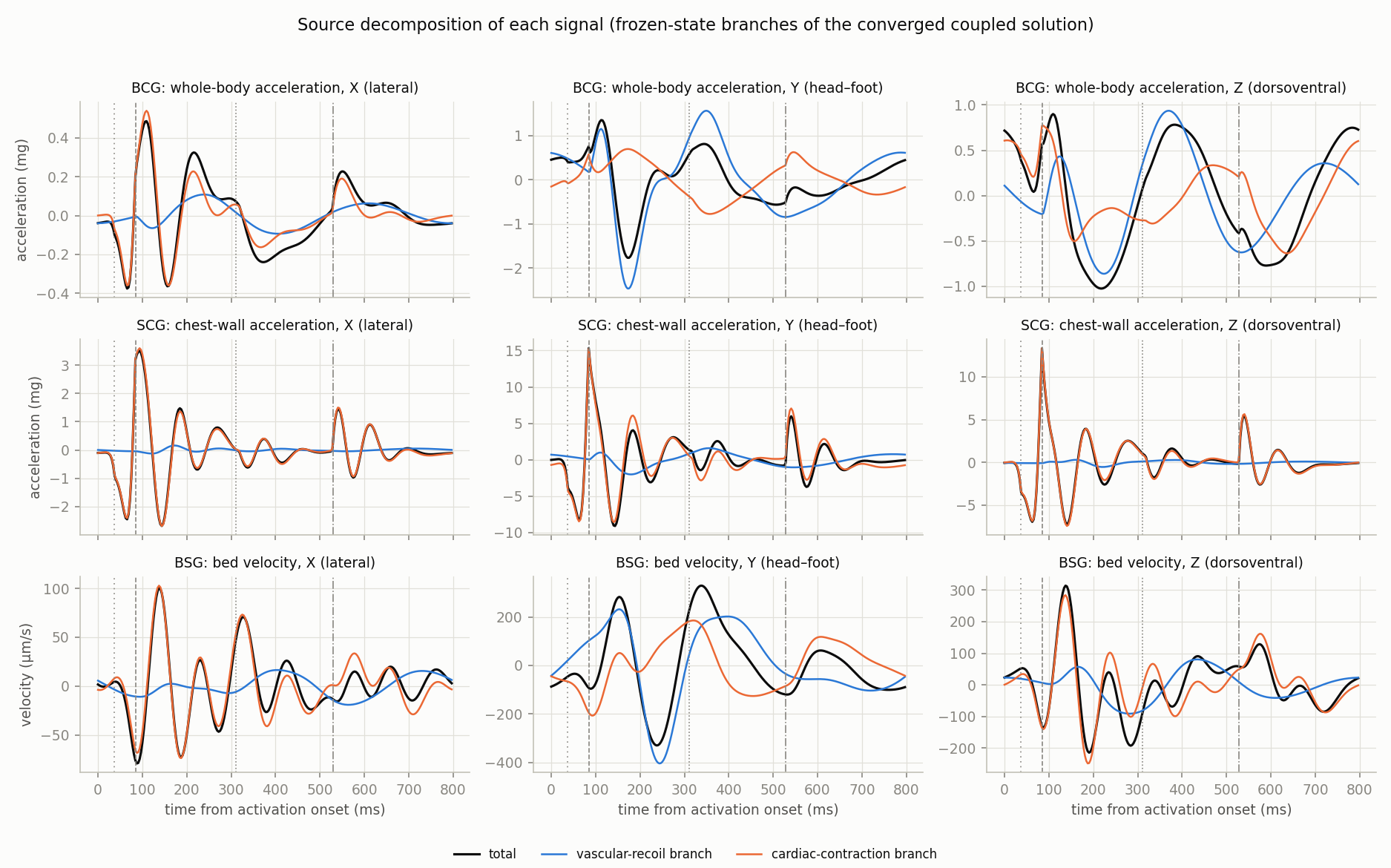}
\caption{BCG, SCG, and BSG from the same periodic coupled solution. Rows show body acceleration, chest-wall acceleration, and bed velocity; columns are $x$ (transverse), $y$ (longitudinal), and $z$ (vertical). Black: total response; blue: vascular recoil; orange: cardiac contraction. BCG/SCG are in mg and BSG in $\mu$m/s, so their plotted amplitudes have different units. The load histories are held at the converged coupled state while the linear mechanical responses are decomposed. Their sums reproduce the total to approximately 0.1\%, 1.1\%, and 0.2\% for BCG, SCG, and BSG, respectively. The figure demonstrates source mixing and distinct observation locations; it is not a validation against measured BCG or SCG.}
\label{fig:modalities}
\end{figure}

In Figure~\ref{fig:modalities}, vascular recoil dominates BCG on $y$ and $z$, while cardiac contraction dominates local SCG. BSG receives both through the body and contacts: the recoil and cardiac-contraction contributions have RMS velocities of $150$ versus $98$ $\mu$m/s on $y$, $45$ versus $91$ on $z$, and $10$ versus $33$ on $x$. Opposing contributions can cancel; these magnitudes are not additive energy fractions. BCG therefore cannot be identified exclusively with recoil, nor SCG exclusively with cardiac contraction \citep{kim2016mechanism,rabineau2021closedloop}.

The source decomposition holds the coupled load histories fixed. Removing a source and re-solving would also change feedback and need not give additive responses. The chest-mode frequencies (12 and 17 Hz), sensor location, apex attachment, and dimensionless chest-wall participation $\chi$ are unvalidated assumptions. Combining modalities adds observation rows and may help constrain source versus transmission effects, but no practical recovery improvement is established here; placement, skin coupling, respiration, and sensor calibration require separate assessment.
\FloatBarrier

\FloatBarrier
\Needspace{8\baselineskip}
\section{Observability and the Value of Multiple Axes}
\label{sec:observability}
\subsection{Concepts and Target Recovery}
Observability here concerns whether the vibration determines a chosen target, even if some individual parameters remain uncertain. Its noiseless first-order criterion is
\begin{equation}
\boxed{\Null(J_A)\subseteq\Null(G)
\quad\Longleftrightarrow\quad\Row(G)\subseteq\Row(J_A).}
\label{eq:observable}
\end{equation}
\textbf{Spaces, directions, and their roles.} $\Null$ denotes a matrix's null space: the parameter changes it maps to zero. Thus $\Null(J_A)$ contains changes invisible to the observation, and $\Null(G)$ contains changes that preserve the target to first order. The symbol $\Row$ means \emph{row space}, the set of all linear combinations of a matrix's rows. Each row of $J_A$ gives the parameter sensitivities of one observed axis/time sample; their linear combinations describe which parameter combinations the observation can distinguish to first order. Rows of $G$ give the corresponding target sensitivities. The inclusion $\Row(G)\subseteq\Row(J_A)$ means that every target row can be formed from observation rows. Equivalently, there is a linear map $L_g$ such that $G=L_gJ_A$ and $\delta g=L_g\delta y_A$ in the noiseless linearization; this map may still amplify noise strongly. The symbols $\subseteq$ and $\Longleftrightarrow$ mean inclusion and logical equivalence. $\Col$ denotes column space, the possible output changes generated by a matrix, and $\operatorname{span}$ denotes all linear combinations of specified vectors. A \emph{direction} is a simultaneous pattern of parameter changes, not a spatial sensing axis. A nuisance direction is defined relative to the selected target and is target-preserving only when the target constraint is satisfied.

An observation-invisible change that changes the target violates Eq.~\eqref{eq:observable}. Large waveform sensitivity alone does not establish recovery: different parameter changes can produce the same waveform change. Adding axes intersects their null spaces,
\begin{equation}
\Null(J_{xyz})=\Null(J_x)\cap\Null(J_y)\cap\Null(J_z),
\label{eq:stacking}
\end{equation}
where $J_x,J_y,J_z$ are the single-axis observation Jacobians and $J_{xyz}$ stacks all their rows. The intersection $\cap$ retains only parameter changes invisible on every axis, so additional axes can remove ambiguities. Three axes do not imply three independent physiological degrees of freedom, nor does a larger score establish that every pair is inadequate.

\begin{table}[!htbp]
\centering\small
\caption{Distinct roles of the observability diagnostics. Every conclusion is conditional on the declared model, parameter family, and measurement convention.}
\label{tab:metric-roles}
\begin{tabularx}{\linewidth}{@{}P{0.23\linewidth}P{0.35\linewidth}Y@{}}
\toprule Quantity & Question addressed & Interpretation\\\midrule
Local target observability & Can a parameter change leave the observation unchanged while changing the target? & Eq.~\eqref{eq:observable} is an exact first-order criterion, not a finite-range recovery guarantee.\\[3pt]
Structural projection residual & How much of the target derivative lies outside the observation row space? & Zero is ideal. This residual is different from the compensation residual below.\\[3pt]
Coverage $\rho_{\rm soft}$ & How strongly are target derivatives covered by regularized observation directions? & Larger means stronger coverage under fixed scaling; it is not prediction accuracy.\\[3pt]
Compensation fraction $\bar r_A^{\mathcal C}$ & How much target-induced observation remains after the best allowed target-preserving compensation? & Larger means less cancellation by the specified family $\mathcal C$. It is not a data-fitting error.\\[3pt]
Finite ambiguity $\mathcal B_A(\epsilon)$ & How different can targets be when admissible states have indistinguishable observations? & Smaller means less worst-case ambiguity under the stated parameter bounds and observation tolerance.\\\bottomrule
\end{tabularx}
\end{table}
\FloatBarrier

\subsection{Soft Coverage and Target-Preserving Compensation}
Comparisons require explicit scales because the parameters and outputs have different units. Let $D_\vartheta$ be a diagonal matrix of positive parameter step sizes, with each diagonal entry in the corresponding parameter's units. Write $\delta\vartheta=D_\vartheta\delta\xi$, where $\delta\xi$ is a dimensionless parameter perturbation. The observation weighting matrix $S_y$ converts waveform differences to dimensionless, optionally noise-weighted values; $S_g$ similarly scales target differences. Their dimensions match the observation and target arrays. Define
\begin{equation}
\widetilde J_A=S_yJ_AD_\vartheta,\qquad
\widetilde G=S_gGD_\vartheta,\qquad
M_A=\widetilde J_A^\T\widetilde J_A.
\label{eq:scaling}
\end{equation}
The matrices $\widetilde J_A$ and $\widetilde G$ are dimensionless observation and target Jacobians. The square matrix $M_A$ measures observation sensitivity in parameter space; it is distinct from the physical inertia matrix $M$. A regularized coverage score is
\begin{equation}
\rho_{\rm soft}^2=
\frac{\tr\!\left[\widetilde G M_A(M_A+\tau_\rho^2I)^{-1}\widetilde G^\T\right]}
{\|\widetilde G\|_F^2},\qquad \widetilde G\ne0.
\label{eq:soft}
\end{equation}
Here $\rho_{\rm soft}$ is a dimensionless coverage score between zero and one. The matrix $I$ is the identity in parameter space, $\tau_\rho>0$ is a dimensionless regularization level, and superscript $-1$ denotes a matrix inverse. The trace $\tr$ adds diagonal entries; the squared Frobenius norm $\|\widetilde G\|_F^2$ adds the squares of all target-Jacobian entries. The score weights target-changing directions according to how strongly they appear in the observations. Its value depends on these scales and on $\tau_\rho$, which is a diagnostic regularization scale rather than a measured noise level. Under fixed conventions, stacking observations cannot decrease the score; changing the uncertainty family is a different operation.

For comparison, a structural projection residual can be written as $r_{\rm struct}=\|\widetilde G(I_\vartheta-\widetilde J_A^\dagger\widetilde J_A)\|_F/\|\widetilde G\|_F$, with $I_\vartheta$ the identity in parameter space and $\dagger$ the Moore--Penrose pseudoinverse. The product $\widetilde J_A^\dagger\widetilde J_A$ projects onto the observation row space. Thus $r_{\rm struct}=0$ is the ideal structural case, whereas a larger \emph{compensation} fraction below means stronger separation. A numerical pseudoinverse retains singular directions according to its declared rank cutoff. Neither a small structural residual nor high $\rho_{\rm soft}$ alone establishes stable recovery in noise.

\Needspace{8\baselineskip}
\textbf{Raw and normalized compensation residuals.} For one scalar target, let $b$ be its nonzero row in $\widetilde G$, the derivative of the scaled target with respect to the scaled parameters. The column $h_g=b^\T/(bb^\T)$ is the smallest Euclidean-norm parameter direction giving a unit target change, since $bh_g=1$. A normalized compensation residual is
\begin{equation}
r_{A,\rm raw}^{\mathcal C}=\min_{v\in\mathcal C}\|\widetilde J_A(h_g+v)\|,
\qquad
\bar r_A^{\mathcal C}=\frac{r_{A,\rm raw}^{\mathcal C}}{\|\widetilde J_Ah_g\|},
\quad \mathcal C\subseteq\Null(b),\quad \|\widetilde J_Ah_g\|>0.
\label{eq:residual}
\end{equation}
Here $r_{A,\rm raw}^{\mathcal C}$ is the unnormalized norm in the declared scaled observation metric; the numerical results below report the normalized fraction $\bar r_A^{\mathcal C}$. The vector $v$ is a dimensionless compensating parameter-change vector, and $\mathcal C$ is the allowed linear subspace of target-preserving changes, so $bv=0$. The unqualified norm $\|\cdot\|$ is the Euclidean norm, the square root of the sum of squared vector entries. The minimum chooses the allowed compensator that leaves the least observable change. The fraction $\bar r_A^{\mathcal C}$ is dimensionless and lies between zero and one. A small value means that allowed target-preserving directions can imitate most of the target-changing observation response. 

\textbf{Allowed compensation families.} The all-direction family takes $\mathcal C=\Null(b)$; a mechanical family restricts the allowed changes. Since $b(h_g+v)=1$, compensation hides the observation response while retaining the imposed target change. With the same $b$, $\widetilde J_A$, and normalization, enlarging $\mathcal C$ cannot increase the residual. This is a statement about the compensation family, not a monotonicity theorem for normalized residuals when adding sensing axes. The ratio is undefined if the uncompensated response is zero; that direction is already observation-invisible. For a scalar target with a nonzero gradient, a strictly positive all-direction residual indicates local first-order separation, while a small value indicates weak separation rather than an exact impossibility result. 

With the remaining coordinates fixed, $\delta g=G_\psi\delta\psi+G_\eta\delta\eta$. A mechanical perturbation needs a feasible physiological compensator satisfying
\begin{equation}
G_\psi\delta\psi=-G_\eta\delta\eta,
\qquad J_{\eta\mid g}=J_\eta-J_\psi G_\psi^\dagger G_\eta,
\label{eq:conditional}
\end{equation}
Here $G_\psi,G_\eta$ and $J_\psi,J_\eta$ are the physiological and mechanical column blocks of the target and selected-axis observation Jacobians. The symbol $\dagger$ denotes the Moore--Penrose pseudoinverse. The matrix $J_{\eta\mid g}$ describes the remaining waveform response to mechanical changes after the particular minimum-norm physiological compensation $\delta\psi=-G_\psi^\dagger G_\eta\delta\eta$; the vertical bar means ``at fixed target.'' This compensator is feasible only if $-G_\eta\delta\eta$ lies in $\Col(G_\psi)$, the space of target changes attainable through physiological perturbations. Other target-preserving physiological changes can be added when the compensation family allows them. Waveform targets require their full multicomponent target subspace, rather than a single arbitrary pressure sample.

\subsection{Tolerance, Pressure Reference, and Finite Uncertainty}
Exact null-space conditions use mathematical zero. Numerical tests require an explicit cutoff. Write the full singular value decomposition as $\widetilde G=U_g\Sigma_gV_g^\T$. The matrices $U_g$ and $V_g$ contain orthonormal target-space and scaled-parameter-space directions; $\Sigma_g$ contains the nonnegative singular values $\sigma_k$ in decreasing order. Each column $v_k$ of $V_g$ is a parameter direction, and $\sigma_k$ measures the target change along it. Exact right-null directions have $\sigma_k=0$, including any additional zero directions of a wide matrix. If the largest singular value $\sigma_1>0$, approximate target preservation uses
\begin{equation}
\mathcal N_{\epsilon_g}=\operatorname{span}\{v_k:\sigma_k\le\epsilon_g\sigma_1\},
\qquad \|\widetilde Gv\|\le\epsilon_g\sigma_1\|v\|\quad(v\in\mathcal N_{\epsilon_g}).
\label{eq:tolerance}
\end{equation}
Here $\epsilon_g$ is a dimensionless relative target-preservation tolerance. The space $\mathcal N_{\epsilon_g}$ contains all linear combinations (the span) of the retained $v_k$, and $v$ is any scaled parameter perturbation in that space. The inequality bounds the local target change per unit scaled parameter displacement. It is not a clinical tolerance or a bound on arbitrarily large compensation. The numerical comparisons distinguish a relative target tolerance $\epsilon_g=10^{-3}$, a stricter numerical rank cutoff $10^{-9}$, and an absolute soft regularization $\tau_\rho\simeq1.16\times10^{-4}$. The latter acts on the scaled observation Jacobian, not on target preservation. Tightening the target cutoff leaves the reported scalar residuals unchanged but raises all-direction waveform residuals from $0.01$--$0.15$ to $0.35$--$0.79$ across the evaluated cases. This reflects a changed admissible compensation family.

An explicit reporting offset provides a simple structural obstruction. Let $p_{\rm internal}$ be the model's pressure at the selected site and $p_{\rm reported}=p_{\rm internal}+c_{\rm ref}$ the reported pressure, all expressed in the same pressure unit. If $c_{\rm ref}$ does not enter the dynamics, then $\partial_{c_{\rm ref}}y_A^{(0)}=0$ while $\partial_{c_{\rm ref}}MAP=1$ for MAP computed from reported pressure. Absolute pressure cannot then be recovered mechanically without a pressure reference. PP and centered/normalized waveforms remove this particular offset, although other ambiguities remain.

Local projections allow unbounded coefficients. Practical recovery requires a finite domain $\Omega$ of admissible physiology and mechanics and a noise-calibrated observation metric $d_A$. Define
\begin{equation}
\mathcal B_A(\epsilon)=
\sup_{\substack{\vartheta_1,\vartheta_2\in\Omega\\
d_A(f_A(\vartheta_1),f_A(\vartheta_2))\le\epsilon}}
 d_g\!\left(\mathfrak g(\vartheta_1),\mathfrak g(\vartheta_2)\right).
\label{eq:bounded}
\end{equation}
Here $\vartheta_1,\vartheta_2$ are any two admissible parameter vectors in $\Omega$. The metric $d_A$ measures their predicted waveform difference under the declared noise weighting, and $\epsilon$ is the largest observation difference treated as indistinguishable, in the same units as $d_A$. The target metric $d_g$ measures target difference, for example absolute SV difference in mL; it is distinct from the direction $h_g$ above. The supremum $\sup$ is the least upper bound over all qualifying pairs. Thus $\mathcal B_A(\epsilon)$ is the worst-case target ambiguity, in the units of $d_g$. This observation tolerance $\epsilon$ is distinct from the relative target tolerance $\epsilon_g$ and soft regularization $\tau_\rho$. A claim of three-axis necessity requires xyz to meet a declared target-error criterion while every fixed pair fails under the same domain and measurement conditions.

\subsection{How the Numerical Indicators Are Calculated}
\label{sec:calculation}
The numerical comparison uses the explicit wall-mechanics closure of Section~\ref{sec:cardiac}, three uncertain body--bed contacts, and deformation modes. The simulated operating point has $HR=75$ beats/min, $SV=66.85$ mL, $MAP=87.63$ mmHg, and $PP=47.3$ mmHg. All seven axis combinations are evaluated at this same point.

\textbf{Which quantities may change.} The primary uncertainty vector has 35 coordinates, grouped in Table~\ref{tab:compensation-parameters}. The 23 mechanical coordinates comprise the body/contact/bed and local-mode group (8), multi-contact group (5), deformation group (3), and cardiac attachment/orientation group (7). All-direction compensation admits joint changes across the full declared vector, subject to the selected target constraint. Mechanical compensation restricts the candidate changes to the mechanical family and requires an explicit rule for enforcing target preservation under feedback. Equation~\eqref{eq:conditional} gives one admissible construction using a physiological correction; the group counts and tabulated fractions alone do not specify the exact numerical compensation basis. Arbitrary physiological or residual-contraction directions are not automatically admitted.

\begin{table}[!htbp]
\centering\small
\caption{Parameter groups admitted to the primary compensation analysis. These are uncertain coefficients or timing parameters, not independently fitted samples of pressure, force, or displacement.}
\label{tab:compensation-parameters}
\begin{tabularx}{\linewidth}{@{}P{0.25\linewidth}P{0.06\linewidth}Y@{}}
\toprule Group & Count & Meaning and permitted role\\\midrule
Physiology & 7 & $HR,c,V_{\rm str},R_{\rm sys},C_{\rm art},\kappa_{\rm prox},\zeta_{\rm term}$, defined in Table~\ref{tab:psi}. They may change jointly in all-direction compensation; a specified subset or combination can correct target changes caused by mechanics.\\[3pt]
Body/contact/bed and local-mode parameters & 8 & Changes in mechanical transmission and local thorax response. Only group-level parameterization is specified here; a full componentwise coordinate and scale map is needed for replication.\\[3pt]
Multi-contact parameters & 5 & Uncertain properties of the three-contact representation. Individual coordinate identities and scales are not fully enumerated here.\\[3pt]
Deformation parameters & 3 & Uncertain properties of the additional deformation representation. These parameter coordinates must be distinguished from the time-varying modal displacement states; their componentwise scale map is not fully specified.\\[3pt]
Cardiac attachment/orientation & 7 & Three connection stiffnesses, one common connection-damping parameter, chest-wall participation, and heart azimuth and elevation. These are mechanical candidates; their feedback can also change the target.\\[3pt]
Residual cardiac contraction & 4 & Calcium rise time, calcium decay time, activation span, and a cross-bridge-rate factor. All-direction compensation permits their joint variation. Each can change SV or pressure and is not individually target-preserving.\\[3pt]
Pressure-reference offset & 1 & $c_{\rm ref}$ changes reported absolute pressure but not the dynamics. It is silent for SV and PP and obstructs absolute-pressure recovery when uncalibrated.\\\bottomrule
\end{tabularx}
\end{table}
\FloatBarrier

The declared cardiac attachment scales are $\pm30\%$ for stiffness, damping, and participation, and $\pm10^\circ$ for orientation. Residual contraction scales are $\pm20$ ms for calcium rise, $\pm80$ ms for calcium decay, $\pm30$ ms for activation span, and $\pm0.25$ for the dimensionless cross-bridge-rate factor. These are analysis assumptions, not measured population uncertainties. In the unbounded local projection they define sensitivity scales, not enforced box constraints on the optimizing coefficients. Independent pulse amplitude/timing coefficients and an additional discrepancy-potential family are outside this 35-coordinate calculation. Adding such freedom defines a different comparison.

\textbf{What is optimized and what is held fixed.} For a scalar target, the optimizer chooses coefficients of target-preserving parameter directions; it does not adjust every pressure or BSG sample independently. The operating point, sampled Jacobians, target definition, phase grid, parameter scales, and observation weights remain fixed during this local least-squares calculation. Pressure, flow, cardiac motion, and bed velocity respond through the coupled derivatives. With SV as the selected target, the compensator preserves SV but may change PP, CO, and other targets; with PP selected, it preserves PP but may change SV and CO. Joint target constraints require the full target Jacobian and a newly constructed compensation space.

The calculation proceeds as follows.
\begin{enumerate}[leftmargin=*,itemsep=3pt,topsep=4pt]
\item \textbf{Solve the common periodic state.} Integrate the circulation, cardiac mechanics, and reciprocal body--bed system jointly to a repeating beat. The numerical realization uses fourth-order Runge--Kutta integration with eight substeps per output sample, with 256 phase samples per beat. Evaluate the targets and bed-velocity waveforms from the same solution and align all derivatives on the same normalized beat-phase grid.
\item \textbf{Differentiate the complete solution.} Forward-mode automatic differentiation gives $J_A$ and $G$, including feedback and the implicit ventricular-pressure solve. Selecting $x$, $y$, $z$, $xy$, $xz$, $yz$, or $xyz$ changes the observation rows, while retaining the same parameter coordinates and target definition. The derivatives are total derivatives, not derivatives of a fixed prescribed pressure input.
\item \textbf{Apply one declared scaling.} Form $\widetilde J_A$ and $\widetilde G$ with Eq.~\eqref{eq:scaling}. Keep parameter scales, target scales, per-channel observation weights, and $\tau_\rho=1.16\times10^{-4}$ fixed across axis combinations. The threshold follows a common reference convention converted to velocity by the reference-Jacobian singular-value ratio $0.0323$ s. It is not independently retuned for each axis. The numerical observation metric is a scaled Euclidean metric, not an empirically noise-whitened covariance model.
\item \textbf{Calculate coverage.} Form $M_A=\widetilde J_A^\T\widetilde J_A$, evaluate Eq.~\eqref{eq:soft}, and take its nonnegative square root. The plotted value is $\rho_{\rm soft}$, not $\rho_{\rm soft}^2$. It measures regularized target-sensitivity coverage; it is not a correlation coefficient, prediction accuracy, or percentage of clinically recovered information.
\item \textbf{Calculate compensation separately.} Construct the allowed target-preserving family, with the target cutoff in Eq.~\eqref{eq:tolerance}, and solve the least-squares cancellation problem in Eq.~\eqref{eq:residual}. The mechanical family restricts the candidate coordinates and must specify how target preservation is enforced under feedback; Eq.~\eqref{eq:conditional} provides a conditional construction. The all-direction family admits every allowed target-preserving latent change. These choices define different residuals.
\end{enumerate}

For a scalar target, let $N_{\mathcal C}$ have columns forming an orthonormal basis of the allowed scaled parameter subspace $\mathcal C$. Write $v=N_{\mathcal C}w$, where $w$ is the vector of freely optimized real coefficients. Then
\begin{align}
 s_A&=\widetilde J_Ah_g,\qquad H_{\mathcal C}=\widetilde J_AN_{\mathcal C},\nonumber\\
 w_*&=-H_{\mathcal C}^{\dagger}s_A,\qquad
 \bar r_A^{\mathcal C}
 =\frac{\|(I_y-H_{\mathcal C}H_{\mathcal C}^{\dagger})s_A\|}{\|s_A\|}.
 \label{eq:residual-computation}
\end{align}
Here $s_A$ is the dimensionless response to the unit target-changing direction $h_g$, $H_{\mathcal C}$ contains the compensating observation responses, and $I_y$ is the identity in observation space. The pseudoinverse gives a minimum-norm least-squares coefficient vector $w_*$. The product $H_{\mathcal C}H_{\mathcal C}^{\dagger}$ projects onto the compensating-response space. The actual physical parameter changes represented by this linear construction are $\delta\vartheta_{\rm comp}=D_\vartheta N_{\mathcal C}w_*$ and $\delta\vartheta_{\rm total}=D_\vartheta(h_g+N_{\mathcal C}w_*)$. The compensator preserves the target to the specified tolerance; the total change retains the imposed target perturbation.

The numerator in Eq.~\eqref{eq:residual-computation} is the response remaining after the best cancellation, and the denominator is its uncompensated magnitude. For a nonzero compensating-response space, the scalar fraction equals $\sin\theta_A$, where $\theta_A$ is the smallest angle between $s_A$ and $\Col(H_{\mathcal C})$; near alignment permits strong cancellation. A residual fraction of $0.509$ means that $50.9\%$ of this response norm remains under the specified family, not that SV is estimated with $50.9\%$ accuracy. No coefficient bound or prior penalty is imposed in this projection. Thus a small residual identifies a local confounding mechanism, but does not establish that its minimizing parameter changes lie within physiological bounds. A bounded nonlinear fit or a prior-weighted analysis is a separate calculation.

\textbf{Numerical checks and replication scope.} The integration check reports bed-waveform normalized RMS error $8.9\times10^{-6}$ between eight and 32 substeps. Coupled acceleration-Jacobian spot checks agree with central differences within $3.3\times10^{-6}$ relative across the three contact models. For velocity, the checks agree within $3\times10^{-3}$ for heart rate (step-size limited) and $2\times10^{-6}$ for contractility, with the beat-duration correction $-\mathbf v_p/HR$ included in the heart-rate derivative at fixed normalized beat phase. These are numerical checks at tested states, not measured recovery errors. Exact independent replication additionally requires the 35-entry coordinate and scale ledger, phase sampling, target-waveform representation, Jacobian arrays, mechanical-family selection and compensation basis, and numerical rank policy. The reported scalar residuals do not include their minimizing coefficient vectors or componentwise parameter shifts, so they cannot identify which particular parameter changes achieved the cancellation. The waveform residual rows below retain the reported multicomponent summaries; their precise aggregation norm is not fully specified by the available methodological description. They must not be reconstructed by applying the scalar formula to a single pressure sample.

The primary synthetic family assumes a known beat-phase origin and fixed global amplitude calibration; an independent unknown global time shift or global gain/body-mass scale is not included. ``All-direction'' therefore means all target-preserving directions of the declared 35-coordinate family, not every source of uncertainty in an uncalibrated measurement. These observation-side quantities require independent calibration or explicit additional coordinates before practical recovery is assessed.

\subsection{Results across Scalar and Waveform Targets}
\label{sec:all-target-results}
\begin{figure}[!htbp]
\centering\includegraphics[width=\linewidth]{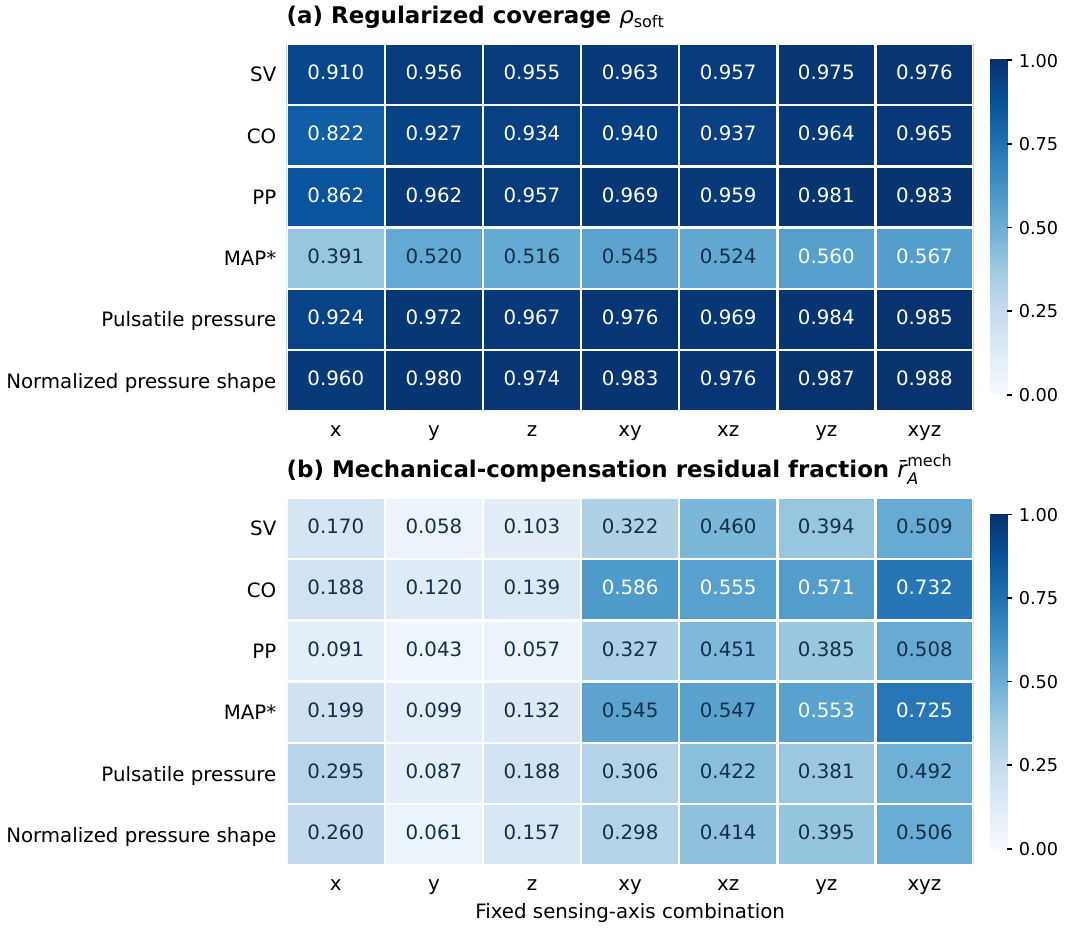}
\caption{Six-target comparison across all seven fixed axis sets for the same 35-coordinate coupled model. (a) $\rho_{\rm soft}$ from Eq.~\eqref{eq:soft}, at $\tau_\rho=1.16\times10^{-4}$. (b) Mechanical-compensation residual fractions, with relative target-preservation tolerance $\epsilon_g=10^{-3}$; scalar rows follow Eq.~\eqref{eq:residual-computation}, and waveform rows are the reported multicomponent summaries described in Section~\ref{sec:calculation}. All entries are dimensionless; both color scales run from zero to one. Larger values mean stronger regularized coverage in (a), and less cancellation by the specified mechanical family in (b). *MAP depends on the assumed scale of the mechanically silent pressure-reference offset; its values are not directly comparable with offset-invariant targets. Numerical values are transcribed at their reported three-decimal precision.}
\label{fig:all-target-axes}
\end{figure}

Figure~\ref{fig:all-target-axes} includes all six targets available in the matched primary comparison: SV, CO, PP, MAP, pulsatile pressure $p_c$, and normalized pressure shape $u$. It separates the coverage score from the mechanical-compensation residual rather than combining them under one symbol.

For SV, CO, and PP, $xyz$ coverage is $0.976$, $0.965$, and $0.983$, respectively; for pulsatile pressure and normalized shape it is $0.985$ and $0.988$. The best coverage pair is $yz$ for all six targets. The longitudinal axis $y$ gives the highest single-axis coverage for SV, PP, MAP, and both waveform targets, while $z$ leads for CO. It also carries the largest baseline bed velocity (RMS $148$ $\mu$m/s, versus $33$ on $x$ and $97$ on $z$). A normalized residual measures the surviving fraction, not its absolute amplitude or strength relative to noise. Strong sensitivity does not imply weak confounding: $y$ has the smallest mechanical residual for every target.

The best mechanical-compensation pair depends on the target: $xz$ for SV, PP, and both waveform targets, $xy$ for CO, and $yz$ for MAP. Relative to these best pairs, the third axis adds $0.049$ for SV, $0.146$ for CO, $0.057$ for PP, $0.172$ for MAP, $0.070$ for pulsatile pressure, and $0.092$ for normalized shape. Coverage and compensation therefore rank the axes differently. MAP has $xyz$ coverage $0.567$ and a mechanical residual $0.725$; the latter does not remove the separate mechanically silent pressure-offset ambiguity.

When all target-preserving latent changes are allowed, $xyz$ residuals fall to $0.014$ for SV and $0.034$ for PP (Table~\ref{tab:scalar-compensation-comparison}). The broader family can reproduce responses left distinct by mechanical variation alone. Complete, equally precise all-direction rows are not available for every target and axis set, so they are not inferred from the mechanical table.

\begin{table}[!htbp]
\centering\small
\caption{Matched scalar compensation fractions for bed velocity at the primary operating point. The first three numerical columns share the same mechanical family; the final column permits all target-preserving directions on $xyz$. Larger fractions mean less cancellation.}
\label{tab:scalar-compensation-comparison}
\begin{tabularx}{\linewidth}{@{}P{0.10\linewidth}YYYY@{}}
\toprule Target & Best single, mechanical & Best pair, mechanical & $xyz$, mechanical & $xyz$, all-direction\\\midrule
SV & $0.170$ ($x$) & $0.460$ ($xz$) & $0.509$ & $0.014$\\[3pt]
PP & $0.091$ ($x$) & $0.451$ ($xz$) & $0.508$ & $0.034$\\\bottomrule
\end{tabularx}
\end{table}

\textbf{Implication for SV and PP recovery.} Only $1.4\%$ and $3.4\%$ of the respective uncompensated response norms remain under all-direction compensation. This indicates weak local separation from concurrent unknown physiology and residual cardiac-contraction variation, even with three axes. These are not estimation-error percentages, and nonzero fractions do not establish a failure of exact first-order uniqueness. Unbounded compensation without empirical noise whitening also does not establish a feasible finite-amplitude ambiguity or a clinical recovery error.

\textbf{What three axes contribute.} The mechanical comparison supports complementary spatial constraints: adding $y$ to the best pair $xz$ raises the residual by $0.049$ for SV and $0.057$ for PP. All channels constrain the same compensator. A new fit over three axes can nevertheless find another compensator, so detecting a discrepancy on an initially held-out axis does not prove unique recovery. Neither universal three-axis necessity nor reliable uncalibrated bed-only SV/PP recovery follows from these local scores.

\FloatBarrier

\textbf{Separate finite-range and measured-data checks.} Target-constrained bounded searches with residual contraction freedom and an additional discrepancy-potential family retain all-axis waveform mismatches of approximately $3.7$--$4.5\%$ for imposed $10\%$ SV/PP changes. Their parameter family and finite optimization differ from the primary 35-coordinate linear projection; these percentages must not be substituted for the residual fractions above. Local search also does not certify a global optimum. In a six-subject comparison with measured bed signals, fitting 23 mechanical coordinates reduces mean calibration waveform NRMSE from $0.96$ to $0.66$, but held-out NRMSE with fixed calibration gains is $1.12$, versus $1.04$ with baseline mechanics. Undocumented sensor units and polarities, unstable fitted gains, and mechanical parameters reaching their bounds prevent this fit from validating the source-to-bed transfer. These empirical results concern waveform prediction, not demonstrated SV/PP recovery.
\FloatBarrier

\Needspace{13\baselineskip}
\subsection{Conclusions Supported by the Theory and Figures}
Shared physiology determines the targets and both cardiovascular loading pathways. Their transmission through body, contacts, and bed mixes the measured signatures. The longitudinal bed-velocity response is strongest at the stated baseline, but is also the easiest for mechanical variation to imitate. Multiple axes improve separation from the specified mechanical family, with the largest gain from one axis to two and further target-dependent gains from the third. The $xyz$ all-direction residuals of $0.014$ for SV and $0.034$ for PP still reveal strong local confounding. Three-axis BSG provides complementary spatial constraints; it does not by itself establish reliable uncalibrated recovery. Practical sufficiency requires justified uncertainty bounds, calibrated observation gain, polarity, timing and noise, and independent validation of the coupled mechanical transfer. The comparison must retain the best fixed dual-axis alternative under the same target-error criterion.

\FloatBarrier

\FloatBarrier
\clearpage
\begin{bsgpartpage}{Numerical Modeling and Validation of BSG}
\part[Numerical Modeling and Validation of BSG]{Numerical Modeling\\and Validation of BSG}
\label{part:dt}
A digital twin (DT) is a computational representation of a physical system whose parameters or state are conditioned on measurements of that system. For BSG, the representation links cardiovascular dynamics to body--bed transmission and the sensor response. The analytical framework in Sections~\ref{sec:framework} and~\ref{sec:observability} supplies the shared-physiology forward model and target-observability conditions. This chapter describes specific computational realizations of the source, transmission, and observation.

This chapter develops a BSG digital-twin framework organized around four-chamber cardiac mechanics, cardiac-to-body excitation, and reduced body--bed transmission. The following sections examine waveform fitting and stroke-volume-related feature transfer, then summarize the implications for BSG reconstruction and physiological interpretation.
\end{bsgpartpage}

\section{A Common Architecture for the BSG Digital Twin}
\label{sec:coupled}
The BSG digital twin connects cardiac physiology to recorded motion through the forward chain
\[
\vct\theta_{\rm cardiac}\xrightarrow{\mathcal G}s(t)
\xrightarrow{\mathcal B}q_{\rm exc}(t)
\xrightarrow{\mathcal H_{\rm BB}}\widehat y_{\rm BSG}(t).
\]
Cardiac physiology supplies the mechanical source histories through $\mathcal G$. The operator $\mathcal B$ constructs and maps the cardiac-to-body excitation, and $\mathcal H_{\rm BB}$ solves body--bed propagation and sensor observation. For the force-driven branch, $q_{\rm exc}$ supplies $\vct f_{\rm body}$. The alternative branch supplies prescribed interface motion $\vct u_\Gamma$. The complete source-to-measurement operator is $\mathcal H=\mathcal H_{\rm BB}\circ\mathcal B$.

The hierarchy separates cardiac source generation, $\mathcal G$, from source-to-measurement transmission, $\mathcal H$:
\[
\vct\theta_{\rm cardiac}
\xrightarrow{\mathcal G} s(t)
\xrightarrow{\mathcal H(\vct\theta_{\rm body},\vct\theta_{\rm bed})}
\widehat y_{\rm BSG}(t).
\]
Here $s(t)$ collects the selected mechanical source components, and $\mathcal H$ includes the specified sensor observation. Section~\ref{sec:physiological-twin} develops the cardiac mechanics underlying $\mathcal G$, and Section~\ref{sec:body-bed-excitation} constructs its mechanical excitation outputs. Section~\ref{sec:simplified-body-bed} estimates effective finite impulse response (FIR) and reduced mechanical forms of $\mathcal H$. The participant-conditioned FIR, $\mathcal H_{\rm subject}^{\rm empirical}$, and the four-region FEM, $\mathcal H_{\rm eff}^{\rm 4R}$, are complementary descriptions of this mapping. The component evaluations in this chapter test cardiac source generation and effective transmission. Section~\ref{sec:numerical-conclusions} summarizes the resulting evidence for BSG reconstruction and physiological interpretation.

\begin{center}
\small
\begin{tabularx}{\linewidth}{@{}P{0.12\linewidth}Y P{0.25\linewidth}@{}}
\toprule
Section & Core question & Output / role\\
\midrule
\ref{sec:physiological-twin} & What mechanical histories does the cardiac model produce? & $s(t)$; cardiac source calculation\\
\ref{sec:body-bed-excitation} & How is that source applied to the body? & $q_{\rm exc}$; excitation interface $\mathcal B$\\
\ref{sec:simplified-body-bed} & Can reduced transfer explain measured BSG? & $\mathcal H_{\rm eff}$; effective realization\\
\ref{sec:numerical-conclusions} & What do the numerical studies establish? & Conclusions for BSG reconstruction and physiological interpretation\\
\bottomrule
\end{tabularx}
\end{center}

In this notation, $\mathcal H_{\rm eff}^{\rm 4R}$ denotes the four-region FEM realization of the source-to-measurement operator, including its excitation mapping $\mathcal B$. The operator $\mathcal H_{\rm BB}$ alone starts at the body excitation. The force-proxy and recoil constructions in Section~\ref{sec:body-bed-excitation} make explicit how the cardiac histories become usable load channels. The reduced models in Section~\ref{sec:simplified-body-bed} combine excitation and propagation at a coarser mechanical resolution while retaining the same source-to-measurement endpoint.

The analytical model in Section~\ref{sec:framework} defines a common set of state variables, interface loads, and observations. Their computational assembly gives the architecture shared by the model representations in this chapter. Following Chapter~\ref{part:theory}, let $\vct r\in\mathbb R^{n_H}$ collect the retained cardiac coordinates and $\vct q\in\mathbb R^{n_B}$ collect the body, thorax, and bed coordinates. Here $n_H$ and $n_B$ are the respective numbers of coordinates. A reduced left-ventricular (LV) realization can assemble these coordinates and their velocities with the valve flows and vascular states as
\begin{equation}
\vct\chi_d=
\left[\vct r^{\mathsf T},\dot{\vct r}^{\mathsf T},Q_{\rm MV},Q_{\rm Ao},
P_{\rm ao},\{P_j,Q_j\}_{j=1}^{N_a},
\vct q^{\mathsf T},\dot{\vct q}^{\mathsf T}\right]^{\mathsf T}.
\label{eq:coupled_state}
\end{equation}
$Q_{\rm MV}$ and $Q_{\rm Ao}$ are the mitral inflow and aortic outflow, in m$^3$/s, with $Q_{\rm Ao}$ corresponding to $Q_{\rm AV}$ in Eq.~\eqref{eq:prelim_hemodynamics}. $P_{\rm ao}$ is aortic-root pressure, in Pa. The $N_a$ downstream arterial units retain pressures $P_j$ (Pa) and flows $Q_j$ (m$^3$/s), with the root pressure listed separately. Each entry of $\vct r$ or $\vct q$ has the units of its generalized coordinate, such as m for translation, rad for rotation, or m$^3$ for cavity volume, and its time derivative has those units per second. Thus $\vct q$ and $\dot{\vct q}$ explicitly contain the mechanical positions and velocities, using the same notation as Section~\ref{sec:mechanics}.

LV pressure, interface displacement, and interface force can be retained as algebraic variables $\vct\chi_a$. Chamber volume is computed from the cardiac coordinates, $V=V(\vct r)$, or retained as an algebraic variable with that constraint. The combined vector $\vct\chi=[\vct\chi_d^{\mathsf T},\vct\chi_a^{\mathsf T}]^{\mathsf T}$ instantiates the complete state $X$ of Eq.~\eqref{eq:state}. With prescribed activation and boundary inputs $\vct u_{\rm in}(t)$ and the parameter vector $\vartheta$ of Eq.~\eqref{eq:theta}, the structure is a differential--algebraic system:
\begin{equation}
\mat E_d(\vct\chi)\dot{\vct\chi}_d=
\vct f_d(\vct\chi_d,\vct\chi_a,\vct u_{\rm in};\vartheta),\qquad
\vct 0=\vct g(\vct\chi_d,\vct\chi_a,\vct u_{\rm in};\vartheta),
\label{eq:dae}
\end{equation}
Here $\mat E_d$ is the coefficient matrix of the time derivatives, $\vct f_d$ collects the corresponding balance terms, and $\vct g=\vct 0$ imposes the algebraic constraints. The output is
\begin{equation}
\vct y_{\rm mech}=
\vct h(\vct\chi_d,\vct\chi_a,\dot{\vct\chi}_d),\qquad
\vct y_{\rm recorded}=\mathcal A\{\vct y_{\rm mech}\}+\vct n .
\label{eq:complete_output}
\end{equation}
$\mathcal A$ denotes the calibrated acquisition operator in its linear range. Valve states make the governing equations piecewise defined. Initial values must satisfy the chamber-volume, flow-compatibility, and interface constraints.

In this coupled representation, LV pressure acts as a hydraulic constraint force compatible with myocardial mechanics and net valve flow. The aortic-root pressure returns the arterial load to the ventricle. Body motion changes the relative cardiac--thorax attachment displacement in Eq.~\eqref{eq:link}. Its stacked interface forces $\vct F_\Gamma$ have units N, with signs defined as forces on the thorax. The heart receives the generalized reaction $-\mat B_H^{\mathsf T}\vct F_\Gamma$, while the thorax receives $+\mat B_T^{\mathsf T}\vct F_\Gamma=\vct z_{\rm cc}$. These generalized loads have the work-conjugate units defined in Section~\ref{sec:cardiac}, including N for translational coordinates and N\,m for rotational coordinates. The pressure and attachment feedback paths connect activation, pressure, flow, recoil, and structural vibration. Consistent recoil timing follows from this shared state alongside pressure and flow.

The four-chamber representation in Section~\ref{sec:physiological-twin} replaces the reduced cardiac coordinates with spatial tissue mechanics and four chamber states.

The physical signal path can be written compactly as
\begin{equation}
\boxed{\begin{gathered}
\text{activation}\ \longrightarrow\
\text{pressure, flow, and wall motion}\ \longrightarrow\
\text{signed interface loads}\\
\longrightarrow\ \text{body--bed motion}\ \longrightarrow\ \text{recorded BSG}
\end{gathered}}
\label{eq:complete_chain}
\end{equation}
with the feedback defined above. BSG amplitude and morphology consequently depend on source dynamics, directional propagation, and measurement response together. Physiological interpretation uses the specified source, propagation, and measurement operators together with their parameters.

\section{Four-Chamber Cardiac Mechanics and Wall Motion}
\label{sec:physiological-twin}
A four-chamber heart model resolves the spatial source behind the analytical cardiovascular balances in Section~\ref{sec:targets}. Its physical sequence is anatomy and fiber orientation, electrical activation, active contraction, chamber loading, and pressure--volume feedback. Our cardiac model couples three-dimensional mechanics to hemodynamics \citep{zhang2027heartintheloop}. The upstream ANSYS/LS-DYNA reference calculation is denoted \emph{Gold}. Its geometry, material state, and active-stress history supply inputs to a JAX mechanical model coupled to a lumped circulation.

\begin{figure}[!htbp]
\centering
\includegraphics[width=\linewidth]{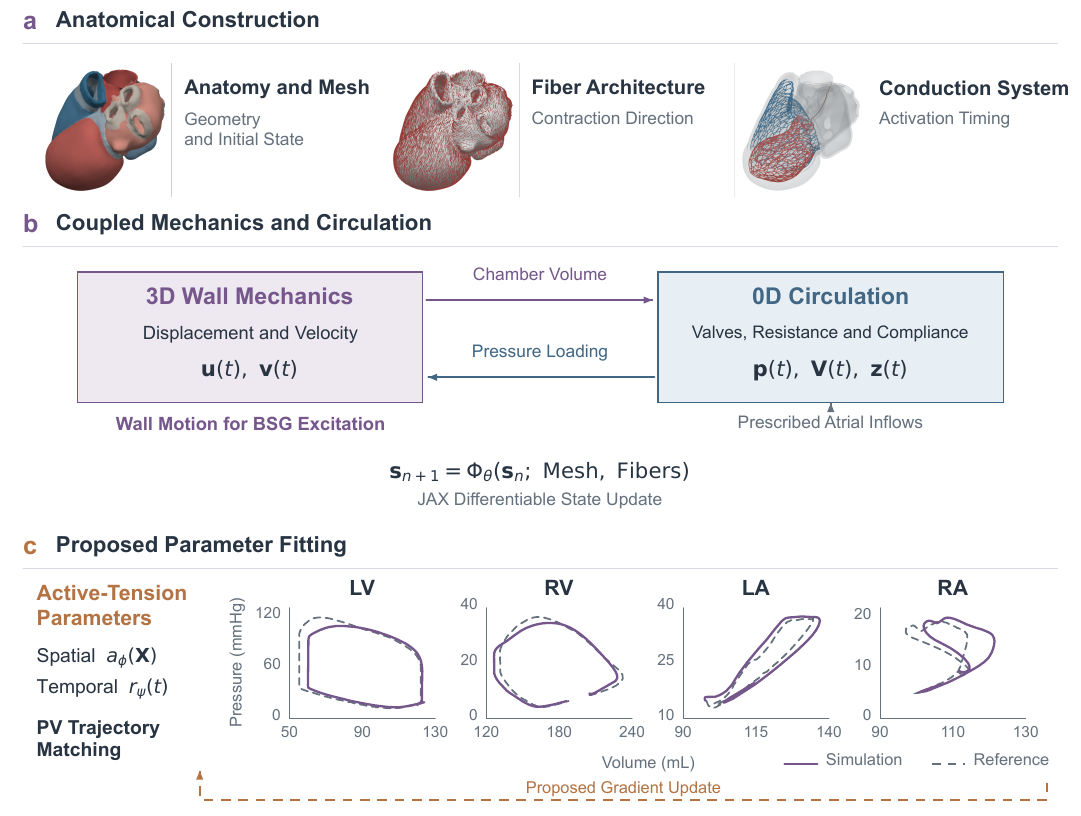}
\caption{Cardiac Model Overview. (a) Anatomical construction. (b) Coupled wall motion and circulation. (c) Proposed fitting from pressure--volume trajectories. Purple and blue arrows show volume and pressure feedback. The dashed ochre path indicates proposed parameter updates \citep{zhang2027heartintheloop}.}
\label{fig:heart_core_overview}
\end{figure}

Figure~\ref{fig:heart_core_overview} connects these components in the cardiac core. Anatomical construction supplies the mesh, fiber directions, and activation history. Three-dimensional mechanics evolves wall displacement and velocity. The enclosed chamber volumes enter the reduced circulation, whose pressures load the moving wall. This feedback generates both the spatial motion needed for a mechanical source and the chamber-level pressure--volume (PV) trajectories used to characterize pumping. Prescribed atrial inflows form the circulation boundary conditions.

The learning notation in the figure distinguishes a spatial active-tension multiplier $a_\phi(\vct X)$ from a temporal drive multiplier $r_\psi(t)$. In the grouped replay used below, $a_\phi$ reduces to the chamber/septal scale $s_{g(e)}$, while $r_\psi(t)=1$. Geometry, prestress, and upstream activation histories are prescribed inputs. Figure~\ref{fig:heart_core_overview}(c) illustrates how selected parameters could be fitted to time-aligned pressure and volume, with wall motion retained as a separate prediction to assess. The displayed PV loops summarize the trajectories. A time-resolved objective also preserves their phase information. Forward comparisons evaluate the cardiac response, while parameter learning requires evaluating the corresponding gradients \citep{zhang2027heartintheloop}. The downstream mapping from wall motion to body--bed excitation is developed in Section~\ref{sec:body-bed-excitation}.

\subsection{Anatomical Geometry and the Loaded Initial State}
A mechanically consistent heart model aligns the mesh coordinates, tissue labels, fiber directions, and initial stresses. The model contains 331,365 four-node tetrahedral cells (TET4) and 72,158 independent solid nodes. Ten material partitions distinguish the left and right ventricles, left and right atria, aorta, pulmonary artery, septum, atrial stiff rings, ventricular base, and atrioventricular isolation region. Active contraction is assigned to the four chambers and septum, comprising 256,051 cells.

\begin{figure}[!htbp]
\centering
\includegraphics[width=0.97\linewidth]{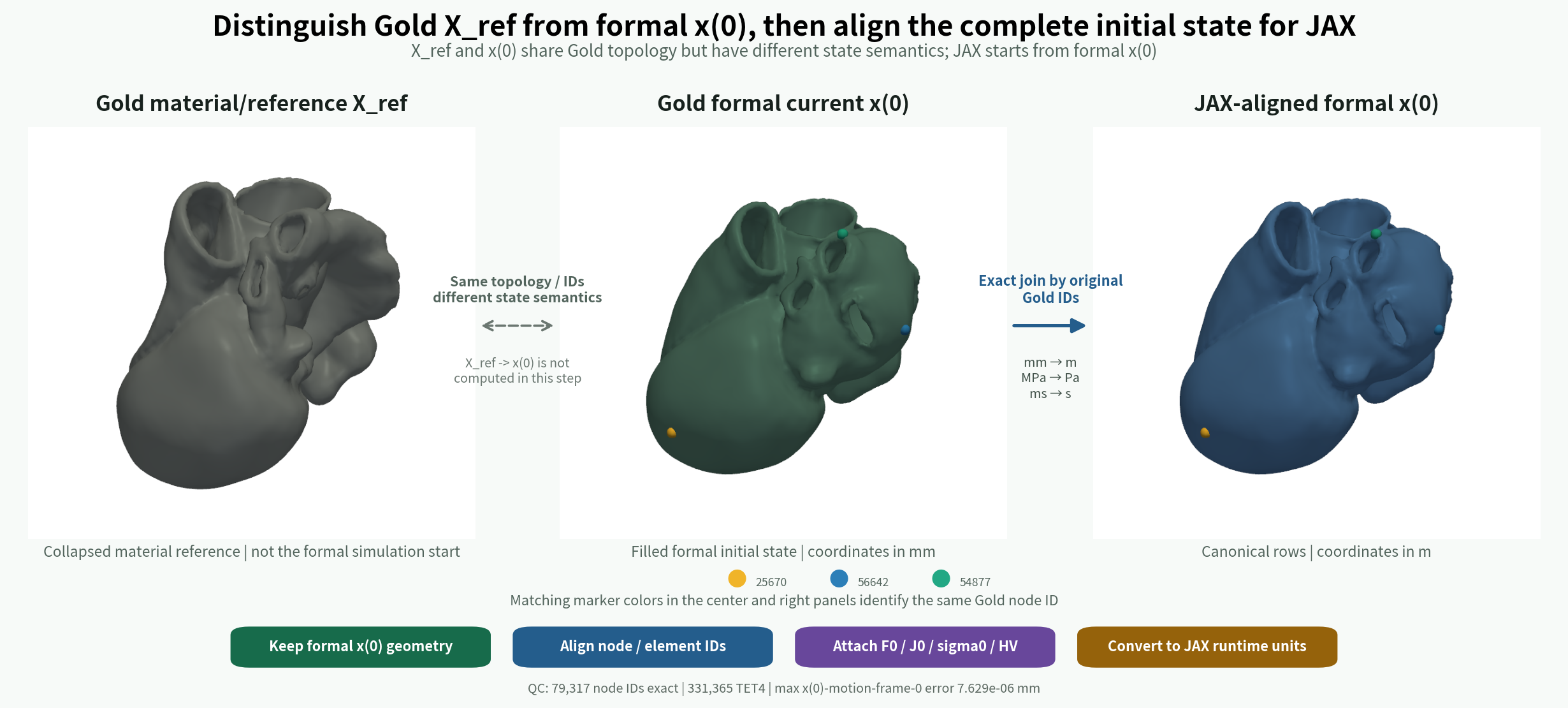}\\[5pt]
\begin{minipage}[c]{0.59\linewidth}
\includegraphics[width=\linewidth]{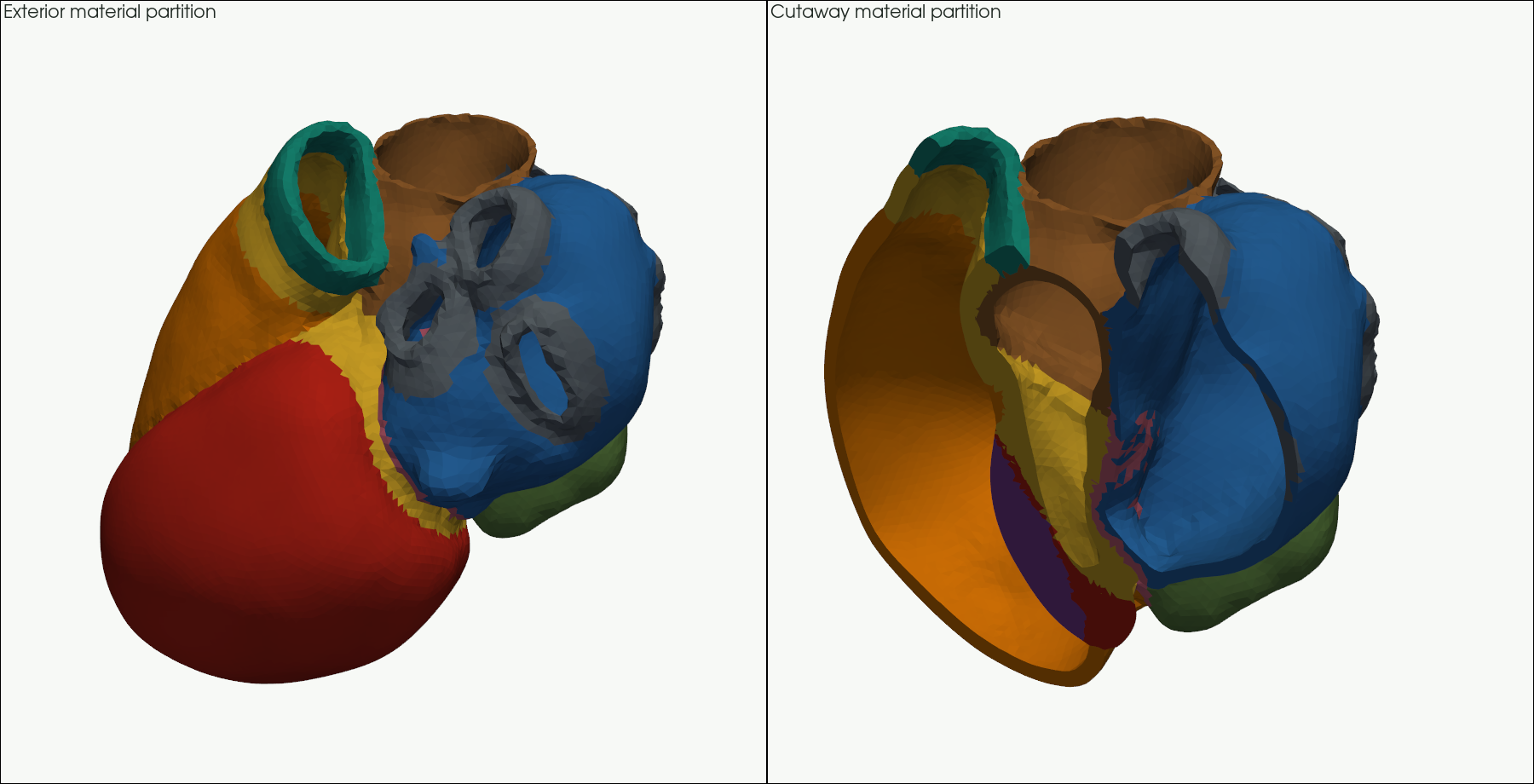}
\end{minipage}\hfill
\begin{minipage}[c]{0.39\linewidth}
\includegraphics[width=\linewidth]{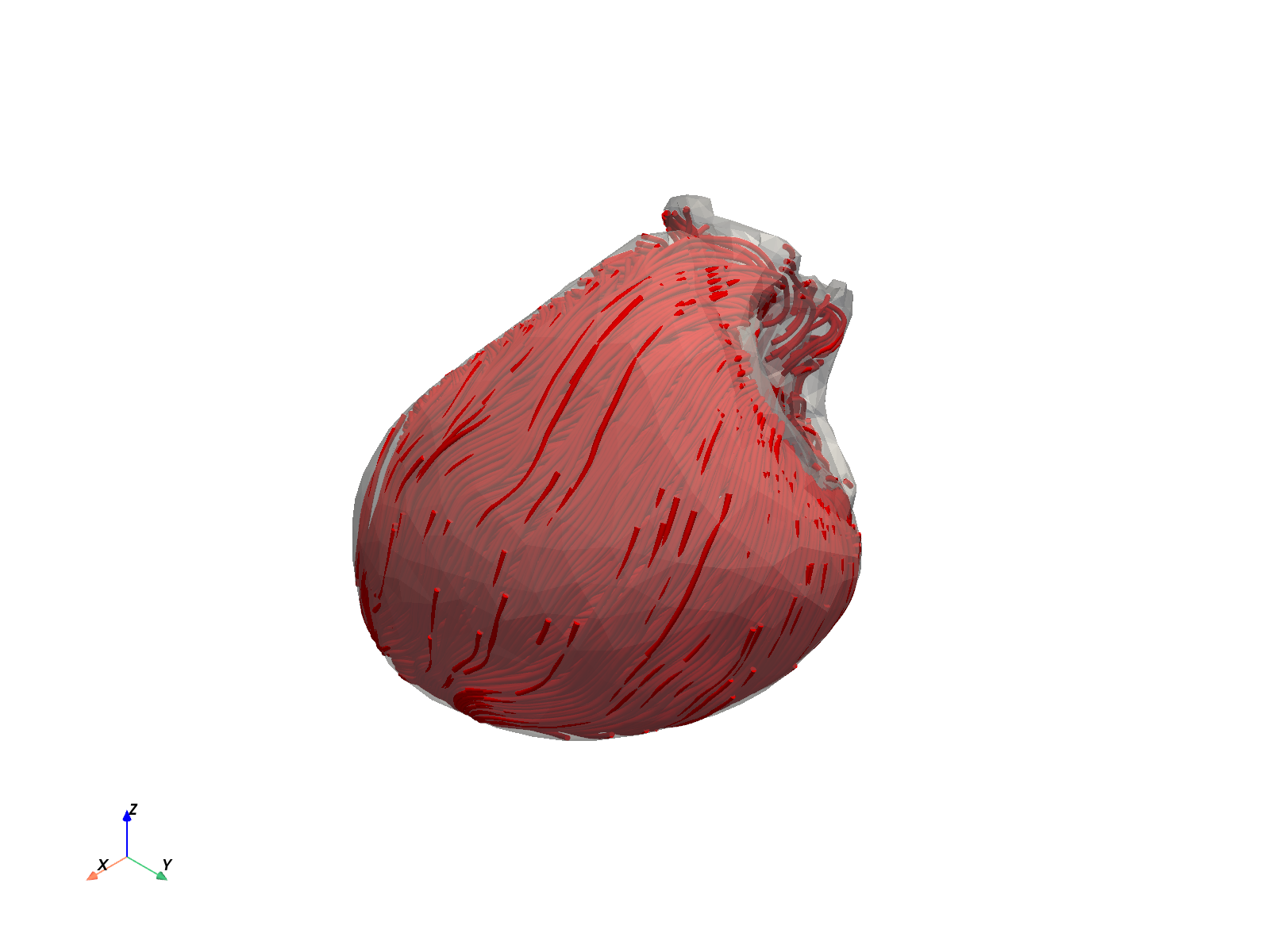}
\end{minipage}
\caption{Heart Geometry and Fibers. Top: reference and loaded geometries. Bottom left: tissue partitions. Bottom right: LV fiber streamlines.}
\label{fig:heart_anatomy}
\end{figure}

Figure~\ref{fig:heart_anatomy} distinguishes the material/reference coordinates $\vct X_{\rm ref}$, the filled geometry $\vct x_0=\vct x(0)$, and the aligned initial geometry used by JAX. The reference coordinates have no established stress-free interpretation. The filled geometry carries deformation and stress from the upstream preload calculation, so transferring this loaded state requires both its shape and its initial stress.

The deformation used in the constitutive law therefore combines the imported initial deformation gradient $\mat F_0$ with subsequent motion:
\begin{equation}
\vct x(t)=\vct x_0+\vct u_H(t),\qquad
\mat F_{\rm total}(t)=\mat F_{\rm incr}(\vct x_0\rightarrow\vct x(t))\,\mat F_0.
\label{eq:heart_initial_kinematics}
\end{equation}
Node and element fields are matched by their original identifiers. Coordinates are converted from millimeters to meters, stresses from MPa to Pa, and input times from milliseconds to seconds at the integration boundary. Tensor orientation is part of this mapping: transposing $\mat F_0$ changes the fiber stretch and hence the mechanical response. The initial Cauchy stress $\bm\sigma_0$ is imported as a separate field because chamber pressure alone does not determine the tissue stress.

\subsection{Fiber Activation, Active Stress, and Prestress}
Myocardial fibers determine the direction in which active tension acts. The fiber, sheet, and universal ventricular coordinate fields describe this anatomical organization. The JAX mechanics uses the formal initial fiber field stored in history variables HV10--12. Sheet directions and ventricular coordinates provide anatomical context and consistency checks. They do not enter its current constitutive update.

\begin{figure}[!htbp]
\centering
\includegraphics[width=0.95\linewidth]{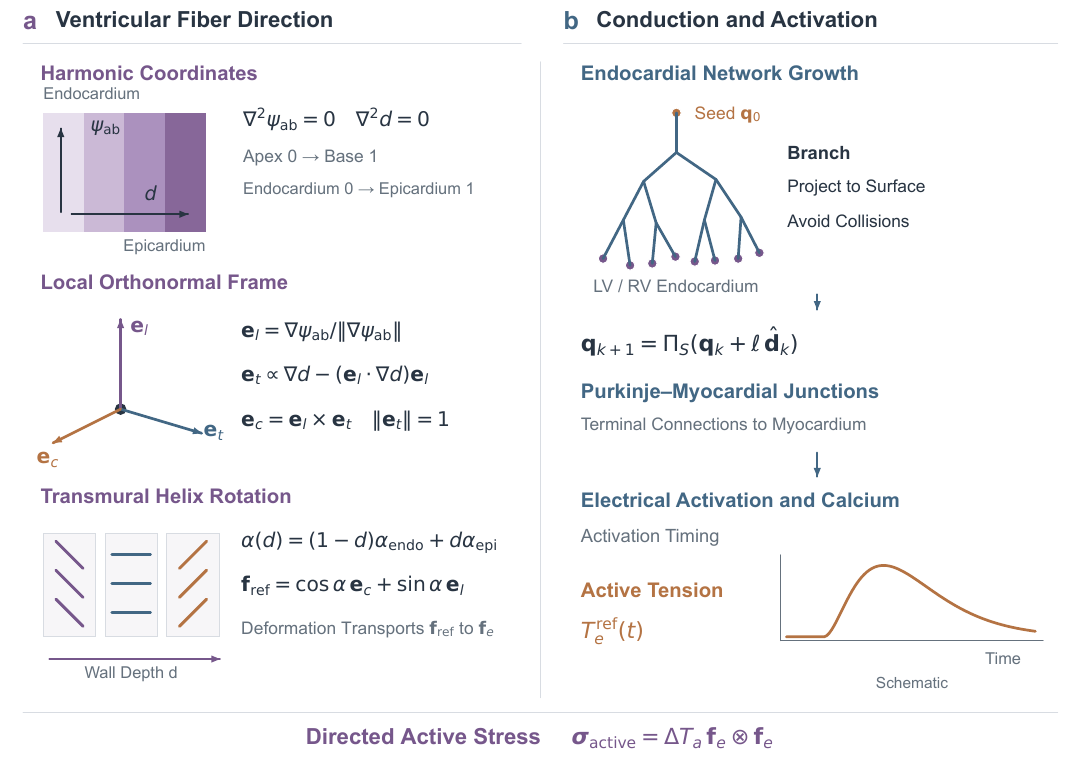}
\caption{Fiber Orientation and Activation. (a) Ventricular fiber construction. (b) Purkinje conduction and active-stress generation \citep{zhang2027heartintheloop}.}
\label{fig:heart_fiber_purkinje}
\end{figure}

Figure~\ref{fig:heart_fiber_purkinje} separates the construction of contraction direction from the construction of activation timing. In Figure~\ref{fig:heart_fiber_purkinje}(a), a Laplace--Dirichlet rule defines an apicobasal coordinate $\psi_{\rm ab}$ and a normalized transmural coordinate $d$ on the construction mesh \cite{bayer2012fibers}. Both satisfy Laplace's equation: $\psi_{\rm ab}=0$ at the apical anchor and $1$ at the base, while $d=0$ on the endocardium and $1$ on the epicardium, with zero normal flux on unconstrained boundaries. Normalizing $\nabla\psi_{\rm ab}$ gives the longitudinal axis $\vct e_l$. Removing its component from $\nabla d$ and normalizing gives the transmural axis $\vct e_t$. The circumferential direction is $\vct e_c=\vct e_l\times\vct e_t$. The helix angle rotates the reference fiber across the wall:
\begin{equation}
\alpha(d)=(1-d)\alpha_{\rm endo}+d\alpha_{\rm epi},\qquad
\vct f_{\rm ref}=\cos\alpha(d)\,\vct e_c+\sin\alpha(d)\,\vct e_l.
\label{eq:heart_fiber_construction}
\end{equation}
The schematic illustrates a ventricular wall. The anatomical construction also uses septal information and separate atrial rules \citep{zhang2027heartintheloop}. Deformation transports the reference fiber direction into the current unit vector $\vct f_e(t)$, which sets the direction of active stress in Equation~\eqref{eq:fourch_stress}.

Figure~\ref{fig:heart_fiber_purkinje}(b) illustrates conduction construction \cite{sahli2016purkinje}. Seeds on the LV and RV endocardium initiate branches. Each step combines the preceding direction with branch repulsion and projects back onto the surface. Branching and collision/boundary checks form the network. Terminal Purkinje--myocardial junctions (PMJs) connect it to myocardium.

The upstream electrophysiology calculation represents the conduction system, including Purkinje pathways, and determines activation timing. Calcium-sensitive contraction then converts the local cellular state, fiber stretch, and material history into active tension. Calcium concentration and active stress are different physical quantities: the former triggers contraction, while the latter supplies a directed mechanical stress \cite{land2017contraction}. The source calculation stores calcium in HV25 and first-fiber active stress in HV22.

In the represented JAX calculation, HV22 is the time-dependent mechanical input. Its values are aligned by element identifier, interpolated in time, and converted to Pa. Activation timing is already contained in this history. Multiplying it by a second onset gate would apply that timing twice. The active-stress increment and the total stress before nodal-pressure averaging are
\begin{align}
\Delta T_{a,e}(t)
&=s_{g(e)}\big[T^{\rm Gold}_{a,e}(t)-T^{\rm Gold}_{a,e}(0)\big],
\label{eq:heart_active_replay}\\
\bm\sigma_e(t)
&=\bm\sigma_{0,e}
+\big[\bm\sigma_{{\rm passive},e}(\mat F_{\rm total})
-\bm\sigma_{{\rm passive},e}(\mat F_0)\big]
+\Delta T_{a,e}(t)\,\vct f_e(t)\otimes\vct f_e(t).
\label{eq:fourch_stress}
\end{align}
Here $e$ is an element, $g(e)$ its chamber or septal group, $s_g$ a dimensionless active-tension multiplier, and $\vct f_e(t)$ the current unit fiber direction. Inactive tissues have zero active increment. The subtraction at $t=0$ preserves the imported prestress while subsequent passive deformation and active tension change the stress. The formulation contains myocardial and passive-tissue constitutive contributions. Their material and active scales have distinct physical roles.

\begin{figure}[!htbp]
\centering
\includegraphics[width=0.48\linewidth]{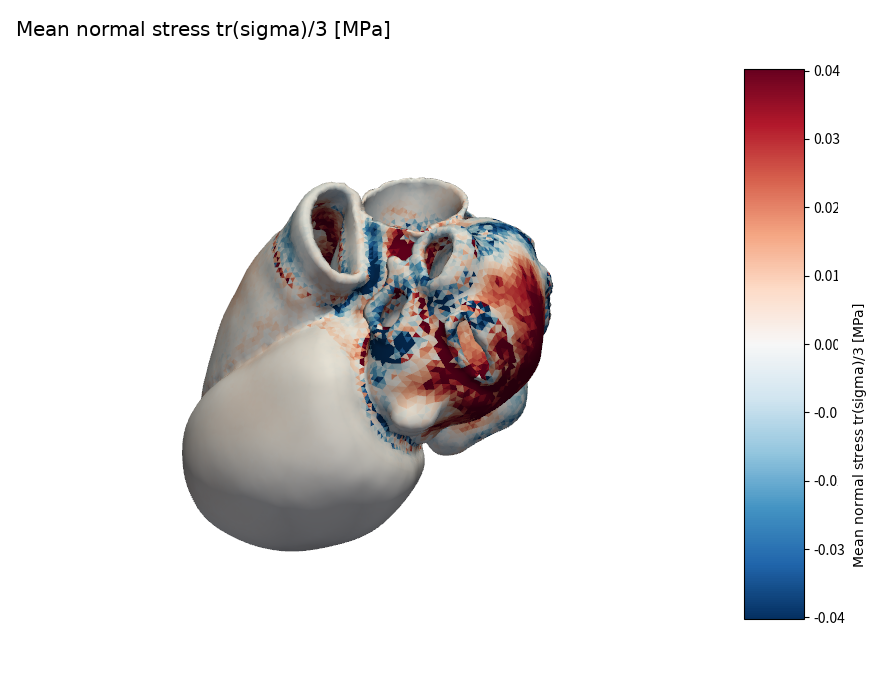}\hfill
\includegraphics[width=0.48\linewidth]{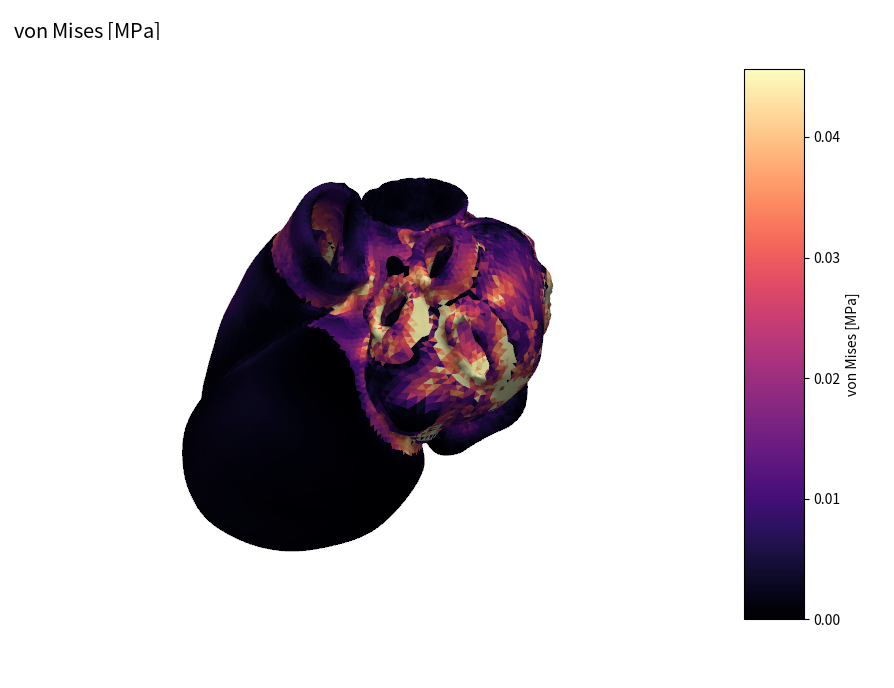}
\caption{Initial Myocardial Stress. Left: mean normal stress. Right: von Mises stress. Both maps use MPa.}
\label{fig:heart_prestress}
\end{figure}

\begin{figure}[!htbp]
\centering
\includegraphics[width=0.97\linewidth]{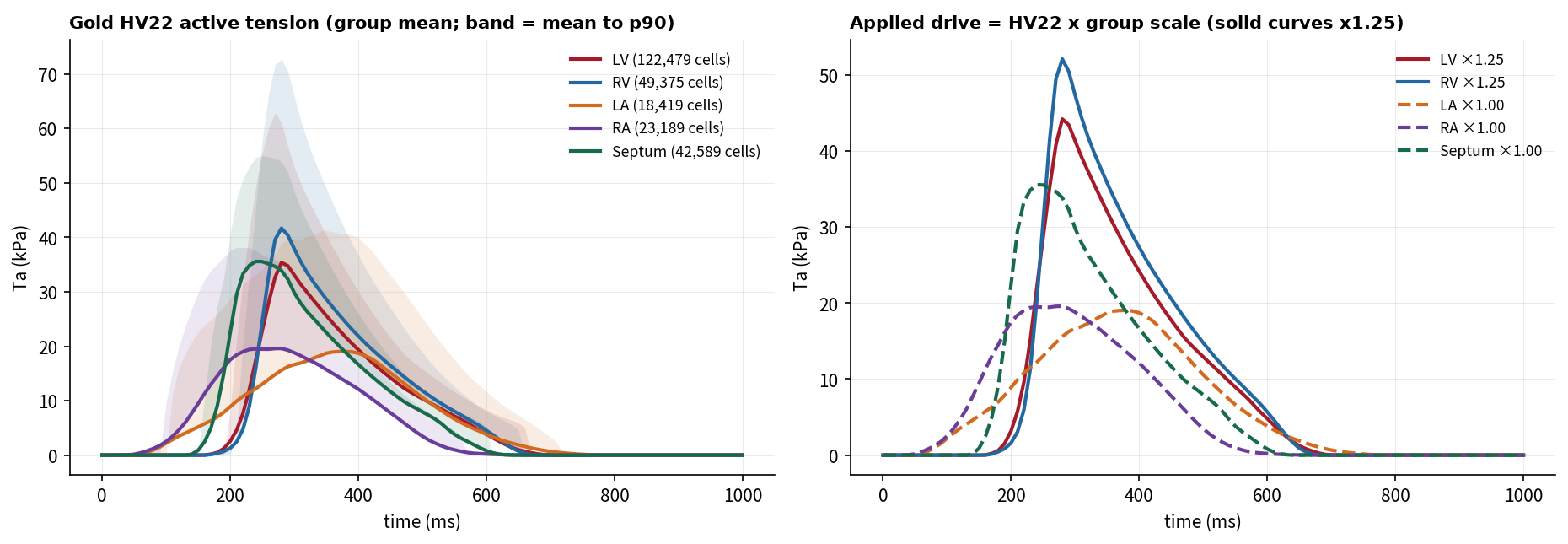}
\caption{Active-Tension Input. Left: reference histories. Right: scaled, baseline-subtracted histories applied to the cardiac model.}
\label{fig:heart_active_drive}
\end{figure}

Figure~\ref{fig:heart_prestress} shows why the initial state requires more than a cavity volume. Hydrostatic and deviatoric tissue stresses coexist in the filled wall. Figure~\ref{fig:heart_active_drive} shows the prescribed active-tension history and its grouped scaling. In this example the LV and RV multipliers are 1.25, while the atrial and septal multipliers are 1.0. These multipliers were selected for the illustrated reference comparison.

\subsection{Cavity Loading, Finite Supports, and Nearly Incompressible Mechanics}
Each chamber is represented by an oriented closed surface. Valve-plane caps complete the cavity topology, with 13 dependent cap nodes tied to the corresponding rims. For outward cavity normal $\vct n_\ell$, the volume follows from the divergence theorem,
\begin{equation}
V_\ell(t)=\frac{1}{3}\int_{\Gamma_\ell(t)}
\vct x\cdot\vct n_\ell\dd A,
\qquad \ell\in\{LA,LV,RA,RV\}.
\label{eq:heart_cavity_volume}
\end{equation}
Pressure produces a follower load on the moving endocardial surface. Its direction changes with the wall normal, while deformation changes the enclosed volume. The initial volumes in this configuration are LV 120.3, RV 205.0, LA 97.6, and RA 111.8 mL.

Finite spring and damper supports allow motion while exerting restoring and dissipative forces. The model uses 6,973 springs and 14,039 dampers, with their zero-force state referenced to $\vct x_0$ and no hard base clamp. These supports specify cardiac loading. Body--bed transmission is represented by the separate operators in Section~\ref{sec:simplified-body-bed}.

\begin{figure}[!htbp]
\centering
\includegraphics[width=0.73\linewidth]{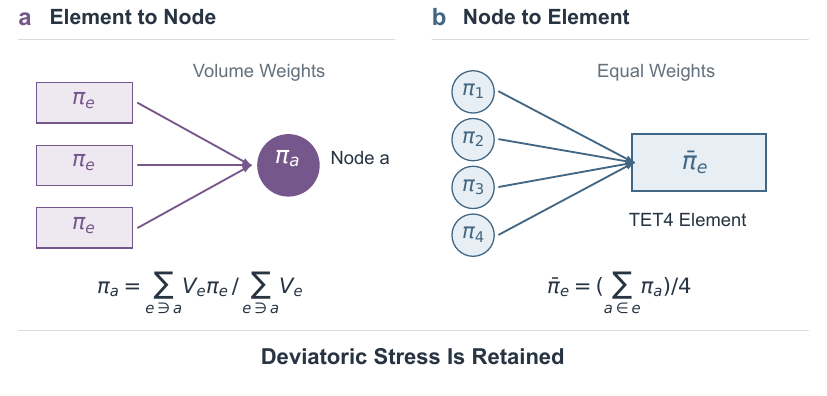}
\caption{Nodal Pressure Averaging. Mean tissue stress is projected (a) from elements to nodes and (b) back to elements, while deviatoric stress is retained \citep{zhang2027heartintheloop}.}
\label{fig:heart_anp_tet4}
\end{figure}

Nearly incompressible tissue presents a numerical difficulty for first-order tetrahedral displacement elements: element-wise volumetric constraints can create locking and oscillatory pressure. The implementation uses averaged nodal pressure (ANP), following the principle of nodal-volume averaging for linear tetrahedra \cite{bonet1998anp}. Figure~\ref{fig:heart_anp_tet4} illustrates the two projections. The mean normal stress is accumulated at shared nodes with volume weights and then mapped back to each element:
\begin{equation}
\begin{gathered}
\pi_e=\tfrac13\operatorname{tr}(\bm\sigma_e),\qquad
\pi_a=\frac{\sum_{e\ni a}(V_e/4)\,\pi_e}
{\sum_{e\ni a}V_e/4},\qquad
\bar\pi_e=\tfrac14\sum_{a\in e}\pi_a,\\
\bm\sigma_e^{\rm used}=\operatorname{dev}(\bm\sigma_e)+\bar\pi_e\mat I.
\end{gathered}
\label{eq:heart_anp}
\end{equation}
$V_e$ is the element-volume weight, distinct from a chamber volume $V_\ell$. The variable $\pi$ denotes tensile-positive mean normal stress, so it need not have the sign of a compression-positive fluid pressure. All 331,365 cells share one nodal averaging pool, including nodes on material interfaces. This defines the JAX discrete stress operator. The upstream and JAX operators share the nodal-averaging principle. Their discrete equivalence has not been established.

This projection changes the volumetric part of the discrete tissue response while preserving the element's deviatoric contribution. It enters the internal forces that determine wall motion. In a differentiated implementation, both projections belong to the mechanical update through which parameter sensitivities propagate \citep{zhang2027heartintheloop}.

\subsection{Coupled Three-Dimensional Mechanics and Zero-Dimensional Circulation}
The cardiac cycle follows from the interaction between tissue mechanics and a lumped pressure--flow model. The three-dimensional structural equations and four chamber-volume balances have the form
\begin{align}
\mat M_H\ddot{\vct u}_H+\mat C_H\dot{\vct u}_H
+\vct f_{\rm int}(\vct u_H,t)
&=\vct f_{\rm pressure}(\vct u_H,\vct p)
+\vct f_{\rm support}(\vct u_H,\dot{\vct u}_H)+\vct f_{\rm bal}(t),
\label{eq:heart_mechanics}\\
\dot V_\ell&=\sum Q_{{\rm in},\ell}-\sum Q_{{\rm out},\ell}.
\label{eq:fourch_mass}
\end{align}
Here $\mat M_H$ is the structural mass, $\mat C_H$ contains the runtime damping, and $\vct f_{\rm support}$ contains the separately represented spring/damper reactions. The temporary force $\vct f_{\rm bal}$ compensates the initial discrete imbalance and is released over 20 ms as a numerical correction. Valve pressure differences select open and closed flow branches. Windkessel resistance and compliance represent the arterial load. The configuration also specifies atrial inflow at its circulation boundary. The term \emph{closed-loop} denotes feedback between the structural and pressure--flow states, with systemic and pulmonary return represented through the stated boundary conditions.

The coupling is partitioned and semi-implicit. Within a circulation time step, explicit central-difference substeps advance the structure. Trial pressure perturbations estimate the four-cavity compliance matrix $\partial\vct V/\partial\vct p$. A four-pressure solve combines that compliance with valve and circulation equations, and the structure is advanced again under the accepted pressures. The calculation uses a fixed number of pressure iterations without an outer iteration to converge the entire mechanics--circulation interface.

Reference pressure initializes the loop once. Thereafter, the coupled mechanics and circulation generate pressure, volume, and displacement from the prescribed HV22 active tension and circulation boundaries. Subsequent reference trajectories serve as comparison targets rather than imposed histories.

Contractility, preload, and afterload enter different places in these equations. Contractility changes active force generation. Preload sets filling, initial stress, and myocardial stretch. Afterload opposes ventricular ejection through the circulation \cite{augustin2021closedloop}. Stroke volume is an integral output,
\begin{equation}
SV=\int_{\rm ejection}Q_{AV}(t)\dd t,
\label{eq:sv}
\end{equation}
where $Q_{AV}$ denotes aortic-valve flow. Wall vibration and vascular reaction depend on the time-resolved motion, force, and flow acceleration. Matching $SV$ alone therefore does not determine a unique BSG waveform.

\subsection{Four-Chamber Pressure--Volume Agreement}
The coupled model reproduces the four chambers' pressure--volume trajectories over the illustrated one-second beat. Pressure RMSE values are 4.92, 1.73, 2.27, and 1.11 mmHg for LV, RV, LA, and RA, respectively, with corresponding volume RMSE values of 4.07, 5.64, 3.94, and 4.10 mL. This simulation-to-simulation evaluation uses active-tension multipliers selected from the same reference pressure--volume histories. It quantifies agreement after that adjustment and tests the coupled chamber response.

\begin{figure}[p]
\centering
\includegraphics[width=0.93\linewidth,height=0.84\textheight,keepaspectratio]{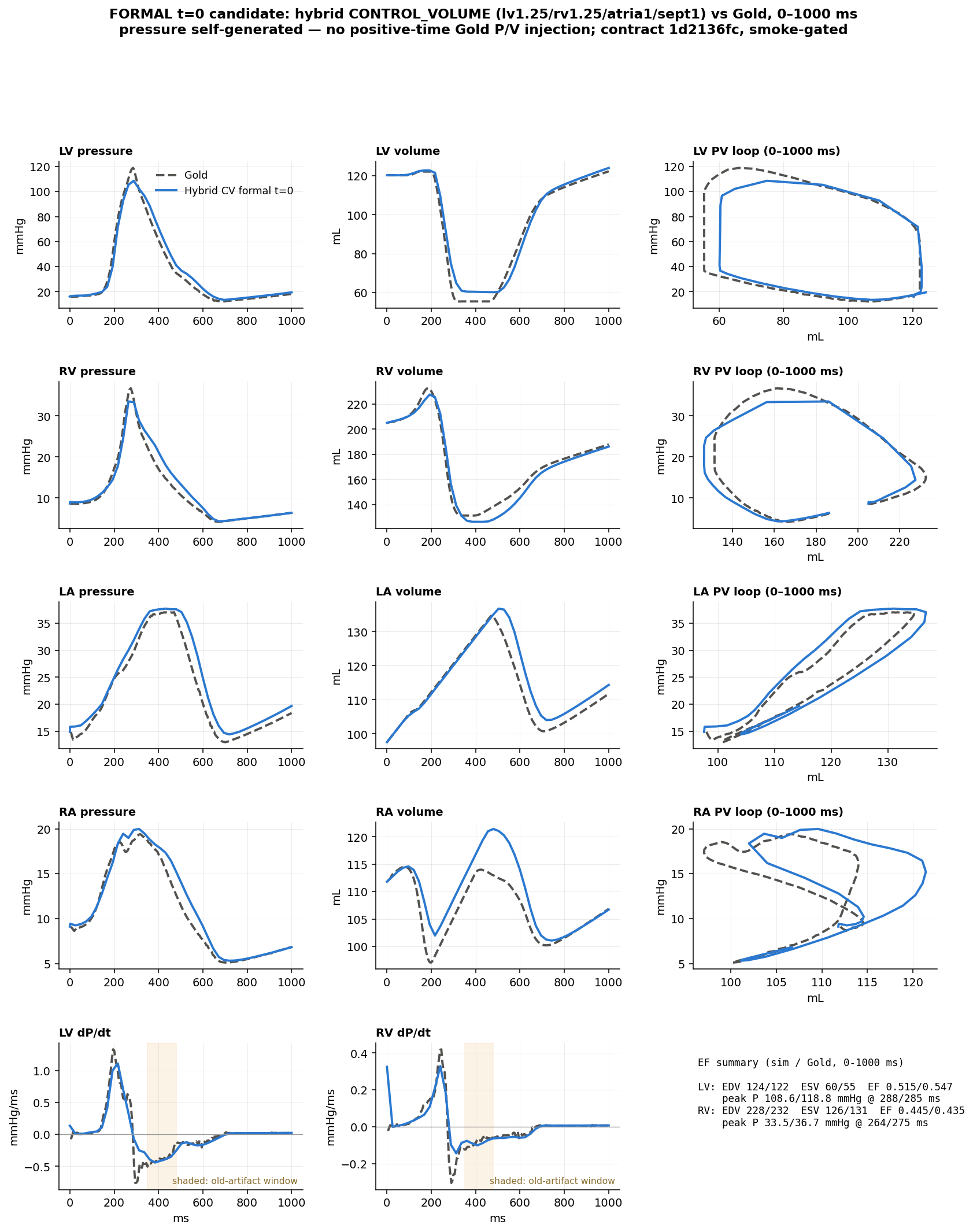}
\caption{Four-Chamber Pressure--Volume Agreement. Solid blue: model. Dashed gray: reference simulation. Rows show chamber pressure, volume, and PV trajectories, followed by ventricular pressure derivatives. Shading marks diagnostic windows. Pressure/PV scoring excludes the reference placeholder at $t=0$.}
\label{fig:heart_four_chamber}
\end{figure}

Figure~\ref{fig:heart_four_chamber} connects this agreement to the pumping cycle. Following mitral closure, both LV valves can be closed while pressure rises at nearly constant volume. Aortic opening permits ejection. Aortic closure begins relaxation with nearly constant volume, and mitral opening permits filling. The detailed valve-flow law and numerical leakage determine how closely a represented interval is isovolumic. The remaining chamber rows and ventricular pressure derivatives place the LV cycle in the context of the coupled heart.

The evaluation targets chamber-scale pressure and volume. Numerical mass damping and the temporary initial-balance force serve stabilization roles, distinct from identified tissue damping or physical support forces. Small-amplitude vibration, high-frequency motion, and bed-recorded BSG require separate evaluation through the source-extraction and transmission models.

\FloatBarrier

\subsection{Local Cardiac Wall Motion and the Proxy SCG}
\label{sec:local-cardiac-motion}
The simulated wall-motion field provides a local acceleration observable for comparison with the reference calculation. The following extraction defines its coordinate frame and sampling region. The same nodal motion and associated flow histories supply the force and recoil excitations developed for the body--bed models in Section~\ref{sec:body-bed-excitation}.

\subsubsection{Estimating the Body Axes from the Heart Mesh}
The local proxy SCG is extracted from simulated motion at three cardiac surface patches. Its acquisition has four steps: establish a body frame, select fixed surface patches, differentiate their displacement, and project the resulting acceleration onto the corresponding axes. Following the right-handed frame in Chapter~\ref{part:theory}, the frame has $+X$ toward the subject's left, $+Y$ toward the feet, and $+Z$ outward toward the anterior chest. For a supine body, Y runs along the bed and Z is the vertical chest-normal direction.

The mesh does not carry a measured torso or sensor registration, so the extractor estimates these directions from initial cardiac anatomy. Material labels and imported lumped nodal masses define mass-weighted centroids of the LV, RV, both atria, and the ventricular groups including septum. Let $\vct c_V$ denote the combined ventricular centroid, $\vct c_A$ the combined atrial centroid, and $\vct d=\vct c_{RV}-\vct c_{LV}$. With $\operatorname{unit}(\vct v)=\vct v/\|\vct v\|$, the axes in mesh coordinates are
\begin{equation}
\begin{aligned}
\vct\ell&=\operatorname{unit}(\vct c_V-\vct c_A),&
\vct e_Z&=\operatorname{unit}\!\left[\vct d-(\vct d\cdot\vct\ell)\times\vct\ell\right],\\
\vct e_Y&=\operatorname{unit}\!\left[\vct\ell-(\vct\ell\cdot\vct e_Z)\times\vct e_Z\right],&
\vct e_X&=\vct e_Y\times\vct e_Z.
\end{aligned}
\label{eq:scg_anatomical_frame}
\end{equation}
The base-to-apex direction provides the approximate head--foot cue, while the LV-to-RV direction supplies the anterior cue. For the illustrated baseline, the mesh-to-body rotation, with these unit vectors as rows, is
\begin{equation}
\mat R_0=\begin{bmatrix}\vct e_X^{\mathsf T}\\\vct e_Y^{\mathsf T}\\\vct e_Z^{\mathsf T}\end{bmatrix}
\simeq
\begin{bmatrix}
-0.349&+0.304&-0.886\\
+0.778&-0.433&-0.455\\
-0.522&-0.849&-0.085
\end{bmatrix}.
\label{eq:scg_frame_matrix}
\end{equation}
Body Z thus has a $-0.849$ projection onto mesh-global Y. Selecting a patch simply by maximum mesh-global $+Y$ samples a different side of this heart. The anterior SCG patch must instead be selected using the estimated body Z direction. The recoil implementation later applies an assumed $20^\circ$ rotation about body $+Z$ to register the cardiac frame with its illustrated torso. This changes X and Y but preserves the Z projection.

\subsubsection{Selecting the Cardiac Surface Sampling Patches}
\label{sec:three-patch-sampling}
\begin{figure}[!htbp]
\centering
\includegraphics[width=0.95\linewidth]{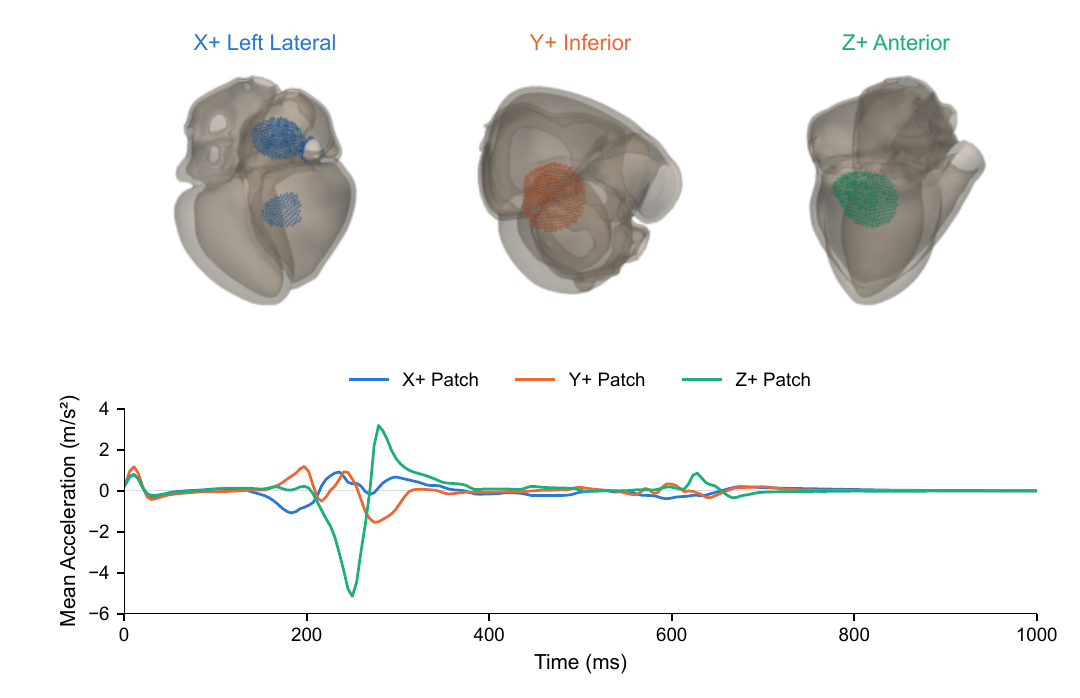}
\caption{Directional Cardiac Surface Sampling. Top: fixed left-lateral, inferior, and anterior patches, with 500 nodes per region. Bottom: mean acceleration projected onto $X$, $Y$, and $Z$ from their respective patches during one simulated cardiac cycle. Colors link each region to its waveform.}
\label{fig:scg_three_patches}
\end{figure}

We extract proxy SCG from the three cardiac surface patches shown in Figure~\ref{fig:scg_three_patches}, using the left-lateral patch for $X$, the inferior patch for $Y$, and the anterior patch for $Z$. Figure~\ref{fig:scg_three_patches} shows left-lateral, inferior, and anterior sampling regions in the same body frame defined above, with $+X$ toward the subject's left, $+Y$ toward the feet, and $+Z$ toward the anterior chest. For each positive axis, the extractor selects the 500 whole-heart boundary nodes with the largest initial-coordinate projection along that axis. These three fixed sets are disjoint in the illustrated geometry. Coordinate-based selection gives two separated areas for the left-lateral set.

Let $\Gamma_k$ denote the selected set for axis $k\in\{X,Y,Z\}$ and $N_k=500$ its node count. The displayed acceleration is
\begin{equation}
a_k(t)=\vct e_k^{\mathsf T}\frac{1}{N_k}
\sum_{a\in\Gamma_k}\frac{\mathrm d^2\vct u_{H,a}(t)}{\mathrm dt^2}.
\label{eq:three-patch-acceleration}
\end{equation}
The calculation differentiates displacement at the recorded snapshot times, averages nodes with equal weights, and projects onto the corresponding fixed body axis. Each trace therefore comes from its own sampling region. The three-patch curves use the same simulated beat and retain the numerical derivatives without additional filtering.

The directional curves share the systolic timing while differing in amplitude and polarity, with the anterior $Z$ response showing the largest excursion in this beat. They describe local myocardial motion before chest-wall and body--bed transmission. The excitation mappings in Section~\ref{sec:body-bed-excitation} connect such cardiac motion to the loads or prescribed motion applied to the body.
\FloatBarrier

\subsubsection{From Displacement Snapshots to the Z-Axis Proxy SCG}
The illustrated extraction uses 211 full-field displacement snapshots over one second, with a median interval of 4.785 ms. Each displacement $\vct u_{H,a}(t)$ is expressed in meters relative to the loaded initial geometry $\vct x_{0,a}$. Time is converted from milliseconds to seconds before differentiation. The extractor applies the numerical time-gradient operator twice on the actual, slightly nonuniform snapshot times, averages over the fixed patch $\mathcal P$, and rotates into the estimated body frame:
\begin{equation}
\begin{aligned}
\vct v_{H,a}(t_n)&=D_t\vct u_{H,a}(t_n),&
\vct a_{H,a}(t_n)&=D_t\vct v_{H,a}(t_n),\\
\vct a_{\rm SCG}^{\rm proxy}(t_n)
&=\mat R_0\times\frac{1}{500}\sum_{a\in\mathcal P}\vct a_{H,a}(t_n),&
a_{{\rm SCG},Z}^{\rm proxy}(t_n)
&=\begin{bmatrix}0&0&1\end{bmatrix}\times\vct a_{\rm SCG}^{\rm proxy}(t_n).
\end{aligned}
\label{eq:scg_patch_extraction}
\end{equation}
Here $D_t$ is the implemented time-axis \texttt{numpy.gradient} operation. The patch average uses equal node weights. All three components have units of m/s$^2$, and the third component is the local Z-axis proxy SCG. Since the node set and frame are fixed, averaging and rotation can equivalently precede differentiation. The extracted acceleration retains both overall cardiac translation and local wall deformation without fitted rigid-body subtraction.

The local Z-axis proxy SCG has peak absolute amplitude 5.15 m/s$^2$ in this configuration. Its main systolic complex lies around 200--300 ms, with a later diastolic complex around 600--650 ms. These are interpreted in relation to ejection and filling. Assigning a particular peak to AO or MO requires the accompanying valve or flow history. The 0--20 ms response overlaps the numerical initial-balance release in Eq.~\eqref{eq:heart_mechanics}, so the reference comparison scores morphology over 30--1000 ms.

Snapshot spacing and differentiation set the proxy's temporal resolution. The roughly 5-ms interval gives a nominal sampling ceiling near 100 Hz. Two numerical derivatives change the frequency response and amplify displacement errors, so acceleration accuracy requires frequency-specific assessment. They provide no anti-aliasing filtering. Sternum SCG and bed BSG additionally depend on mechanical transmission and sensor response.

\subsubsection{Reference Agreement in Local Wall Motion}
The comparison samples the same patch and axes in the upstream ANSYS/LS-DYNA calculation. Subtracting the reference coordinates at $t=0$ from the current coordinates aligns the displacement origin with JAX. Both displacement series are unfiltered, and scoring spans 30--1000 ms. For acceleration, the reference solver's native field is sampled at 10-ms intervals and processed with a fourth-order 30-Hz Butterworth zero-phase low-pass filter. The JAX series is differentiated from displacement snapshots without an additional low-pass filter and linearly interpolated onto the reference time grid. This processing defines the acceleration comparison. Post-sampling filtering leaves any pre-existing aliasing unresolved.

The model preserves the timing of local wall displacement, with Z-axis PCC $0.959$ and RMSE $2.30$ mm over 30--1000 ms. Its excursion is 9.45 mm, compared with 7.13 mm in the reference. The high correlation and larger excursion distinguish temporal agreement from amplitude agreement. Under the stated acceleration processing, Z-axis PCC is $0.715$, with peak absolute accelerations of 5.15 and 6.02 m/s$^2$ for JAX and the filtered reference, respectively. These results establish reference agreement for the local myocardial source, upstream of chest-wall and body--bed transmission.

\FloatBarrier

JAX can differentiate operations represented in its computational graph. The short-window computational graph contains active and passive stress, structural integration, cavity geometry, and the pressure--flow update. Active-tension multipliers, material scales, and selected resistances and compliances are parameterized. The mesh, initial state, and HV22 activation table remain fixed. Changing a tension multiplier does not recompute electrophysiology or calcium. Derivatives are local to a valve branch and can change discontinuously at switching events. Reliable full-beat gradients on the full mesh and unified inversion through the downstream body and bed require further validation.

The cardiac calculation supplies wall motion, chamber pressure and volume, and flow histories. Section~\ref{sec:body-bed-excitation} constructs mechanical excitations from these outputs and defines how they enter the body--bed transmission models.

\FloatBarrier
\section{Cardiac-to-Body Mechanical Excitation}
\label{sec:body-bed-excitation}
Body--bed transmission begins with a defined mechanical excitation. The four-chamber model in Section~\ref{sec:physiological-twin} supplies wall motion and blood-flow histories. This section converts those outputs into a cardiac inertial force proxy, distributed vascular-recoil loads, or prescribed interface motion. Each source retains its location, coordinate frame, units, and force convention. These excitation descriptions feed the simplified transmission models in Section~\ref{sec:simplified-body-bed}.

\subsection{Mechanical Source Interface}
\label{sec:source-observation-interface}
This section defines $\mathcal B$, the conversion from a cardiac mechanical source to an excitation accepted by the body model. The source histories include wall displacement $\vct u_H(t)$, nodal acceleration $\vct a_H(t)$, and the derived inertial force proxy $\vct f_{H,\rm proxy}(t)$. The associated flow histories supply distributed vascular loads. Each source representation has defined units and spatial support, which determine its mechanical interface.
\[
 s(t)=\{\vct u_H(t),\vct a_H(t),\vct f_{H,\rm proxy}(t),
          \{\Delta\vct F_{A,i}(t)\}_i\},\qquad
 q_{\rm exc}(t)=\mathcal B[s](t)
 =\boxed{[\vct f_{\rm body}(t),\vct u_\Gamma(t)]}.
\]
The bracketed excitation specifies a choice of force-driven or prescribed-motion loading for each interface degree of freedom. Each branch supplies one member of the conjugate force--displacement pair. The common frame and local motion histories come from Section~\ref{sec:local-cardiac-motion}. The force-source constructions below retain the cardiac and vascular definitions used by the implemented models.

The 27-participant reduced fits use scalar source proxies, with $Z$ defined by the whole-heart inertial force proxy and $X+Y$ by the planar source combination. Their fitted gains set an effective loading scale. The traction and nodal-force maps specify the alternative force-driven interface. Acceleration alone is a local motion observable and requires an appropriate mechanical mapping before it can be used as a body load.

Local wall motion excites the SCG pathway. The distributed vascular source in Section~\ref{sec:cardiac-recoil-sources} adds the momentum-reaction pathway from the associated flow state and vascular geometry. The Z-axis heart-solid proxy and the spatial aortic loads can enter one observation framework while retaining their distinct source definitions. The cardiac-core and source-to-signal comparisons evaluate components of the forward model. A joint fit of cardiac mechanics, vascular recoil, body, bed, and sensor to measured triaxial BSG remains to be evaluated.

\subsection{Prescribed-Motion Excitation}
\label{sec:prescribed-excitation}
For displacement-driven excitation, the heart-to-body map includes coordinate rotation, spatial interpolation, and the selected interface nodes:
\[
 \vct u_\Gamma(t)=\mathcal M_u\vct u_H(t).
\]
The receiving interface displacement $\vct u_\Gamma$ is denoted $\vct u_b$ in the work-conjugate map below. With compatible coordinates, $\mathcal M_u$ is represented by $\mat T$, which includes any frame rotation.

\subsubsection{Displacement Mapping and Mechanical Work}

If a cardiac interface mesh and a body mesh use different nodes, a displacement interpolation is written
\begin{equation}
\vct u_b=\mat T\vct u_H,\qquad
\vct f_H^{\,\rm eq}=\mat T^{\mathsf T}\vct f_b,\qquad
\vct f_b^{\mathsf T}\dot{\vct u}_b
=(\vct f_H^{\,\rm eq})^{\mathsf T}\dot{\vct u}_H .
\label{eq:work_map}
\end{equation}
The transpose is the work-conjugate force map for the same represented load. The physical reaction on the other subdomain has the opposite sign. Preservation of rigid translations and rotations in the interpolation supports conservation of resultant force and moment. A map that only matches nearby points can alter the effective moment arm or interface stiffness.

The source definition includes the source surface or vessel segment, coordinates and polarity, SI units, and the time origin. A displacement input and a traction input impose different boundary conditions. The vascular reaction is included once, with a control-volume definition compatible with the cardiac interface load.

\subsubsection{Equivalent Excitation on the Unconstrained Coordinates}
The mapped interface history supplies the prescribed coordinates $\vct d_c(t)$. Eliminating those coordinates defines the excitation terms to be passed to the propagation solver:
Partition the mechanical degrees of freedom into unknown values $\vct d_f$ and prescribed heart-interface motion $\vct d_c(t)$. The free equations become
\begin{equation}
\mat M_{ff}\ddot{\vct d}_f+\mat C_{ff}\dot{\vct d}_f+\mat K_{ff}\vct d_f
=\vct f_f-\mat M_{fc}\ddot{\vct d}_c
-\mat C_{fc}\dot{\vct d}_c-\mat K_{fc}\vct d_c.
\label{eq:prescribed_motion}
\end{equation}
Thus a prescribed displacement history excites the body through inertial, damping, and stiffness coupling. The constrained equations determine the reaction required to impose that motion. Applying this reaction again as an additional independent force would double-count the excitation.

A traction-driven formulation instead places a load on the right-hand side and solves for interface motion. Displacement and traction are conjugate variables: prescribing both arbitrarily on the same degree of freedom generally overconstrains the interface.

\subsection{Force-Driven Excitation}
A force-driven interface maps the selected source components into a spatially distributed body load. For the heart-solid contribution, the notation is
\[
 \vct f_{\rm body,H}(t)=\mathcal M_f\vct f_{H,\rm proxy}(t),\qquad
 \vct f_{\rm body}(t)=\vct f_{\rm body,H}(t)
       +\sum_i\mathcal M_{A,i}\Delta\vct F_{A,i}(t).
\]
The maps specify force convention, rotation, receiving nodes, and spatial load weights. Their construction requires the interface balance appropriate to the selected source proxy. The following cardiac and vascular source models define the histories being mapped. A net force requires a load distribution, whereas a resolved traction is integrated directly.

\subsubsection{Cardiac Inertial Force as a Body--Bed Excitation}
The nodal acceleration field from Section~\ref{sec:local-cardiac-motion} provides the whole-heart inertial force proxy. It uses every solid node and the imported lumped nodal masses, whose sum is 0.323795 kg in the documented configuration. In column-vector notation,
\begin{equation}
\vct f_{H,\rm proxy}(t)=\mat R_{H\rightarrow B}
\sum_a m_a\ddot{\vct u}_{H,a}(t),
\qquad s_{H,Z}(t)=\begin{bmatrix}0&0&1\end{bmatrix}\vct f_{H,\rm proxy}(t).
\label{eq:heart_inertial_source}
\end{equation}
The rotation maps mesh coordinates into the common body frame. The extraction defines the positive inertial sum as its force proxy. A physical traction applied to the surrounding body must follow the opposite-side reaction convention and the complete interface force balance. The inertial sum alone does not separate support, pressure, or numerical stabilization forces. The initial balancing force and numerical damping discussed in Section~\ref{sec:physiological-twin} can therefore affect the extracted waveform.

The two cardiac observables serve different mechanical roles: $a_{{\rm SCG},Z}^{\rm proxy}$ in Eq.~\eqref{eq:scg_patch_extraction} is local patch acceleration in m/s$^2$, while $s_{H,Z}$ in Eq.~\eqref{eq:heart_inertial_source} is the whole-heart inertial force proxy in N. The 27-participant source-to-bed analysis in Section~\ref{sec:simplified-body-bed} uses the latter as its Z input. Each observable requires a transfer operator with matching input units.

\subsubsection{Vascular Recoil as a Distributed Body--Bed Excitation}\label{sec:cardiac-recoil-sources}
\begin{figure}[!htbp]
\centering
\includegraphics[width=0.98\linewidth]{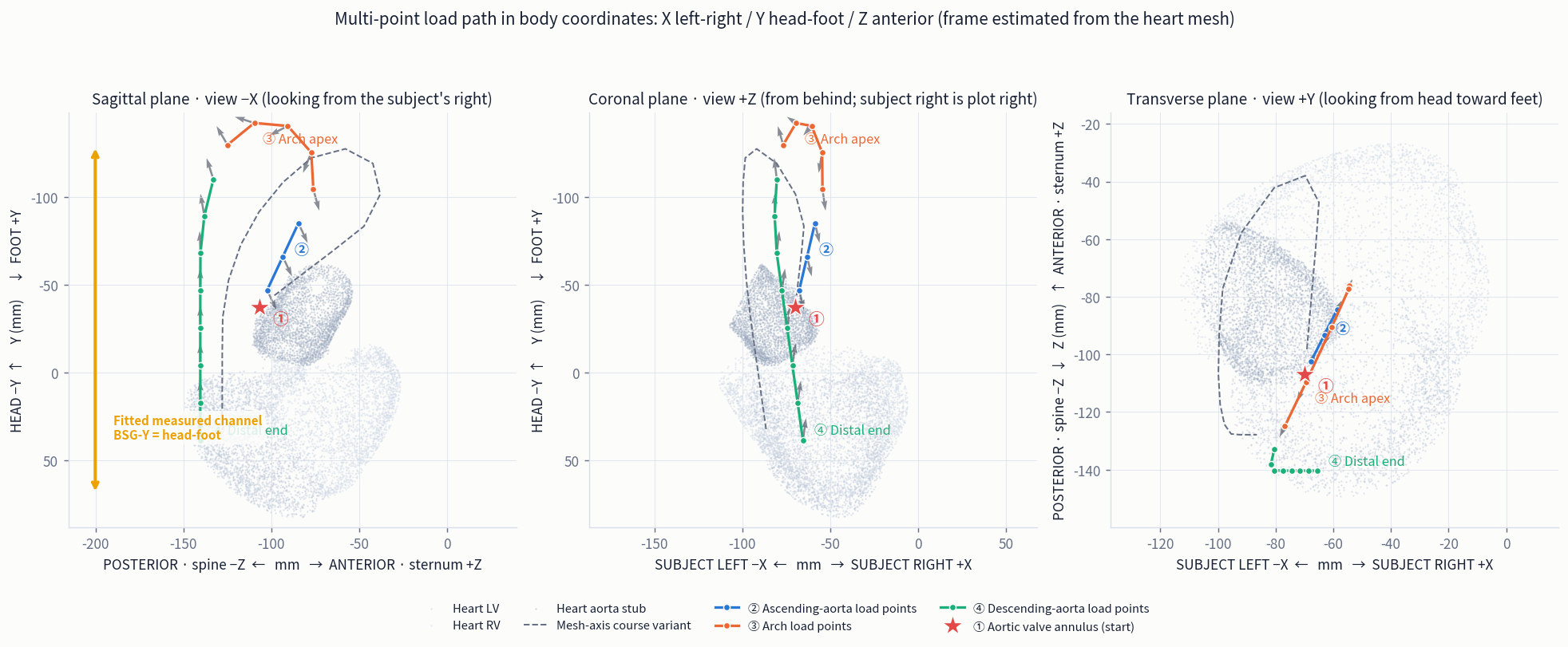}
\caption{Aortic Recoil Geometry. Blue, orange, and green indicate ascending, arch, and descending load points. Arrows show local tangents. The red star marks the valve anchor. The dashed curve shows an alternative course. BSG-Y is the assumed head--foot direction.}
\label{fig:recoil_centerline}
\end{figure}

\paragraph{Flow-Driven Loads Along the Aortic Centerline.}
The recoil model distributes cardiac output along a parameterized aortic path (Fig.~\ref{fig:recoil_centerline}). The valve-annulus anchor comes from the heart mesh's aortic-root region. Prescribed anatomical approximations define the primary ascending direction, arch, and descending course. The illustrated path uses 16 load points, a 70-mm ascending segment, a 30-mm-radius semicircular arch, and a 180-mm descending region that includes a straightening bend. Cross-sectional area tapers from 6.0 through 4.0 to 3.0 cm$^2$. This prescribed geometry supplies a spatial source model, with the participant's complete aorta outside the available anatomical reconstruction.

LV cavity volume is reconstructed from 211 displacement snapshots of the same simulated beat. Aortic outflow is approximated by
\begin{equation}
Q_{ao}(t)\simeq\max\!\left(0,-\frac{\mathrm d V_{LV}}{\mathrm dt}\right).
\label{eq:recoil_volume_flow}
\end{equation}
This reduction assumes negligible LV inflow during ejection. In general, chamber balance gives $Q_{ao}=Q_{MV}-\dot V_{LV}$. The implemented derivative is low-pass filtered at 30 Hz and clipped to nonnegative flow. Its integral is 62.82 mL, compared with a 63.93-mL volume excursion, a 1.7\% difference from filtering and clipping. These values characterize the simulated source SV.

At arc-length position $s_i$, a pulse-wave delay gives the prescribed local mean blood speed
\begin{equation}
\tau_i=\frac{s_i}{c_{\rm pw}},\qquad
v_i(t)=\frac{\chi_i\times Q_{ao}(t-\tau_i)}{A_i},
\label{eq:recoil_waveform_delay}
\end{equation}
where $A_i$ is the local cross-section and $\chi_i$ the retained flow fraction. The primary source uses $\chi_i=1$. A branch-runoff variant decreases it to 0.7 downstream of the arch. The illustrative pulse-wave speed is 5.0 m/s. Pulse-wave delay transports the prescribed waveform, while blood-particle transport enters the convective acceleration below. These are distinct physical time scales in the prescribed-flow approximation, which leaves pulsatile fluid--structure interaction unresolved.

\paragraph{Unsteady, Convective, and Curvature Contributions.}
For each load point, $\rho_b$ is blood density, $\rho_b A_i\Delta s_i$ is the slice mass, $\vct t_i$ is the flow tangent, $\vct n_i$ points toward the center of curvature, and $\kappa_i$ is the local centerline curvature. The modeled body-recoil load is
\begin{equation}
\Delta\vct F_{A,i}(t)=-\rho_b\times A_i\times\Delta s_i
\left[
\left(\frac{\partial v_i}{\partial t}
+v_i\times\left.\frac{\partial v}{\partial s}\right|_{s_i}\right)\vct t_i
+\kappa_i\times v_i^2\,\vct n_i
\right].
\label{eq:distributed_aorta_recoil}
\end{equation}
The three contributions are temporal acceleration, acceleration associated with spatial changes in speed, and acceleration required to turn the flow. The minus sign gives the modeled reaction opposite to blood acceleration. Blood occupies the ascending, arch, and descending regions during the same ejection event. Their different tangents, curvature, and delayed flow histories generate different signed loads, rather than an imposed reversal of one point force.

\begin{figure}[!htbp]
\centering
\includegraphics[width=0.98\linewidth]{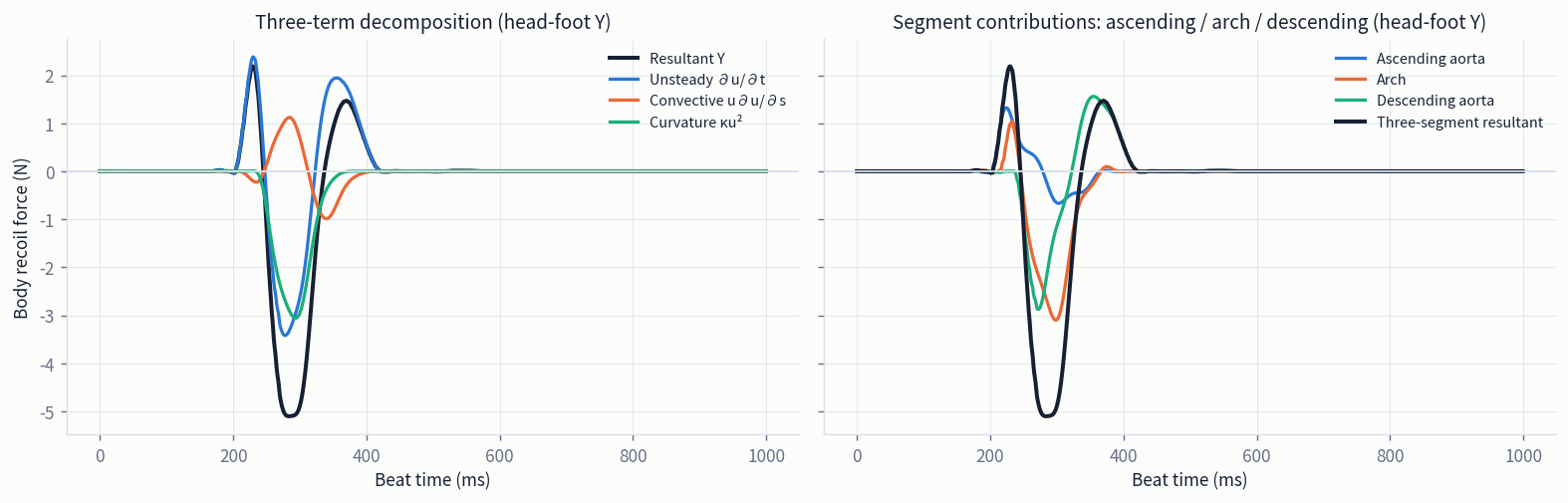}
\caption{Simulated Aortic Recoil. Left: unsteady, convective, and curvature contributions. Right: ascending, arch, and descending contributions. Both decompositions sum to the same longitudinal force.}
\label{fig:recoil_decomposition}
\end{figure}

For a straight, constant-area tube with a common undelayed flow, the unsteady term reduces to $\vct F_A=-\rho_b\times L\times\dot Q\,\vct t$. For steady flow through a constant-area 180-degree bend, integration gives reaction magnitude $2\times\rho_b\times A\times v^2$. These limits connect the load points to momentum balance. The code also retains a pressure-gradient alternative, $-A_r\times(p_{\rm in}-p_{\rm out})\,\vct t_r$, for each region. This pressure-based alternative and the momentum-derived reaction describe overlapping mechanisms. A combined formulation requires control-volume accounting to include each mechanism once.

The source can retain its resultant and moment about a reference location $\vct r_0$:
\begin{equation}
\vct F_A=\sum_i\Delta\vct F_{A,i},\qquad
\vct M_A=\sum_i(\vct r_i-\vct r_0)\times\Delta\vct F_{A,i}.
\label{eq:aorta_resultant_moment}
\end{equation}
The resultant and moment summarize global loading. The spatial support of the distributed loads additionally determines the local deformation modes excited in a detailed body model.

\paragraph{Traction and Equivalent Nodal Forces.}
The heart model supplies chamber pressures and volumes, flow histories, and spatial wall motion. To drive a body model, its mechanical output is expressed as either displacement or traction at a specified cardiac interface. A traction-to-load map preserves the distributed source:
\begin{equation}
\vct F_H(t)=\int_{\Gamma_H}\mat N_b^{\mathsf T}(\vct x)\,
\vct t_H(\vct x,t)\dd A.
\label{eq:traction_export}
\end{equation}
$\mat N_b$ interpolates the receiving body's interface motion. Replacing this spatial load with one resultant removes its moment and distribution unless they are retained separately. Signed components, source position, coordinate transforms, and units are therefore part of the BSG source definition.

\subsection{Coupling Assumption and Excitation Handoff}
The prescribed-source implementation uses one-way coupling: the cardiac trajectory drives body deformation under the chosen mechanical boundary conditions. Cardiac surface displacement supplies the myocardial pathway, and separately accounted vascular loads represent reaction from blood acceleration throughout the vascular network (Section~\ref{sec:recoil}). The prescribed-motion equations define this forward transmission problem independently of a measured-data fit.

Two-way coupling additionally returns body traction to the cardiac solve so that body loading can alter cardiac motion. This feedback formulation is distinct from the one-way implementation evaluated here, consistent with the separation between the general analytical framework and the reduced-model assumptions in Section~\ref{sec:framework}.

The output of this section is $q_{\rm exc}$ with its spatial support, coordinates, polarity, units, and time origin. Section~\ref{sec:simplified-body-bed} provides the reduced/effective source-to-measurement realization in which excitation mapping and propagation can be combined into one fitted operator.
\FloatBarrier

\FloatBarrier
\section{Simulated Body--Bed System}
\label{sec:simplified-body-bed}
The simulated body--bed system connects cardiac mechanics to measured waveforms and identifies how transmission reshapes physiological information. Three realizations are evaluated against recordings from 27 participants: the low-order analytical model (Section~\ref{sec:mechanics}), a multilayer finite-element model (FEM), and a finite impulse response (FIR) operator. The mechanical models expose transmission with different spatial detail, while FIR describes an effective input--output relation. The matched-window experiment tests calibrated mechanical reconstruction. Controlled cardiac perturbations through frozen FIR models test how signal features track stroke volume (SV) after transmission. Our VibroSV model estimates SV changes from BSG \citep{zhang2027vibrosv}.

\paragraph{Reduced Realizations of the Source-to-Measurement Map.}
The operator in this section combines the excitation interface and the subsequent transmission:
\[
 \mathcal H=\mathcal H_{\rm BB}\circ\mathcal B,\qquad
 \widehat y(t)=\mathcal H_{\rm eff}[s](t),\qquad
 \mathcal H_{\rm eff}\in
 \{\mathcal H_{\rm eff}^{\rm analytical},\mathcal H_{\rm eff}^{\rm 4R},\mathcal H_{\rm eff}^{\rm FIR}\}.
\]
The analytical model, multilayer FEM, and FIR are alternative realizations of the source-to-measurement mapping. They retain different structural assumptions and fitting procedures.

A reduced body--bed model connects the selected myocardial and recoil inputs $\vct s[n]$ to the three measured axes. The body--bed dynamics and frozen-coefficient response in Section~\ref{sec:mechanics} motivate a matrix-valued finite impulse response (FIR),
\begin{equation}
\vct y[n]=\sum_{k=0}^{K_f-1}\mat H[k]\,\vct s[n-k]+\vct n[n],
\qquad \mat H[k]\in\mathbb R^{3\times m}.
\label{eq:fir}
\end{equation}
Each axis has its own gains, delays, and source weights. The FIR approximates the combined body, bed, and included acquisition response over a fixed operating condition. Its coefficients describe effective transmission, with tissue material properties represented separately by the mechanical models. The wave-propagation formulation separates physiological excitation from mechanical transmission \citep{zhang2026phybp}.

\subsection{Reduced Transmission and Multilayer FEM}
\label{sec:reduced-body-bed}
The recorded channels are the vertical component $Z$ and the signed planar combination $X+Y$, denoted $XY$. The latter is the dataset's planar-sum channel, including its window scaling. It is not $\sqrt{X^2+Y^2}$. Each native 10-second window is summarized by one 100-sample cardiac-cycle template. Valid J--J cycles are resampled, aligned, and combined by their median, and the final positive peak is placed at phase sample 25. Source cycles receive the corresponding phase normalization and peak anchoring.

The FIR-121 representation is
\begin{equation}
\widehat y_{p,c}[n]
= b_{p,c}+\sum_{k=-60}^{60}h_{p,c}[k]\times
s_c\!\left[(n-k-\ell_{p,c})\bmod 100\right],
\qquad c\in\{Z,XY\},
\label{eq:cycle-fir}
\end{equation}
where $p$ identifies a participant, $h$ is a regularized transfer kernel, $b$ is an intercept, and $\ell$ is a training-derived phase shift. The coefficients are shared across that participant's training windows. Because the 121 taps operate on a periodic 100-sample cycle, they represent repeated phase offsets in an offline morphology operator. Ridge regularization stabilizes the fit. The taps do not identify 121 independent physical delays or causal transit times.

\begin{figure}[!htbp]
\centering
\includegraphics[width=0.97\linewidth]{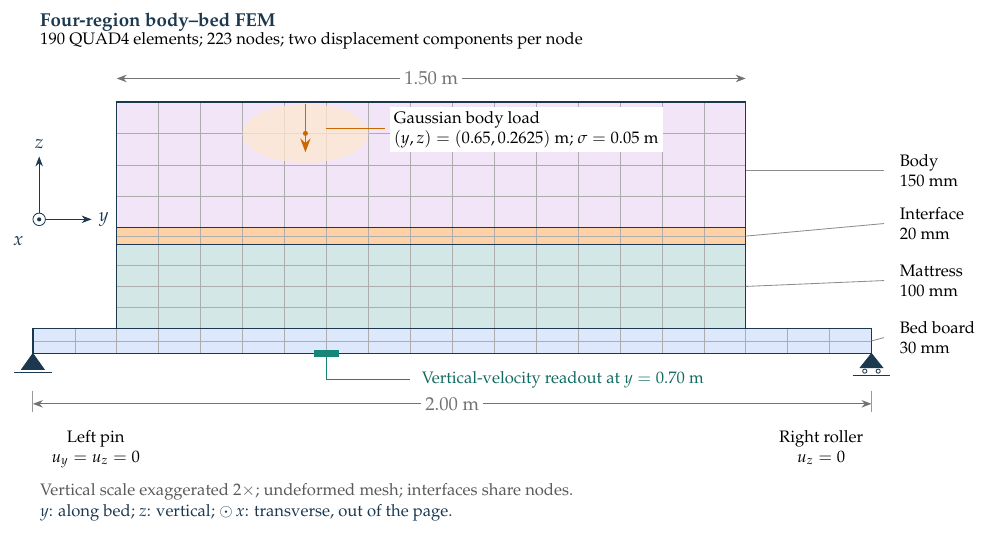}
\caption{Four-Region Body--Bed Model. Longitudinal--vertical mesh with supports, load region, and readout location. Transverse $x$ points out of the page. Vertical dimensions are exaggerated twofold.}
\label{fig:simplified_fem}
\end{figure}

The mechanical alternative is the coarse FEM shown in Fig.~\ref{fig:simplified_fem}. It is a two-dimensional $y$--$z$ section with longitudinal $y$, vertical $z$, and transverse $x$ perpendicular to the section. Four conforming regions represent the body, compliant interface, contact/mattress, and bed board. The mesh contains 190 four-node quadrilateral elements and 223 nodes, each carrying $u_y$ and $u_z$. Shared interface nodes enforce displacement continuity. The left lower board support fixes both components. The right lower support fixes $u_z$ and permits longitudinal motion. No separate skeletal domain is enabled in this reduced configuration.

Normalized cardiac proxies and fitted gains set an effective loading scale, with absolute source-force calibration outside this fit. The matched-window FEM retains horizontal and vertical input/output coupling, with the planar source sum mapped to its horizontal direction. Its two-dimensional representation resolves the vertical and planar-sum channels, rather than three independently validated anatomical axes.

\subsection{Matched-Window Comparison of Analytical, FEM, and FIR Models}
\label{sec:matched-model-comparison}
The matched-window experiment tests whether calibrated body--bed mechanics reconstructs measured waveform morphology across 27 participants. The low-order analytical model and multilayer FEM share the cardiovascular source generator (Section~\ref{sec:targets}), the $XY$ and $Z$ templates, and the calibration and evaluation-window selections. FIR-121 supplies an effective-transmission reference on the same evaluation targets. This comparison evaluates transmission from the reduced cardiovascular source. Section~\ref{sec:sv-transfer} separately evaluates perturbations from the four-chamber source.

The mechanical models fit three physiological controls and four mechanical parameters using one subset of beats and are evaluated on held-out beats from the same recording. Channel gains are profiled during fitting. A participant-specific complex frequency correction is learned from separate calibration data and held fixed during evaluation. Defining this correction in physical frequency allows the same calibration to be applied at different cardiac periods. The corrected analytical and FEM results therefore include an empirical frequency response in addition to their mechanics.

The multilayer FEM uses the four-region geometry in Fig.~\ref{fig:simplified_fem}, with nine stiffness and damping multipliers at calibration. Its horizontal and vertical velocity responses retain cross-directional coupling. The excitation combines the analytical vascular forces (Section~\ref{sec:recoil}) and thorax/myocardial branch (Section~\ref{sec:cardiac}). A projection of the assembled FEM matrices accelerates the periodic frequency-domain solution. During window fitting, body stiffness, board stiffness, body viscosity, and mattress viscosity may vary around their calibrated values. The observation includes the sensor response and the 0.7--20 Hz readout filter.

FIR is trained and tested on separate recording segments and is compared with the mechanical models on the same held-out target waveforms. The $XY$ comparison uses targets unaffected by polarity correction. Performance is summarized by each participant's median PCC and then across all 27 participants. The common targets support a descriptive comparison of waveform reconstruction under the models' respective fitting procedures.

\begin{figure}[!htbp]
\centering
\includegraphics[width=0.97\linewidth]{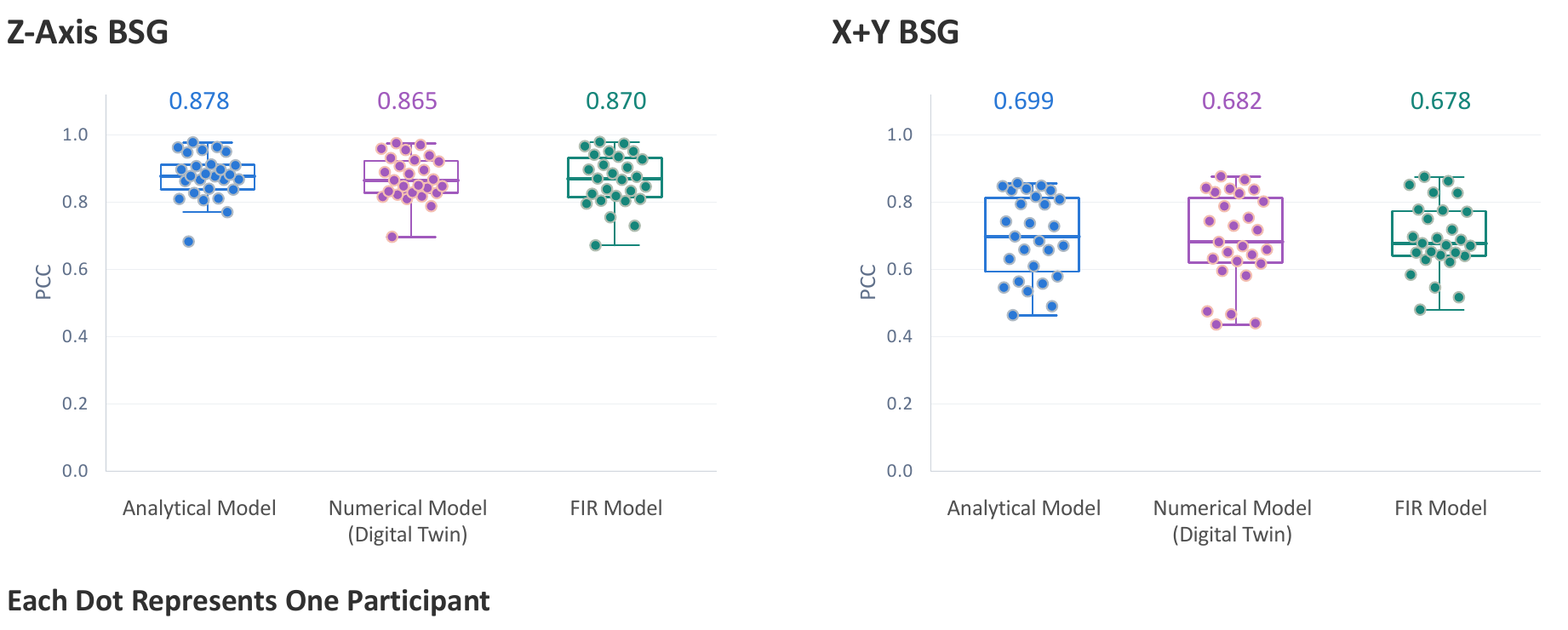}
\caption{BSG Waveform Reconstruction. Analytical, numerical, and FIR models use identical test waveforms. Each dot is one participant's median PCC ($n=27$). Boxes show medians and interquartile ranges, with whiskers extending to observations within 1.5 interquartile ranges. Numbers are cohort medians. Mechanical models include separately calibrated frequency correction.}
\label{fig:matched_model_comparison}
\end{figure}

\begin{table}[!htbp]
\centering\small
\caption{Matched-Window PCC Before and After Frequency Correction. Values are the median and interquartile range across 27 participant medians. FIR has no additional correction beyond its fitted operator.}
\label{tab:matched-model-comparison}
\begin{tabular}{llcc}
\toprule
Model & Frequency Correction & $Z$ PCC [IQR] & $X+Y$ PCC [IQR]\\
\midrule
Analytical Model & No & 0.807 [0.755, 0.838] & 0.501 [0.404, 0.608]\\
Analytical Model & Yes & 0.878 [0.839, 0.911] & 0.699 [0.594, 0.813]\\
Multilayer FEM & No & 0.838 [0.802, 0.868] & 0.644 [0.546, 0.704]\\
Multilayer FEM & Yes & 0.865 [0.828, 0.923] & 0.682 [0.621, 0.814]\\
FIR-121 & Not Applied & 0.870 [0.815, 0.931] & 0.678 [0.642, 0.774]\\
\bottomrule
\end{tabular}
\end{table}

Both mechanical realizations reconstruct measured waveforms after participant-specific frequency calibration. The multilayer FEM reaches median PCCs of 0.865 for $Z$ and 0.682 for $XY$, while the analytical model reaches 0.878 and 0.699. FIR gives 0.870 and 0.678 on the same targets. The uncorrected FEM reaches 0.838 and 0.644, and the uncorrected analytical model reaches 0.807 and 0.501. Frequency correction thus contributes to reconstruction in both mechanical models, whose reported responses combine physical dynamics with an empirical frequency response.

The paired comparisons place the mechanical and FIR reconstructions in a similar observed PCC range. For FEM minus FIR, mean differences in participant-median PCC are $+0.005$ for $Z$ (95\% bootstrap interval $[-0.017,0.029]$) and $-0.005$ for $XY$ ($[-0.043,0.033]$). Corresponding differences for the analytical model are $+0.010$ ($[-0.014,0.036]$) and $+0.003$ ($[-0.033,0.039]$). Every interval spans both signs, leaving the model ordering unresolved without establishing equivalence. The experiment supports calibrated waveform reconstruction with either mechanical realization. Physiological target recovery is a distinct endpoint from this waveform comparison.

\FloatBarrier

\subsection{Transmission Shapes Feature Sensitivity to Stroke Volume}
\label{sec:sv-transfer}
Controlled cardiac perturbations test how transmission changes the direction and strength of feature sensitivity to SV. Active-tension control $T_a$ and peripheral resistance $R_p$ vary contractile drive and afterload. The state bank contains one reference, six additional contractility states, seven additional afterload states, and five combined perturbations. Simulated SV ranges from $44.38$ to $84.58$ mL, with a reference of $63.90$ mL. Each control changes the complete cardiac trajectory and source waveform, so feature--SV associations are evaluated across distinct contractility and afterload trajectories.

The vertical input is the whole-heart inertial force proxy $s_{H,Z}$ in Eq.~\eqref{eq:heart_inertial_source}, and the planar input is aortic recoil $X+Y$. The local RV-patch acceleration in Eq.~\eqref{eq:scg_patch_extraction} is a separate observable. The same 19 simulated source pairs pass through every participant's frozen FIR-121 pair. These operators come from real-data waveform fits and retain their calibrated polarity, lag, gain, and offset during the perturbations. The experiment therefore combines simulated SV changes with transmission models fitted to measured recordings. Participant SV responses were not measured in this analysis.

The analysis separates the conditioned cardiac source, the phase-normalized FIR input, and the calibrated FIR output to locate changes introduced by preprocessing and transmission. All signals use a sampling rate of 100 Hz ($\Delta t=10$ ms). Conditioning applies a taper over the first 30 ms and removes the mean. Cycle standardization and peak anchoring then produce the FIR input. Frozen calibration and the transmission kernel both act on the extracted features.

For a feature functional $\phi$, transmission generally does not commute with feature extraction:
\begin{equation}
\phi\!\left[\mathcal H_p s(\vct\theta)\right]
\ \text{need not have the same SV dependence as}\ 
\phi\!\left[s(\vct\theta)\right].
\label{eq:feature-transmission}
\end{equation}
Frequency-dependent attenuation redistributes spectral power, and phase dispersion changes peak shape and width. In the illustrated analysis, the planar feature $C$ is the power-weighted spectral centroid of a Hann-windowed single cycle over 1--20 Hz. The vertical feature $W$ is the J-peak full width at half maximum returned by the existing detector applied to six repeated cycles. This detector supplies a waveform-width proxy, with valve-event and ejection timing outside its manual verification. Relative to each stage's reference features $C_0$ and $W_0$, the declared scores are
\begin{equation}
q_{XY}=-100\times\frac{C-C_0}{|C_0|},\qquad
q_Z=+100\times\frac{W-W_0}{|W_0|},\qquad
q_{\rm fusion}=q_{XY}+\beta\times q_Z.
\label{eq:sv-feature-fusion}
\end{equation}
These dimensionless scores quantify relative SV sensitivity. Their signs are fixed across participants, with positive scores assigned to an SV increase from the reference. Signed correlation evaluates agreement with this declared direction, while correlation magnitude $|r|$ measures association strength irrespective of direction. An absolute SV scale in mL is not fitted.

\begin{figure}[!htbp]
\centering
\includegraphics[width=0.97\linewidth]{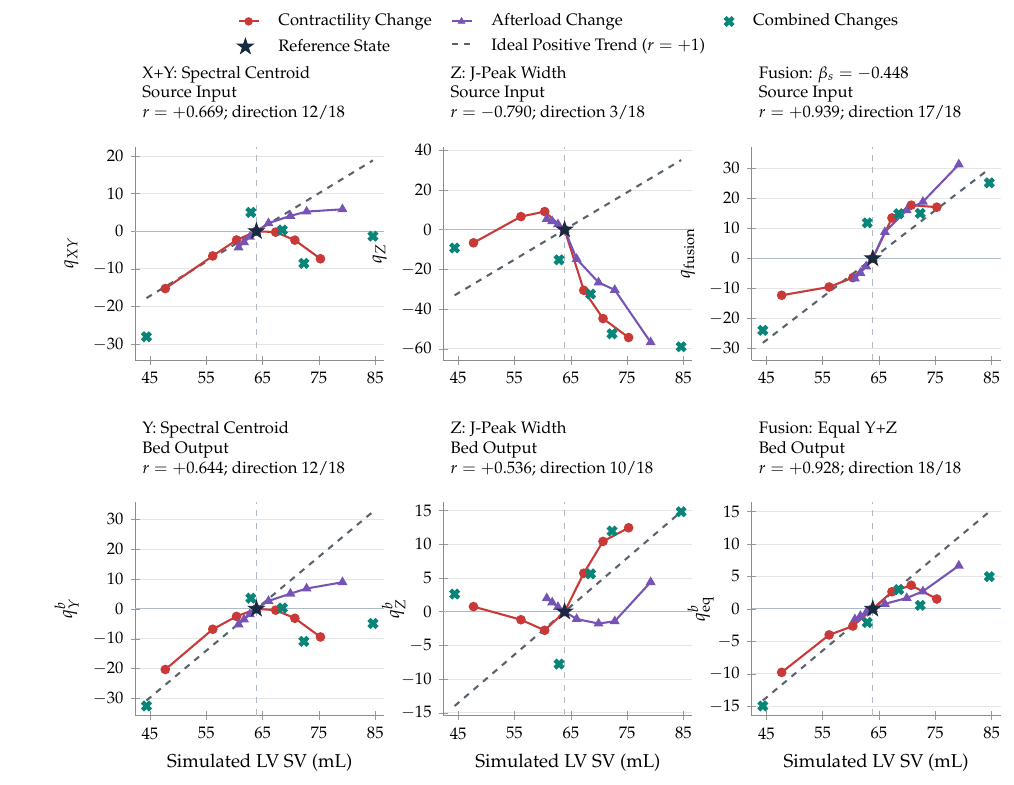}
\caption{Feature Fusion and Stroke Volume. Top: source features. Bottom: transmitted features from one representative bed. Fusion rules differ between rows. Colors distinguish contractility, afterload, and combined changes across 19 simulated states. Stars mark the reference. Dashed diagonals are trend guides. Each plot reports correlation and correctly signed SV changes.}
\label{fig:sv_feature_transfer}
\end{figure}

For the shared FIR input in the upper row of Fig.~\ref{fig:sv_feature_transfer}, $\beta_s=-0.448046$ (displayed as $-0.448$) maximizes signed Pearson correlation between $q_{XY}+\beta_s\times q_Z$ and simulated SV over the 18 nonreference input states. The coefficient of $q_{XY}$ is fixed at one, with no fitted intercept or absolute SV scale. Calibration and scoring use the same states, so this experiment evaluates source association. The calibrated coefficient gives PCC $+0.939$, compared with $+0.883$ when the bed-selected coefficient $-1.405$ is applied to the source.

Figure~\ref{fig:sv_feature_transfer} compares feature behavior at the cardiac-source FIR input with feature complementarity after transmission through a representative bed. The transmission model is shared across two different beds, with calibration fixed separately for each bed and channel. The same 19 simulated cardiac states drive the comparison. The illustrated bed's horizontal channel $Y$ is driven by aortic recoil along body $Y$, whereas the source analysis uses $X+Y$. Both use the whole-heart inertial force proxy along $Z$ for the vertical channel. The illustrated bed's calibration PCCs are $0.937$ for $Y$ and $0.794$ for $Z$. The perturbation experiment evaluates the resulting feature--SV association.

For the illustrated bed, let $C_b$ and $W_b$ denote the output spectral centroid and J-peak width, with subscript $0$ identifying the reference state. We apply the same sign convention to their relative changes and combine them with an equal-weight rule:
\begin{equation}
q_Y^b=-100\times\frac{C_b-C_{b,0}}{|C_{b,0}|},\qquad
q_Z^b=+100\times\frac{W_b-W_{b,0}}{|W_{b,0}|},\qquad
q_{\rm eq}^b=\frac{q_Y^b+q_Z^b}{2}.
\label{eq:bedA-equal-fusion}
\end{equation}
The equal-weight rule evaluates feature complementarity within the illustrated bed. The upper row uses source-calibrated $\beta_s$, and the 27-participant analysis below uses $\beta=-1.405$. Because the rows differ in horizontal source and fusion rule, their numerical difference combines those choices with transmission and does not isolate transmission alone.

Source fusion aligns spectral-centroid and width information with simulated SV, reaching PCC $+0.939$ and the correct direction for 17/18 changes at the FIR input. At the conditioned source, PCCs are $+0.632$ for $q_{XY}$ and $-0.790$ for $q_Z$, with fusion PCC $+0.934$ using input-calibrated $\beta_s$. The corresponding FIR-input values are $+0.669$, $-0.790$, and $+0.939$. The combined low-contractility/low-resistance state has an incorrect fusion direction. These results connect the source association to the signed contribution of each feature.

The two bed-level features recover complementary SV information: their equal-weight mean reaches PCC $+0.928$ and Spearman $\rho=+0.897$, with all 18 change directions correct, including all five combined perturbations. The horizontal and vertical scores separately give PCCs $+0.644$ and $+0.536$, with 12/18 and 10/18 correct directions. Fusion therefore resolves the direction errors present in each single score for this bed configuration. These descriptive results use features identified in preceding analyses and the stated simulated state bank. Feature selection and absolute SV calibration are separate from this evaluation.

Participant-specific transmission changes the signed feature--SV relationship. With $\beta=-1.405$ held fixed before and after transfer, FIR-input fusion has PCC $+0.883$. Across the 27 participant transfer models, median output PCCs are $+0.679$ for $q_{XY}$, $-0.322$ for $q_Z$, and $+0.695$ for their fusion. The vertical association reverses sign relative to the FIR input in 8 of 27 participants. This observed sign change makes the transmitted feature response, as well as the cardiac source, part of SV interpretation.

\subsection{Calibration Determines the Benefit of Feature Fusion}
\label{sec:sv-fusion}
The shared fusion weight tests how a coefficient calibrated on a subset behaves across participant-specific transfer models. The coefficient $\beta=-1.405$ in Eq.~\eqref{eq:sv-feature-fusion} maximizes mean signed PCC across five models. Their mean rises from $0.686$ for the planar score to $0.812$ for fusion. The weight is then frozen for all 27 models, including the other 22 outside coefficient selection. This exploratory selection is distinct from the source-calibrated $\beta_s$ and the equal-weight bed-output rule in Fig.~\ref{fig:sv_feature_transfer}.

\begin{figure}[!htbp]
\centering
\includegraphics[width=0.97\linewidth]{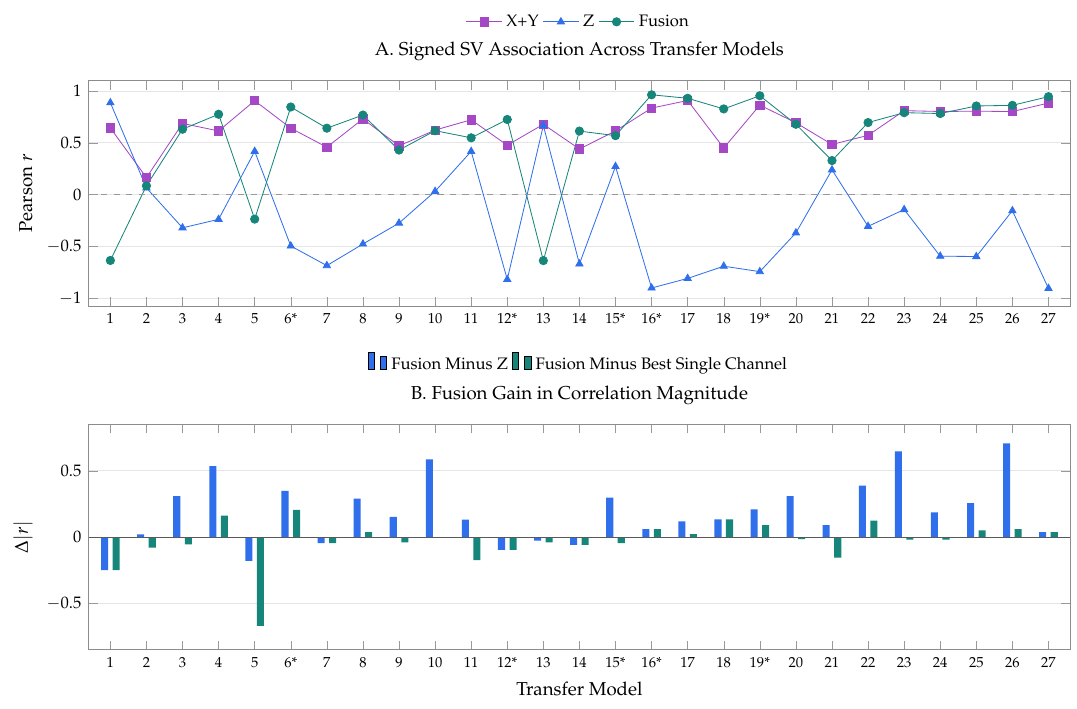}
\caption{Fusion Benefit Varies Across Transmission Models. (A) Signed SV correlations. (B) Correlation-magnitude gains over single channels. All 27 models use one shared fusion weight across 18 nonreference simulated states. Asterisks mark the five models used to select that weight.}
\label{fig:sv_fusion_cohort}
\end{figure}

A shared fusion weight improves the vertical-only association in most transfer models, while the planar score supplies a stronger single-channel reference. Across all 27 models, fusion increases $|r|$ relative to $Z$ in 21 and relative to the stronger of $Z$ and $XY$ in 11 (Fig.~\ref{fig:sv_fusion_cohort}). Among the 22 models outside coefficient selection, the counts are 17 and 8. Median gains are $+0.153$ against $Z$ and $-0.019$ against the stronger single channel. The median number of correctly signed changes is 15/18 for fusion, 15/18 for $XY$, and 8/18 for $Z$. Thus the fixed weight chiefly adds to the vertical feature, with participant-specific benefit relative to the best available single-channel feature.

The signed channel responses explain why fusion benefit depends on calibration. In one participant transfer model, $r_{XY}=+0.449$ and $r_Z=-0.693$ combine to give $r_{\rm fusion}=+0.827$, with all 18 changes correctly directed. In another, a positive vertical association of $r_Z=+0.887$ becomes $r_{\rm fusion}=-0.638$ under the same weight. Retaining the signed components reveals whether a shared coefficient combines complementary information or reverses the physiological interpretation. This experiment evaluates $Z$, the planar sum $XY$, and their fusion. The planar sum already combines two measured axes.

Individual coefficient calibration evaluates whether labeled perturbations can adapt fusion to a participant's transfer response. One or two labeled simulated states determine the coefficient under a fixed planar coefficient and prescribed score scale. Calibration states and the reference are excluded from subsequent correlations. The two-state test uses one contractility perturbation and one afterload perturbation for calibration, with the remaining 16 states forming the held-out test set. Allowing an additional unknown output scale makes the one-state problem underdetermined.

With these two calibration states, test PCC increases over $XY$ alone in four of five participants: $0.481\!\rightarrow\!0.736$, $0.555\!\rightarrow\!0.784$, $0.850\!\rightarrow\!0.965$, and $0.856\!\rightarrow\!0.960$. The fifth changes from $0.821$ to $0.671$. The selected calibration states therefore influence how the coefficient carries feature complementarity into the remaining perturbations.

Controlled perturbations connect relative SV sensitivity to both cardiac source dynamics and body--bed transmission. Spectral-centroid and peak-width features supply complementary information, with their combination recovering all 18 change directions in the illustrated bed. The participant analyses show why signed channel responses and coefficient calibration belong to that interpretation. Together, these results establish a framework for studying relative SV information in transmitted waveforms under controlled simulated perturbations.
\FloatBarrier

\subsection{Cardiovascular Source Models Increase Held-Out Waveform Correlation}
\label{sec:segmented-effective-transfer}
The temporal source-to-measurement experiment tests whether cardiovascular source waveforms add information beyond beat timing. It retains the vascular source locations defined in Section~\ref{sec:cardiac-recoil-sources} and uses full temporal waveforms, complementing the cycle-template experiments across 27 participants.

\subsubsection{Preserving Source Locations Through the Observation Model}

Ascending, arch, and descending loads reach the sensor through different mechanical paths. A reduced Y-channel observation can retain three vascular channels alongside the heart-solid component,
\begin{equation}
\widehat y_Y(t)=(h_H*s_{H,Y})(t)
+\sum_{r\in\{\mathrm{asc},\mathrm{arch},\mathrm{desc}\}}
(h_r*F_{A,r,Y})(t),
\label{eq:segmented_recoil_transfer}
\end{equation}
where $*$ denotes convolution and each $h_r$ is a separate effective transfer kernel. Summing the vascular sources before applying one kernel imposes a common transmission path and can discard useful spatial information. Combining myocardial and vascular inputs requires compatible force conventions and interface boundaries. Their physical coupling is specified by those definitions in addition to the source sum.

The temporal experiment uses separate training and held-out test segments. A simulated beat is retimed to the measured heart rate and placed at each detected beat. A 121-tap temporal FIR spans 1.21 s at 100 Hz. The timing control passes beat impulses through the same FIR, testing how much morphology the kernel can reconstruct from timing alone. Measured chest SCG supplies a source-input reference. The simulated source is one beat with uncalibrated absolute bed amplitude. This temporal protocol differs from the circular phase-template comparison in Section~\ref{sec:simplified-body-bed}. Its input comparisons evaluate information beyond timing.

The cardiovascular source model yields higher held-out PCC than the timing control under both tested splits. On held-out segments drawn from the same acquisition sets, the heart-plus-three-segment model reaches median PCC 0.374, compared with 0.317 for the timing control and 0.399 for the measured chest-SCG reference. When an entire acquisition set is withheld from training, the source-model and timing-control medians are 0.210 and 0.179. The source contribution beyond timing persists in both settings, with lower correlations when transferring to a separate acquisition set.

\FloatBarrier

\clearpage
\section{Conclusions and Key Takeaways}
\label{sec:numerical-conclusions}
\textbf{Theory and numerical simulation establish a physical framework for BSG.} The framework connects cardiovascular dynamics, mechanical excitation, body--bed transmission, and sensor observation. The four-chamber model reproduces reference chamber dynamics and local wall-motion timing, while the transmission models reconstruct measured BSG morphology. Across 27 participants, the multilayer FEM reaches median PCCs of $0.865$ for $Z$ and $0.682$ for $X+Y$ after participant-specific frequency correction. These results connect the mechanisms of BSG generation to recorded waveforms.

\textbf{BSG contains complementary SCG and BCG contributions.} Local thoracic deformation and whole-body recoil arise from the coupled cardiovascular state and propagate to the bed through the body and its contacts. The model retains myocardial and vascular loading separately, resolving how their spatial directions and timing combine in the measured BSG waveform. This combined description links local cardiac motion and momentum exchange to the physiological content of BSG.

\textbf{Multiple sensing directions provide complementary physiological information.} Under controlled simulated contractility and afterload changes, combining horizontal spectral-centroid and vertical peak-width features raises PCC from $0.644$ and $0.536$ for the individual features to $0.928$ in the illustrated bed. Fusion recovers all 18 SV-change directions, compared with 12 and 10 for the individual features. Together with the SV, CO, and BP observability analysis in Chapter~\ref{part:theory}, these results demonstrate the physiological value of combining directional information in BSG.

\textbf{Body--bed transmission shapes both BSG morphology and physiological feature sensitivity.} Frequency-dependent attenuation and phase shifts reshape waveform amplitudes, peaks, and spectral content. Across 27 recording-derived transfer models, identical simulated cardiac perturbations produce different feature--SV relationships, including a reversal of the vertical feature's association in 8 models relative to the source input. The body and bed therefore form an essential part of the physiological measurement model. Accounting for their transfer functions connects the observed BSG features to the underlying cardiovascular changes.
\FloatBarrier

\FloatBarrier
\section*{Acknowledgments}
We thank Runnan Si of the University of Georgia for sharing his expertise and supporting the later-stage optimization of the digital twin. We also thank Dr. Nori Nakata for his constructive comments on BSG modeling, particularly the representation of cardiac sources, wave propagation through the body and bed, and sensor response.

\clearpage
\markboth{References}{}
\phantomsection
\addcontentsline{toc}{section}{References}

\begingroup
\small
\linespread{1.0}\selectfont
\setlength{\bibsep}{1pt}

\endgroup

\begin{thebibliography}{78}
\providecommand{\natexlab}[1]{#1}
\providecommand{\url}[1]{\texttt{#1}}
\expandafter\ifx\csname urlstyle\endcsname\relax
  \providecommand{\doi}[1]{doi: #1}\else
  \providecommand{\doi}{doi: \begingroup \urlstyle{rm}\Url}\fi

\bibitem[Song et~al.(2024{\natexlab{a}})Song, Li, Luo, Xie, Phillips, Ke, and
  Song]{song2024engagement}
Yingjian Song, Bingnan Li, Dan Luo, Zaipeng Xie, Bradley~G. Phillips, Yuan Ke,
  and Wenzhan Song.
\newblock Engagement-free and contactless bed occupancy and vital signs
  monitoring.
\newblock \emph{IEEE Internet of Things Journal}, 11\penalty0 (5):\penalty0
  7935--7947, 2024{\natexlab{a}}.
\newblock URL \url{https://pmc.ncbi.nlm.nih.gov/articles/PMC11162756/}.

\bibitem[Zhang et~al.(2026{\natexlab{a}})Zhang, Zhang, Li, Wu, Dou, Yin, An,
  Noh, and Song]{zhang2026phybp}
Yuanyuan Zhang, Yida Zhang, Jiahui Li, Yuyan Wu, Fei Dou, Xiao Yin, Zhenlin An,
  Hae~Young Noh, and Wenzhan Song.
\newblock Physics-constrained deep learning model for contactless blood
  pressure monitoring from triaxial bodyseismography.
\newblock \emph{arXiv preprint arXiv:2608.23562}, 2026{\natexlab{a}}.
\newblock URL \url{https://arxiv.org/abs/2608.23562}.

\bibitem[Zhang et~al.(2026{\natexlab{b}})Zhang, Si, Wu, Biddappa, Zhang, Zeng,
  Chen, Noh, Dou, An, and Song]{zhang2027heartintheloop}
Yuanyuan Zhang, Runnan Si, Yuyan Wu, Sahas Biddappa, Yida Zhang, Zixuan Zeng,
  Jiayu Chen, Hae~Young Noh, Fei Dou, Zhenlin An, and Wenzhan Song.
\newblock {Heart in the Loop: Learning Coupled Dynamics through Differentiable
  Cardiac Mechanics and Hemodynamics}.
\newblock submitted to {ICLR} 2027, 2026{\natexlab{b}}.

\bibitem[Li et~al.(2018)Li, Clemente, and Song]{li2018nonintrusive}
Fangyu Li, Jose Clemente, and WenZhan Song.
\newblock Non-intrusive and non-contact sleep monitoring with seismometer.
\newblock In \emph{2018 IEEE GlobalSIP}, 2018.
\newblock URL
  \url{http://sensorweb.engr.uga.edu/wp-content/uploads/2018/09/Dropbox_li2018intrusive.pdf}.

\bibitem[Clemente et~al.(2019)Clemente, Li, Valero, and
  Song]{clemente2019contactless}
Jose Clemente, Fangyu Li, Maria Valero, and WenZhan Song.
\newblock Contactless device for monitoring on-bed activities and vital signs.
\newblock In \emph{2019 IEEE International Conference on Smart Computing
  (SMARTCOMP)}, pages 472--474. IEEE, 2019.

\bibitem[Clemente et~al.(2020)Clemente, Valero, Li, Wang, and
  Song]{clemente2020helena}
Jose Clemente, Maria Valero, Fangyu Li, Chengliang Wang, and WenZhan Song.
\newblock Helena: Real-time contact-free monitoring of sleep activities and
  events around the bed.
\newblock In \emph{2020 IEEE International Conference on Pervasive Computing
  and Communications (PerCom)}, pages 1--10. IEEE, 2020.

\bibitem[Valero et~al.(2021)Valero, Clemente, Li, and Song]{valero2021health}
Maria Valero, Jose Clemente, Fangyu Li, and WenZhan Song.
\newblock Health and sleep nursing assistant for real-time, contactless, and
  non-invasive monitoring.
\newblock \emph{Pervasive and mobile computing}, 75:\penalty0 101422, 2021.

\bibitem[Li et~al.(2021)Li, Valero, Clemente, Tse, and Song]{li2020smart}
Fangyu Li, Maria Valero, Jose Clemente, Zion Tse, and WenZhan Song.
\newblock Smart sleep monitoring system via passively sensing human vibration
  signals.
\newblock \emph{IEEE sensors journal}, 21\penalty0 (13):\penalty0 14466--14473,
  2021.

\bibitem[Song et~al.(2024{\natexlab{b}})Song, Pitafi, Dou, Sun, Zhang,
  Phillips, and Song]{song2024self}
Yingjian Song, Zaid~Farooq Pitafi, Fei Dou, Jin Sun, Xiang Zhang, Bradley~G
  Phillips, and WenZhan Song.
\newblock Self-supervised representation learning and temporal-spectral feature
  fusion for bed occupancy detection.
\newblock \emph{Proceedings of the ACM on Interactive, Mobile, Wearable and
  Ubiquitous Technologies}, 8\penalty0 (3):\penalty0 1--25, 2024{\natexlab{b}}.

\bibitem[Song et~al.(2025)Song, Xiang, Zeng, Chen, Zhang, Pitafi, Yang, Lu,
  Zhang, Phillips, et~al.]{song2025multi}
Yingjian Song, Haotian Xiang, Zixuan Zeng, Jiayu Chen, Yida Zhang, Zaid~Farooq
  Pitafi, He~Yang, Qin Lu, Xiang Zhang, Bradley~G Phillips, et~al.
\newblock Multi-granularity supervised contrastive learning with online
  adaptation for contactless in-bed posture classification.
\newblock \emph{Proceedings of the ACM on Interactive, Mobile, Wearable and
  Ubiquitous Technologies}, 9\penalty0 (2):\penalty0 1--32, 2025.

\bibitem[Song et~al.(2026{\natexlab{a}})Song, Chen, Zeng, Zhang, Pitafi, Das,
  Phillips, Kwon, Healy, Zhang, Dou, and Song]{song2026bodyseismography}
Yingjian Song, Jiayu Chen, Zixuan Zeng, Yida Zhang, Zaid~Farooq Pitafi,
  Deepak~Kumar Das, Bradley~G. Phillips, Younghoon Kwon, William~J. Healy,
  Xiang Zhang, Fei Dou, and Wenzhan Song.
\newblock Contactless sleep apnea detection with bodyseismography.
\newblock \emph{Proceedings of the ACM on Interactive, Mobile, Wearable and
  Ubiquitous Technologies}, 10\penalty0 (2):\penalty0 59:1--59:31,
  2026{\natexlab{a}}.

\bibitem[Song et~al.(2026{\natexlab{b}})Song, Chen, Zeng, Zhang, Pitafi,
  Phillips, Zhang, Dou, and Song]{song2026attention}
Yingjian Song, Jiayu Chen, Zixuan Zeng, Yida Zhang, Zaid~Farooq Pitafi,
  Bradley~G. Phillips, Xiang Zhang, Fei Dou, and WenZhan Song.
\newblock Attention feature fusion with cluster contrastive learning for
  snoring and breath-holding detection using seismic sensing.
\newblock In \emph{Proceedings of the 2026 IEEE International Conference on
  Pervasive Computing and Communications (PerCom)}. IEEE, 2026{\natexlab{b}}.

\bibitem[Song et~al.(2024{\natexlab{c}})Song, Li, Luo, Glasgow, Phillips, Ke,
  and Song]{song2024real}
Yingjian Song, Bingnan Li, Dan Luo, Glenna S~Brewster Glasgow, Bradley~G
  Phillips, Yuan Ke, and WenZhan Song.
\newblock Real-time continuous blood pressure estimation with contact-free
  bedseismogram.
\newblock In \emph{ICC 2024-IEEE International Conference on Communications},
  pages 214--219. IEEE, 2024{\natexlab{c}}.

\bibitem[Zhang et~al.(2025{\natexlab{a}})Zhang, Chen, Song, Zeng, Zhang, Lu,
  Phillips, Xie, and Song]{zhang2025blood}
Yida Zhang, Jiayu Chen, Yingjian Song, Zixuan Zeng, Xiang Zhang, Qin Lu,
  Bradley~G Phillips, Zaipeng Xie, and WenZhan Song.
\newblock Blood pressure estimation from vibration signals via coarse-to-fine
  contrastive learning, feature selection and synthesis.
\newblock In \emph{Proceedings of the ACM/IEEE International Conference on
  Connected Health: Applications, Systems and Engineering Technologies}, pages
  134--138, 2025{\natexlab{a}}.

\bibitem[Zhang et~al.(2026{\natexlab{c}})Zhang, Zhang, Chen, Zeng, Li, Dou,
  Fundora, Maher, and Song]{zhang2026triaxial}
Yuanyuan Zhang, Yida Zhang, Jiayu Chen, Zixuan Zeng, Jiahui Li, Fei Dou,
  Michael Fundora, Kevin Maher, and Wenzhan Song.
\newblock {ID}: 4566283 triaxial bodyseismography for contactless blood
  pressure monitoring using physics-constrained deep learning.
\newblock \emph{Heart Rhythm O2}, Sep. 2026{\natexlab{c}}.

\bibitem[Wu et~al.(2026)Wu, Zhang, Zhang, Li, Dou, Song, and
  Noh]{wu2026contact}
Yuyan Wu, Yuanyuan Zhang, Yida Zhang, Jiahui Li, Fei Dou, WenZhan Song, and
  Hae~Young Noh.
\newblock Contact-free continuous {ABP} waveform reconstruction via triaxial
  bodyseismography.
\newblock In \emph{Proceedings of the 4th International Workshop on
  Human-Centered Sensing, Modeling, and Intelligent Systems}, pages 111--112.
  ACM, 2026.

\bibitem[Zhang et~al.(2026{\natexlab{d}})Zhang, Zhang, Zeng, Chen, Li, Yin,
  Xie, Zhang, An, Dou, and Song]{zhang2026bsg2ecg}
Yida Zhang, Yuanyuan Zhang, Zixuan Zeng, Jiayu Chen, Jiahui Li, Xiao Yin,
  Zaipeng Xie, Xiang Zhang, Zhenlin An, Fei Dou, and Wenzhan Song.
\newblock {BSG2ECG}: Contactless electrocardiogram monitoring with triaxial
  bodyseismography.
\newblock submitted, 2026{\natexlab{d}}.

\bibitem[Zhang et~al.(2025{\natexlab{b}})Zhang, Guan, Li, Yang, Yue, and
  Lim]{zhang2024radarODE}
Yuanyuan Zhang, Runwei Guan, Lingxiao Li, Rui Yang, Yutao Yue, and Eng~Gee Lim.
\newblock {radarODE}: An {ODE}-embedded deep learning model for contactless
  {ECG} reconstruction from millimeter-wave radar.
\newblock \emph{IEEE Transactions on Mobile Computing}, Oct.
  2025{\natexlab{b}}.

\bibitem[Zhang et~al.(2025{\natexlab{c}})Zhang, Yang, Yue, and
  Lim]{zhang2024radarODE-MTL}
Yuanyuan Zhang, Rui Yang, Yutao Yue, and Eng~Gee Lim.
\newblock {radarODE-MTL}: A multitask learning framework with eccentric
  gradient alignment for robust radar-based {ECG} reconstruction.
\newblock \emph{IEEE Transactions on Instrumentation and Measurement}, Apr.
  2025{\natexlab{c}}.

\bibitem[Zhang et~al.(2026{\natexlab{e}})Zhang, Zhao, Xiong, Yang, Lim, and
  Yue]{zhang2025high}
Yuanyuan Zhang, Haocheng Zhao, Sijie Xiong, Rui Yang, Eng~Gee Lim, and Yutao
  Yue.
\newblock From {High-SNR} radar signal to {ECG}: A transfer learning model with
  cardio-focusing algorithm for scenarios with limited data.
\newblock \emph{IEEE Transactions on Mobile Computing}, Apr.
  2026{\natexlab{e}}.

\bibitem[Zhang et~al.(2026{\natexlab{f}})Zhang, Zhang, Zeng, Chen, Li, Yin,
  Xie, Zhang, An, Dou, and Song]{zhang2027vibrosv}
Yida Zhang, Yuanyuan Zhang, Zixuan Zeng, Jiayu Chen, Jiahui Li, Xiao Yin,
  Zaipeng Xie, Xiang Zhang, Zhenlin An, Fei Dou, and Wenzhan Song.
\newblock {VibroSV: Contactless Tracking of Stroke Volume Changes from
  Bodyseismography}.
\newblock submitted to {PerCom} 2027, 2026{\natexlab{f}}.

\bibitem[Chen et~al.(2025)Chen, Song, Zhang, Zeng, Zhang, Pitafi, Xie, Das,
  Dong, Lu, Yin, and Song]{chen2025selfdenoiser}
Jiayu Chen, Yingjian Song, Yida Zhang, Zixuan Zeng, Xiang Zhang, Zaid~Farooq
  Pitafi, Zaipeng Xie, Deepak~Kumar Das, Nishan Dong, Junjie Lu, Xiao Yin, and
  Wenzhan Song.
\newblock {SelfDenoiser}: Self-supervised seismic signal denoiser for
  continuous and contactless cardiac monitoring.
\newblock \emph{Proceedings of the ACM on Interactive, Mobile, Wearable and
  Ubiquitous Technologies}, 9\penalty0 (4):\penalty0 165:1--165:31, 2025.

\bibitem[Li et~al.(2026)Li, Zhang, Zeng, Chen, Song, Xiao, Dong, Lu, Kwon,
  Zhang, et~al.]{li2026peak}
Jiahui Li, Yida Zhang, Zixuan Zeng, Jiayu Chen, Yingjian Song, Yin Xiao, Nishan
  Dong, Junjie Lu, Younghoon Kwon, Xiang Zhang, et~al.
\newblock {Peak-Detector}: Explainable peak detection via instruction-tuned
  large language models in physiological signal.
\newblock \emph{Proceedings of the ACM on Interactive, Mobile, Wearable and
  Ubiquitous Technologies}, 10\penalty0 (2):\penalty0 1--44, 2026.

\bibitem[Zhang et~al.(2023)Zhang, Yang, Yue, Lim, and Wang]{zhang2023overview}
Yuanyuan Zhang, Rui Yang, Yutao Yue, Eng~Gee Lim, and Zidong Wang.
\newblock An overview of algorithms for contactless cardiac feature extraction
  from radar signals: Advances and challenges.
\newblock \emph{IEEE Transactions on Instrumentation and Measurement}, Aug.
  2023.

\bibitem[Zhang et~al.(2025{\natexlab{d}})Zhang, Xiong, Yang, Lim, and
  Yue]{zhang2025horcrux}
Yuanyuan Zhang, Sijie Xiong, Rui Yang, Eng~Gee Lim, and Yutao Yue.
\newblock Recover from {Horcrux}: A spectrogram augmentation method for cardiac
  feature monitoring from radar signal components.
\newblock In \emph{2025 47th Annual International Conference of the IEEE
  Engineering in Medicine and Biology Society (EMBC)}. IEEE, Jul.
  2025{\natexlab{d}}.

\bibitem[Zhang(2025)]{zhang2025robust}
Yuanyuan Zhang.
\newblock \emph{Robust Cardiac Feature Monitoring Based on Millimeter-Wave
  Radar}.
\newblock PhD thesis, University of Liverpool, Sep. 2025.

\bibitem[Song et~al.(2024{\natexlab{d}})Song, Pitafi, Zeng, Chen, Zhang,
  Phillips, Brainard, and Song]{song2024poster}
Yingjian Song, Zaid~Farooq Pitafi, Zixuan Zeng, Jiayu Chen, Yida Zhang,
  Bradley~G Phillips, Benjamin~M Brainard, and WenZhan Song.
\newblock Poster: A contactless health monitoring system for humans and
  animals.
\newblock In \emph{Proceedings of the 22nd ACM Conference on Embedded Networked
  Sensor Systems}, pages 869--870, 2024{\natexlab{d}}.

\bibitem[Pitafi et~al.(2024)Pitafi, Sleda, Moreno, Tompkins, Brainard, and
  Song]{pitafi2024cagedot}
Zaid~Farooq Pitafi, Melissa Sleda, Silvia~NJ Moreno, S~Mark Tompkins,
  Benjamin~M Brainard, and WenZhan Song.
\newblock {CageDot}: Contactless animal activity monitoring system to follow
  infectious disease progress.
\newblock In \emph{ICC 2024-IEEE International Conference on Communications},
  pages 812--817. IEEE, 2024.

\bibitem[Pitafi et~al.(2025)Pitafi, Song, Xie, Brainard, and
  Song]{pitafi2025contactless}
Zaid~Farooq Pitafi, Yingjian Song, Zaipeng Xie, Benjamin Brainard, and WenZhan
  Song.
\newblock Contactless vital signs monitoring for animals.
\newblock \emph{IEEE Internet of Things Journal}, 2025.

\bibitem[Pitafi et~al.(2026)Pitafi, Yang, Chen, Song, Ye, Tse, Song, and
  Song]{pitafi2026testbed}
Zaid~Farooq Pitafi, He~Yang, Jiayu Chen, Yingjian Song, Jin Ye, Zion Tse, Kenan
  Song, and Wenzhan Song.
\newblock A testbed for development and validation of contactless vital signs
  monitoring systems.
\newblock Preprint, 2026.

\bibitem[{RacoTech}(2018)]{rgi45hz}
{RacoTech}.
\newblock {RGI-4.5Hz} low frequency geophone element, 2018.
\newblock URL \url{http://www.racotech.biz/product.php?id=84}.
\newblock Accessed March 8, 2025.

\bibitem[{Deutsches Elektronen-Synchrotron (DESY)}(n.d.)]{desy_geophones}
{Deutsches Elektronen-Synchrotron (DESY)}.
\newblock Geophones.
\newblock Ground Vibrations technical documentation, n.d.
\newblock URL \url{https://vibration.desy.de/equipment/geophones/}.
\newblock Accessed October 12, 2026.

\bibitem[{Safran Sensing Technologies Switzerland}(2022)]{safran_si1003}
{Safran Sensing Technologies Switzerland}.
\newblock \emph{{SI1000}: Single Axis Analog Accelerometer, Datasheet}, 2022.
\newblock URL \url{https://www.safran-group.com/download/media/413488}.
\newblock Document 30S.SI1000.B.04.22; SI1003 typical white noise; accessed
  September 8, 2026.

\bibitem[{Silicon Designs}(n.d.)]{sdi1521}
{Silicon Designs}.
\newblock \emph{{SDI 1521} Surface Mount Accelerometers}, n.d.
\newblock URL
  \url{https://www.silicondesigns.com/product/industrial-grade-surface-mount-accelerometers/1521/}.
\newblock Manufacturer specifications for the 1521-002, accessed September 8,
  2026.

\bibitem[Li et~al.(2024)Li, Huang, Yao, Mu, Zong, and Shen]{li2024sleepbcg}
Yong-Xian Li, Jiong-Ling Huang, Xin-Yu Yao, Si-Qi Mu, Shou-Xin Zong, and
  Yan-Fei Shen.
\newblock A ballistocardiogram dataset with reference sensor signals in
  long-term natural sleep environments.
\newblock \emph{Scientific Data}, 11:\penalty0 1091, 2024.
\newblock URL \url{https://www.nature.com/articles/s41597-024-03950-5}.

\bibitem[Albukhari et~al.(2019)Albukhari, Lima, and
  Mescheder]{albukhari2019bed}
Almothana Albukhari, Frederico Lima, and Ulrich Mescheder.
\newblock Bed-embedded heart and respiration rates detection by longitudinal
  ballistocardiography and pattern recognition.
\newblock \emph{Sensors}, 19\penalty0 (6):\penalty0 1451, 2019.
\newblock URL \url{https://www.mdpi.com/1424-8220/19/6/1451}.

\bibitem[S{\o}rensen et~al.(2018)S{\o}rensen, Schmidt, Jensen, S{\o}gaard, and
  Struijk]{sorensen2018}
Kasper S{\o}rensen, Samuel~E. Schmidt, Ask~S. Jensen, Peter S{\o}gaard, and
  Johannes~J. Struijk.
\newblock Definition of fiducial points in the normal seismocardiogram.
\newblock \emph{Scientific Reports}, 8:\penalty0 15455, 2018.
\newblock URL \url{https://www.nature.com/articles/s41598-018-33675-6}.
\newblock Figure~\ref{fig:prelim_scg}(b) adapted from Figure~4 under
  \href{https://creativecommons.org/licenses/by/4.0/}{CC BY 4.0}.

\bibitem[Taebi et~al.(2017)Taebi, Sandler, Kakavand, and Mansy]{taebi2017}
Amirtaha Taebi, Richard~H. Sandler, Bahram Kakavand, and Hansen~A. Mansy.
\newblock Seismocardiographic signal timing with myocardial strain.
\newblock In \emph{2017 IEEE Signal Processing in Medicine and Biology
  Symposium (SPMB)}, pages 1--2, 2017.
\newblock URL \url{https://arxiv.org/abs/1804.11184}.

\bibitem[Ganti et~al.(2022)Ganti, Gazi, An, et~al.]{ganti2022sv}
Venu~G. Ganti, Asim~H. Gazi, Sungtae An, et~al.
\newblock Wearable seismocardiography-based assessment of stroke volume in
  congenital heart disease.
\newblock \emph{Journal of the American Heart Association}, 11\penalty0
  (18):\penalty0 e026067, 2022.
\newblock URL \url{https://pmc.ncbi.nlm.nih.gov/articles/PMC9683676/}.

\bibitem[Agam et~al.(2022)Agam, S{\o}gaard, Kragholm, Jensen, S{\o}rensen,
  Hansen, and Schmidt]{agam2022diastolic}
Ahmad Agam, Peter S{\o}gaard, Kristian Kragholm, Ask~Schou Jensen, Kasper
  S{\o}rensen, John Hansen, and Samuel Schmidt.
\newblock Correlation between diastolic seismocardiography variables and
  echocardiography variables.
\newblock \emph{European Heart Journal - Digital Health}, 3\penalty0
  (3):\penalty0 465--472, 2022.
\newblock URL \url{https://pmc.ncbi.nlm.nih.gov/articles/PMC9707922/}.

\bibitem[Carlson et~al.(2021)Carlson, Turpin, Suliman, Ade, Warren, and
  Thompson]{carlson2021bed}
Charles Carlson, Vanessa-Rose Turpin, Ahmad Suliman, Carl Ade, Steve Warren,
  and David~E. Thompson.
\newblock Bed-based ballistocardiography: Dataset and ability to track
  cardiovascular parameters.
\newblock \emph{Sensors}, 21\penalty0 (1):\penalty0 156, 2021.

\bibitem[P{\'e}rez-{\'A}vila et~al.(2024)P{\'e}rez-{\'A}vila, Ruiz-Herrera,
  Mart{\'i}nez-Olmos, Carvajal, Capit{\'a}n-Vallvey, L{\'o}pez-Ruiz, and
  Palma]{perezavila2024sleep}
Antonio~J. P{\'e}rez-{\'A}vila, Noelia Ruiz-Herrera, Antonio
  Mart{\'i}nez-Olmos, Miguel~A. Carvajal, Luis~Ferm{\'i}n Capit{\'a}n-Vallvey,
  Nuria L{\'o}pez-Ruiz, and Alberto~J. Palma.
\newblock Smartphone-based and non-invasive sleep stage identification system
  with piezo-capacitive sensors.
\newblock \emph{Sensors and Actuators A: Physical}, 376:\penalty0 115659, 2024.

\bibitem[Zhang et~al.(2025{\natexlab{e}})Zhang, Peng, Dong, Hu, Long, Lyu, and
  Lu]{zhang2025apnea}
Biyong Zhang, Zheng Peng, Chunjiao Dong, Jun Hu, Xi~Long, Tan Lyu, and Peilin
  Lu.
\newblock Non-obtrusive monitoring of obstructive sleep apnea syndrome based on
  ballistocardiography: A preliminary study.
\newblock \emph{Frontiers in Neuroscience}, 19:\penalty0 1549783,
  2025{\natexlab{e}}.

\bibitem[Sandelin et~al.(2025)Sandelin, Lahdenoja, Elnaggar, Rekola, Anzanpour,
  Seifizarei, Kaisti, Koivisto, Lehto, Nuotio, Jaakkola, Relander, Vasankari,
  Airaksinen, and Kiviniemi]{sandelin2025af}
Jonas Sandelin, O.~Lahdenoja, I.~Elnaggar, R.~Rekola, A.~Anzanpour,
  S.~Seifizarei, M.~Kaisti, T.~Koivisto, J.~Lehto, J.~Nuotio, J.~Jaakkola,
  A.~Relander, T.~Vasankari, J.~Airaksinen, and T.~Kiviniemi.
\newblock Bed sensor ballistocardiogram for non-invasive detection of atrial
  fibrillation: A comprehensive clinical study.
\newblock \emph{Physiological Measurement}, 46\penalty0 (3):\penalty0 035003,
  2025.

\bibitem[Xing et~al.(2024)Xing, Li, Chen, Jiang, and Dong]{xing2024bedbp}
Xiaoman Xing, Huan Li, Qi~Chen, Chenyu Jiang, and Wen-fei Dong.
\newblock Blood pressure monitoring with piezoelectric bed sensor systems.
\newblock \emph{Biomedical Signal Processing and Control}, 87:\penalty0 105479,
  2024.

\bibitem[Starr et~al.(1939)Starr, Rawson, Schroeder, and
  Joseph]{starr1939ballistocardiogram}
Isaac Starr, A.~J. Rawson, H.~A. Schroeder, and N.~R. Joseph.
\newblock Studies on the estimation of cardiac output in man, and of
  abnormalities in cardiac function, from the heart's recoil and the blood's
  impacts; the ballistocardiogram.
\newblock \emph{American Journal of Physiology}, 127\penalty0 (1):\penalty0
  1--28, 1939.

\bibitem[Starr and Schroeder(1940)]{starr1940standards}
Isaac Starr and Henry~A. Schroeder.
\newblock Ballistocardiogram. {II}. normal standards, abnormalities commonly
  found in diseases of the heart and circulation, and their significance.
\newblock \emph{Journal of Clinical Investigation}, 19\penalty0 (3):\penalty0
  437--450, 1940.

\bibitem[Starr(1954)]{starr1954essay}
Isaac Starr.
\newblock Essay on the ballistocardiogram.
\newblock \emph{JAMA}, 155\penalty0 (16):\penalty0 1413--1425, 1954.

\bibitem[Starr(1965)]{starr1965progress}
Isaac Starr.
\newblock Progress towards a physiological cardiology: A second essay on the
  ballistocardiogram.
\newblock \emph{Annals of Internal Medicine}, 63\penalty0 (6):\penalty0
  1079--1105, 1965.

\bibitem[Starr(1967)]{starr1967newtonian}
Isaac Starr.
\newblock Ballistocardiography: The place of the ballistocardiogram in a
  newtonian cardiology; and the new light it sheds on certain old clinical
  problems.
\newblock \emph{Proceedings of the Royal Society of Medicine}, 60\penalty0
  (12):\penalty0 1297--1306, 1967.

\bibitem[Kim et~al.(2016)Kim, Ober, McMurtry, Finegan, Inan, Mukkamala, and
  Hahn]{kim2016mechanism}
Chang-Sei Kim, Stephanie~L. Ober, M.~Sean McMurtry, Barry~A. Finegan, Omer~T.
  Inan, Ramakrishna Mukkamala, and Jin-Oh Hahn.
\newblock Ballistocardiogram: Mechanism and potential for unobtrusive
  cardiovascular health monitoring.
\newblock \emph{Scientific Reports}, 6:\penalty0 31297, 2016.

\bibitem[Starr and Schnabel(1954)]{starr1954genesis}
Isaac Starr and Truman~G. Schnabel, Jr.
\newblock Studies made by simulating systole at necropsy: {III}. on the genesis
  of the systolic waves of the ballistocardiogram.
\newblock \emph{Journal of Clinical Investigation}, 33\penalty0 (1):\penalty0
  10--22, 1954.

\bibitem[Rabineau et~al.(2021)Rabineau, Nonclercq, Leiner, van~de Borne,
  Migeotte, and Haut]{rabineau2021closedloop}
Jeremy Rabineau, Antoine Nonclercq, Tim Leiner, Philippe van~de Borne,
  Pierre-Francois Migeotte, and Benoit Haut.
\newblock Closed-loop multiscale computational model of human blood
  circulation: Applications to ballistocardiography.
\newblock \emph{Frontiers in Physiology}, 12:\penalty0 734311, 2021.

\bibitem[Guidoboni et~al.(2019)Guidoboni, Sala, Enayati, Sacco, Szopos, Keller,
  Popescu, Despins, Huxley, and Skubic]{guidoboni2019cardiovascular}
Giovanna Guidoboni, Lorenzo Sala, Moein Enayati, Riccardo Sacco, Marcela
  Szopos, James~M. Keller, Mihail Popescu, Laurel Despins, Virginia~H. Huxley,
  and Marjorie Skubic.
\newblock Cardiovascular function and ballistocardiogram: A relationship
  interpreted via mathematical modeling.
\newblock \emph{IEEE Transactions on Biomedical Engineering}, 66\penalty0
  (10):\penalty0 2906--2917, 2019.

\bibitem[Starr(1959)]{starr1959studies}
Isaac Starr.
\newblock Studies made by simulating systole at necropsy: {XII}. estimation of
  the initial cardiac forces from the ballistocardiogram.
\newblock \emph{Circulation}, 20\penalty0 (1):\penalty0 74--87, 1959.

\bibitem[Starr et~al.(1950)Starr, Horwitz, Mayock, and
  Krumbhaar]{starr1950standardization}
Isaac Starr, O.~Horwitz, R.~L. Mayock, and E.~B. Krumbhaar.
\newblock Standardization of the ballistocardiogram by simulation of the
  heart's function at necropsy; with a clinical method for the estimation of
  cardiac strength and normal standards for it.
\newblock \emph{Circulation}, 1\penalty0 (5):\penalty0 1073--1096, 1950.

\bibitem[Starr et~al.(1953)Starr, Schnabel, and Mayock]{starr1953peripheral}
Isaac Starr, T.~G. Schnabel, Jr., and R.~L. Mayock.
\newblock Studies made by simulating systole at necropsy: {II}. experiments on
  the relation of cardiac and peripheral factors to the genesis of the pulse
  wave and the ballistocardiogram.
\newblock \emph{Circulation}, 8\penalty0 (1):\penalty0 44--61, 1953.

\bibitem[Starr(1955)]{starr1955strokevolume}
Isaac Starr.
\newblock Studies made by simulating systole at necropsy: {VI}. estimation of
  cardiac stroke volume from the ballistocardiogram.
\newblock \emph{Journal of Applied Physiology}, 8\penalty0 (3):\penalty0
  315--329, 1955.

\bibitem[Starr et~al.(1954)Starr, Schnabel, Askovitz, and
  Schild]{starr1954pulsepressure}
Isaac Starr, T.~G. Schnabel, Jr., S.~I. Askovitz, and A.~Schild.
\newblock Studies made by simulating systole at necropsy: {IV}. on the relation
  between pulse pressure and cardiac stroke volume, leading to a clinical
  method of estimating cardiac output from blood pressure and age.
\newblock \emph{Circulation}, 9\penalty0 (5):\penalty0 648--663, 1954.

\bibitem[Starr(1958)]{starr1958dynamics}
Isaac Starr.
\newblock Studies made by simulating systole at necropsy: {XI}. on the higher
  dynamic functions of the heart, and their reflections in the pulse wave.
\newblock \emph{Circulation}, 17\penalty0 (4, Part 1):\penalty0 589--600, 1958.

\bibitem[Starr and Friedland(1946)]{starr1946respiratory}
Isaac Starr and Carl~K. Friedland.
\newblock On the cause of the respiratory variation of the ballistocardiogram,
  with a note on sinus arrhythmia.
\newblock \emph{Journal of Clinical Investigation}, 25\penalty0 (1):\penalty0
  53--64, 1946.

\bibitem[Starr and Hildreth(1952)]{starr1952aging}
Isaac Starr and E.~A. Hildreth.
\newblock The effect of aging and of the development of disease on the
  ballistocardiogram: A study of eighty subjects, originally healthy, followed
  from ten to fourteen years.
\newblock \emph{Circulation}, 5\penalty0 (4):\penalty0 481--495, 1952.

\bibitem[Starr and Wood(1961)]{starr1961twentyyear}
Isaac Starr and Francis~C. Wood.
\newblock Twenty-year studies with the ballistocardiograph: The relation
  between the amplitude of the first record of {"Healthy"} adults and eventual
  mortality and morbidity from heart disease.
\newblock \emph{Circulation}, 23\penalty0 (5):\penalty0 714--732, 1961.

\bibitem[Starr(1964)]{starr1964prognostic}
Isaac Starr.
\newblock Prognostic value of ballistocardiograms: Studies on evaluation of the
  doctor's experience.
\newblock \emph{JAMA}, 187\penalty0 (7):\penalty0 511--517, 1964.

\bibitem[Starr(1960)]{starr1960smoking}
Isaac Starr.
\newblock Significance of electrocardiographic and ballistocardiographic
  changes induced by smoking.
\newblock \emph{Annals of the New York Academy of Sciences}, 90\penalty0
  (1):\penalty0 199--215, 1960.

\bibitem[Zhan et~al.(2024)Zhan, Wu, Fu, Li, Deng, Wei, Zhang, Zhao, Li, Huang,
  Chen, Wang, Li, and Lu]{zhan2024cardiac}
Jing Zhan, Xiaoyan Wu, Xuelei Fu, Chenze Li, Ke-Qiong Deng, Qin Wei, Chao
  Zhang, Tao Zhao, Congcong Li, Longting Huang, Kewei Chen, Qiongxin Wang,
  Zhengying Li, and Zhibing Lu.
\newblock Non-contact assessment of cardiac physiology using {FO-MVSS}-based
  ballistocardiography: A promising approach for heart failure evaluation.
\newblock \emph{Scientific Reports}, 14\penalty0 (1):\penalty0 3269, 2024.
\newblock URL \url{https://www.nature.com/articles/s41598-024-53464-8}.

\bibitem[Herkert et~al.(2021)Herkert, Migeotte, Hossein, Spee, and
  Kemps]{herkert2021kinocardiograph}
Cyrille Herkert, Pierre-Fran{\c c}ois Migeotte, Amin Hossein, Rudolph~Ferdinand
  Spee, and Hareld Marijn~Clemens Kemps.
\newblock The kinocardiograph for assessment of changes in haemodynamic load in
  patients with chronic heart failure with reduced ejection fraction.
\newblock \emph{ESC Heart Failure}, 8\penalty0 (6):\penalty0 4925--4932, 2021.
\newblock URL \url{https://pmc.ncbi.nlm.nih.gov/articles/PMC8712789/}.

\bibitem[Solbiati et~al.(2021)Solbiati, Paglialonga, Costantini, {\v S}imuni{\v
  c}, Pi{\v s}ot, Narici, and Caiani]{solbiati2021bedrest}
Sarah Solbiati, Alessia Paglialonga, Lorenzo Costantini, Bo{\v s}tjan {\v
  S}imuni{\v c}, Rado Pi{\v s}ot, Marco~V. Narici, and Enrico~G. Caiani.
\newblock Assessing cardiac electro-mechanical deconditioning during bed rest
  using smartphone's inertial sensors.
\newblock In \emph{Computing in Cardiology}, volume~48, pages 1--4, 2021.
\newblock URL \url{https://www.cinc.org/archives/2021/pdf/CinC2021-191.pdf}.

\bibitem[S{\o}rensen et~al.(2017)S{\o}rensen, Verma, Blaber, Zanetti, Schmidt,
  Struijk, and Tavakolian]{sorensen2017challenges}
Kasper S{\o}rensen, Ajay~K. Verma, Andrew Blaber, John Zanetti, Samuel~Emil
  Schmidt, Johannes~J. Struijk, and Kouhyar Tavakolian.
\newblock Challenges in using seismocardiography for blood pressure monitoring.
\newblock In \emph{Computing in Cardiology}, volume~44, pages 1--4, 2017.
\newblock URL \url{https://www.cinc.org/archives/2017/pdf/026-452.pdf}.

\bibitem[Zhan et~al.(2025)]{zhan2025dataset}
Jing Zhan et~al.
\newblock A multi-pathology ballistocardiogram dataset for cardiac function
  monitoring and arrhythmia assessment.
\newblock \emph{Scientific Data}, 12:\penalty0 963, 2025.
\newblock URL \url{https://www.nature.com/articles/s41597-025-05287-z}.

\bibitem[Land et~al.(2017)Land, Park-Holohan, Smith, dos Remedios, Kentish, and
  Niederer]{land2017contraction}
Sander Land, So-Jin Park-Holohan, Nicolas~P. Smith, Cristobal~G. dos Remedios,
  Jonathan~C. Kentish, and Steven~A. Niederer.
\newblock A model of cardiac contraction based on novel measurements of tension
  development in human cardiomyocytes.
\newblock \emph{Journal of Molecular and Cellular Cardiology}, 106:\penalty0
  68--83, 2017.

\bibitem[Arts et~al.(1991)Arts, Bovendeerd, Prinzen, and
  Reneman]{arts1991relation}
T.~Arts, P.~H. Bovendeerd, F.~W. Prinzen, and R.~S. Reneman.
\newblock Relation between left ventricular cavity pressure and volume and
  systolic fiber stress and strain in the wall.
\newblock \emph{Biophysical Journal}, 59\penalty0 (1):\penalty0 93--102, 1991.
\newblock URL \url{https://pubmed.ncbi.nlm.nih.gov/2015392/}.

\bibitem[Guccione et~al.(1991)Guccione, McCulloch, and
  Waldman]{guccione1991passive}
J.~M. Guccione, A.~D. McCulloch, and L.~K. Waldman.
\newblock Passive material properties of intact ventricular myocardium
  determined from a cylindrical model.
\newblock \emph{Journal of Biomechanical Engineering}, 113\penalty0
  (1):\penalty0 42--55, 1991.
\newblock URL \url{https://pubmed.ncbi.nlm.nih.gov/2020175/}.

\bibitem[Gurev et~al.(2012)Gurev, Tavakolian, Constantino, Kaminska, Blaber,
  and Trayanova]{gurev2012mechanisms}
Viatcheslav Gurev, Kouhyar Tavakolian, Jason Constantino, Bozena Kaminska,
  Andrew~P. Blaber, and Natalia~A. Trayanova.
\newblock Mechanisms underlying isovolumic contraction and ejection peaks in
  seismocardiogram morphology.
\newblock \emph{Journal of Medical and Biological Engineering}, 32\penalty0
  (2):\penalty0 103--110, 2012.
\newblock URL \url{https://pmc.ncbi.nlm.nih.gov/articles/PMC3480230/}.

\bibitem[Bayer et~al.(2012)Bayer, Blake, Plank, and Trayanova]{bayer2012fibers}
J.~D. Bayer, R.~C. Blake, G.~Plank, and N.~A. Trayanova.
\newblock A novel rule-based algorithm for assigning myocardial fiber
  orientation to computational heart models.
\newblock \emph{Annals of Biomedical Engineering}, 40\penalty0 (10):\penalty0
  2243--2254, May 2012.

\bibitem[Sahli~Costabal et~al.(2016)Sahli~Costabal, Hurtado, and
  Kuhl]{sahli2016purkinje}
Francisco Sahli~Costabal, Daniel~E. Hurtado, and Ellen Kuhl.
\newblock Generating purkinje networks in the human heart.
\newblock \emph{Journal of Biomechanics}, 49\penalty0 (12):\penalty0
  2455--2465, August 2016.

\bibitem[Bonet and Burton(1998)]{bonet1998anp}
J.~Bonet and A.~J. Burton.
\newblock A simple average nodal pressure tetrahedral element for
  incompressible and nearly incompressible dynamic explicit applications.
\newblock \emph{Communications in Numerical Methods in Engineering},
  14\penalty0 (5):\penalty0 437--449, 1998.

\bibitem[Augustin et~al.(2021)Augustin, Gsell, Karabelas, Willemen, Prinzen,
  Lumens, Vigmond, and Plank]{augustin2021closedloop}
Christoph~M. Augustin, Matthias A.~F. Gsell, Elias Karabelas, Erik Willemen,
  Frits~W. Prinzen, Joost Lumens, Edward~J. Vigmond, and Gernot Plank.
\newblock A computationally efficient physiologically comprehensive 3d--0d
  closed-loop model of the heart and circulation.
\newblock \emph{Computer Methods in Applied Mechanics and Engineering},
  386:\penalty0 114092, 2021.

\end{thebibliography}
\end{document}